\documentclass[twocolumn,showpacs,prb,superscriptaddress,floatfix]{revtex4}
\usepackage{graphicx,dcolumn,bm,amsmath,amssymb,latexsym,float}
\renewcommand{\theequation}{\arabic{section}.\arabic{equation}}
\begin{document}
\newcommand{\be}{\begin{equation}}
\newcommand{\ee}{\end{equation}}
\newcommand{\bea}{\begin{eqnarray}}
\newcommand{\eea}{\end{eqnarray}}
\newcommand{\br}{{\bf r}}
\newcommand{\bv}{{\bf v}}
\newcommand{\bq}{{\bf q}}
\newcommand{\bk}{{\bf k}}

\title{
Effect of gauge-field interaction on fermion transport in 2D:\\
Hartree conductivity correction and dephasing
}

\author{T.~Ludwig}
\affiliation{Instituut-Lorentz, Universiteit Leiden, P.O.~Box 9506,
 2300 RA Leiden, The Netherlands
}

\author{I.~V.~Gornyi}
\altaffiliation{
 Also at A.F.~Ioffe Physico-Technical Institute,
 194021 St.~Petersburg, Russia.
}
\affiliation{
 Institut f\"ur Nanotechnologie, Forschungszentrum Karlsruhe,
 76021 Karlsruhe, Germany
}
\affiliation{
Center for Functional Nanostructures,
Universit\"at Karlsruhe, 76128 Karlsruhe, Germany
}

\author{A.~D.~Mirlin}
\altaffiliation{
 Also at Petersburg Nuclear Physics Institute,
 188300 St.~Petersburg, Russia.
}
\affiliation{
 Institut f\"ur Nanotechnologie, Forschungszentrum Karlsruhe,
 76021 Karlsruhe, Germany
}
\affiliation{
 \mbox{Institut f\"ur Theorie der kondensierten Materie,
 Universit\"at Karlsruhe, 76128 Karlsruhe, Germany}
}
\affiliation{
Center for Functional Nanostructures,
Universit\"at Karlsruhe, 76128 Karlsruhe, Germany
}

\author{P.~W\"olf\/le}
\affiliation{
 \mbox{Institut f\"ur Theorie der kondensierten Materie,
 Universit\"at Karlsruhe, 76128 Karlsruhe, Germany}
}
\affiliation{
Center for Functional Nanostructures,
Universit\"at Karlsruhe, 76128 Karlsruhe, Germany
}
\affiliation{
 Institut f\"ur Nanotechnologie, Forschungszentrum Karlsruhe,
 76021 Karlsruhe, Germany
}

\date{\today}

\begin{abstract}

We consider the quantum corrections to the conductivity of fermions interacting via a Chern-Simons gauge field, and concentrate on the Hartree-type contributions. The first-order Hartree approximation is only valid in the limit of weak coupling $\lambda\ll g^{-1/2}$ to the gauge field ($g\gg 1$ is the dimensionless conductance), and results in an antilocalizing conductivity correction \mbox{$\sim \lambda^2g\,{\rm ln}^2T$}. 
In the case of strong coupling, an infinite summation of higher-order terms is necessary, including both the virtual
(renormalization of the frequency) and real (dephasing) processes. At intermediate temperatures, \mbox{$T_0\ll T\ll gT_0$}, where \mbox{$T_0\sim 1/g^2\tau$} and $\tau$ is the elastic scattering time, the $T$-dependence of the conductivity is determined by the Hartree correction, \mbox{$\delta\sigma^H(T)-\delta\sigma^H(g T_0)\propto g^{1/2}-(T/T_0)^{1/2}[1+\ln{(gT_0/T)^{1/2}}]$}, so that $\sigma(T)$ increases with lowering $T$. At low temperatures, \mbox{$T\ll T_0$}, the temperature-dependent part of the Hartree correction assumes a logarithmic form with a coefficient of order unity,
\mbox{$\delta\sigma^H\propto\ln{(1/T)}$}. As a result, the negative exchange contribution \mbox{$\delta\sigma^{\rm ex}\propto-\ln{g}\ln{(1/T)}$} becomes dominant, yielding localization in the limit of \mbox{$T\to 0$}. We further discuss dephasing at strong coupling and show that the dephasing rates are of the order of $T$, owing to the interplay of inelastic scattering and renormalization. On the other hand, the dephasing length is anomalously short,
$L_\varphi\ll L_T$, where $L_T$ is the thermal length.
For the case of composite fermions with long-range Coulomb interaction, the gauge field propagator is less singular. The resulting Hartree correction has the usual sign and temperature-dependence, \mbox{$\delta\sigma^H\propto\ln g\,\ln{(1/T)}$}, and for realistic $g$ is overcompensated by the negative exchange contribution due to the gauge-boson and scalar parts of the interaction. In this case, the dephasing length $L_\varphi$ is of the order of $L_T$
for not too low temperatures and exceeds $L_T$ for $T\lesssim g T_0$.

\end{abstract}


\maketitle

\section{Introduction}
\label{intro}
\setcounter{equation}{0}

The problem of particles interacting with a transverse gauge field was first considered \cite{reizer} in the context of the magnetic interaction of electrons in metals. It was found that such interactions lead to singular contributions to observables, since they are not screened, in contrast to the conventional interaction via a scalar potential. However, for the case of magnetic interactions of the electrodynamic origin, these effects are weak, since they are of relativistic nature. More recently, a two-dimensional (2D) version of the problem has attracted  considerable interest \cite{gauge} in connection with effective theories of strongly correlated systems, where gauge field interactions lead to very strong effects: the gauge theory of high-$T_c$ superconductors \cite{htsc} and, most prominently, the Chern-Simons theory of the half-filled Landau level.

In a field-theoretical description\cite{Lopez_Fradkin_91} of 2D electrons in a strong magnetic field at half-filling of the lowest Landau level, electrons undergo a statistical transformation which transforms them into so-called composite fermions by effectively attaching two flux quanta to each electron\cite{Jain_89}. As a result, the composite fermions interact strongly with a (fictitious) Chern-Simons gauge field. Although this gauge field vanishes on average at half filling, the density fluctuations of the electrons induce fluctuations of the gauge field. A treatment of these fluctuations has been developed in Ref.~\onlinecite{HLR_93} (for reviews see e.g.~Refs.\onlinecite{Lopez_Fradkin_review_97,Simon_review_98}).

Due to the strong coupling of the fermions to the gauge field and the singular properties of the gauge field, the interaction effects can be much stronger and more complex than for Coulomb interaction. For previous work in this context, the reader is referred to Refs.~\onlinecite{Aronov_Mirlin_Woelfle_93,Aronov_Woelfle_PRL_94,Aronov_Woelfle_94,Aronov_Altshuler_Mirlin_Woelfle_EPL_95,Aronov_Altshuler_Mirlin_Woelfle_PRB_95,Stern_Halperin_95,Mirlin_Altshuler_Woelfle_95,Lee_Mucciolo_Smith_96,Khveshchenko_96,Mirlin_Woelfle_97,Woelfle_2000,TL_diss_06}. In particular, Refs.~\onlinecite{Aronov_Woelfle_PRL_94,Aronov_Woelfle_94,Lee_Mucciolo_Smith_96,Woelfle_2000,TL_diss_06} have addressed questions related to dephasing phenomena, finding unusually high dephasing rates, while Refs.~\onlinecite{Mirlin_Woelfle_97,Khveshchenko_96} have considered the conductivity correction due to exchange interaction for such systems, predicting a negative correction to the conductivity varying as \mbox{$\ln{T}$} at low temperatures with a non-universal prefactor logarithmically dependent on the resistivity. The experimental observation of such a correction has been reported in Ref.~\onlinecite{Rokhinson_Su_Goldman_95}.

A new boost to the research in this direction was given by a recent work, Ref.~\onlinecite{Galitski_05}. It was found there that the {\it positive} Hartree contribution to the quantum corrections to the density of states and to the conductivity dominates over the exchange contribution, therefore letting the system remain metallic at low temperatures. Most surprisingly, the Hartree contribution in Ref.~\onlinecite{Galitski_05} diverges in the limit of large systems, \mbox{$L\to\infty$}. If true, this would imply that the conductivity of such a system (in the thermodynamic limit) is infinite for sufficiently low temperatures. The very interesting and partly puzzling findings of Ref.~\onlinecite{Galitski_05} have served as one of the motivations of this work.

This paper presents a systematic analysis of the Hartree correction to the conductivity of a disordered fermion-gauge field system. We start with a calculation of the first-order Hartree correction to conductivity in Section~\ref{Hartree_wo_dephasing}. In Section~\ref{first_order} we derive an effective interaction, which helps us to bring the considered contribution into a form similar to the usual exchange correction. At variance with Ref.~\onlinecite{Galitski_05}, we find a natural low-momentum cutoff set by the diffusive dynamics, which ensures that gauge invariance is obeyed. This leads to a result for the conductivity correction which is finite in the thermodynamic limit and positive, varying as \mbox{$\ln^2T$} with temperature. In Section~\ref{Hartree_by_current_corr} we elucidate the physical meaning of the obtained contribution. We show that it is governed by scattering on static mesoscopic fluctuations of local currents. To demonstrate this, we rederive the gauge-field-induced  correction to the conductivity by using an earlier result for the correlation function of local mesoscopic currents~\cite{Gornyi_Mirlin_Woelfle_01}.

When the interaction coupling constant $\lambda$ is not too small (as e.g.~in the composite-fermion problem, where $\lambda\sim 1$), it is necessary to include higher orders of the interaction. Since the gauge-field interaction leads not only to renormalization but also to anomalously strong dephasing effects and since renormalization and dephasing get mixed in higher orders, we first discuss dephasing of Cooperons and diffusons coupled to a fluctuating gauge field.

In Section~\ref{Cooperon_dephasing}, we discuss the effect of dephasing on weak localization and find very short dephasing lengths, confirming earlier work\cite{Aronov_Woelfle_PRL_94,Aronov_Woelfle_94,Lee_Mucciolo_Smith_96,Woelfle_2000}.
The physics of this strong dephasing, dominated by quasistatic gauge-field configurations is also discussed there. We show a deep relation between dephasing of weak localization and mesoscopic conductance fluctuations in Section~\ref{ucf_dephasing}\cite{TL_diss_06}. Based on these results, the dephasing of diffusons with finite delay times (which arise as elements of diagrams for the interaction-induced conductivity correction) is inspected in Section~\ref{diffuson_dephasing}. In Sec.~\ref{2-loop-dephasing} we discuss the ``true'' dephasing rate governed by inelastic processes (rather than by ensemble averaging) and showing up in the two-loop weak localization correction.

Using the results for dephasing of diffusons, we then construct a scheme to treat interaction effects to all orders. This starts with the treatment of large self-energies \mbox{$\Sigma^Z\sim g\omega$}, which we present in Section~\ref{low_T}. At low temperatures, \mbox{$T\ll T_0$} (where \mbox{$T_0\sim 1/g^2\tau$}, $g$ is the dimensionless conductance, and $\tau$ is the elastic scattering time) dephasing is not important and the strong renormalization effects lead to a low-frequency Hartree correction which is logarithmic in temperature with a coefficient of order unity,
\be
\delta\sigma^H(T)\propto\ln{(1/T)}\:\:.
\ee
This is accompanied by a high-frequency contribution which saturates to a constant (and is smaller than the Drude conductivity).
As a result, the system of disordered fermions that interact through the gauge fields, while showing metallic-like behavior at sufficiently high $T$, eventually gets localized in the limit of lowest temperatures due to the negative exchange contribution finally overcompensating the Hartree contribution.

At intermediate temperatures \mbox{$T_0\ll T\ll gT_0$} (for \mbox{$\lambda\sim 1$}, higher temperatures are outside the diffusive regime, since the dephasing length $L_\varphi$ becomes shorter than the mean free path $l$), dephasing and renormalization effects are both present, and special care is needed to evaluate the Hartree contribution. We develop a proper method in Section~\ref{high_T}. The Hartree correction assumes the form
\be
\delta\sigma^H(T)-\delta\sigma^H(g T_0)
\propto
g^{1/2}-\frac{T^{1/2}}{T_0^{1/2}}\left[1+\frac{1}{2} \ln{\frac{gT_0}{T}}\right],
\ee
with the temperature dependence resulting from strong dephasing removing the contribution of low frequencies.

Taking into account the influence of the renormalization processes upon dephasing at strong coupling, we show in Section~\ref{dephasing-strong-coupling} that for \mbox{$\lambda=1$} the dephasing rates are of the order of $T$. The renormalization of the frequency by virtual processes (inducing a strong $Z$-factor, $Z\sim g$) compensates the large factor of $g$ in the dephasing part of the self-energy. On the other hand, the dephasing length is anomalously short compared to the thermal length, \mbox{$L_\varphi\ll L_T$}.

Finally, in Section~\ref{Coulomb} we turn to the model of composite fermions with an unscreened long-range Coulomb interaction. This suppresses charge fluctuations and leads\cite{HLR_93} to a less singular propagator of the gauge-field (which is induced by the density fluctuations via the Chern-Simons transformation). In this situation, the effect of the gauge-field interaction is much less dramatic: the large parameter $g$ does not appear in the perturbative expressions for the dephasing rate as well as the first-order Hartree correction. As a result, a formalism beyond first order (the resummation of  higher-order gauge-field interaction terms) is not needed for realistic experimental parameters.
Specifically, in Section~\ref{Coulomb_dephasing} we find at not too low temperatures that the dephasing rate is of the order of the temperature and \mbox{$L_\varphi\sim L_T$}, while at the lowest temperatures, \mbox{$L_\varphi\gg L_T$}.
Likewise, in Section~\ref{Coulomb_Hartree} we find that the Hartree correction is positive and has the usual $T$-dependence,
\be
\delta\sigma^H\propto\ln g\,\ln{(1/T)}\:\:,
\ee
with a small numerical prefactor. For realistic $g$, the total interaction correction is dominated by the gauge-field exchange contribution (Ref.~\onlinecite{Mirlin_Woelfle_97}) at intermediate temperatures, and by the scalar part of the interaction (Ref.~\onlinecite{Altshuler_Aronov_85})
at the lowest $T$.

Our results are summarized in Section~\ref{summary}. Technical details are relegated to several appendices. Throughout the paper we set \mbox{$\hbar=1$}.

\section{Small coupling: first-order Hartree correction}
\label{Hartree_wo_dephasing}
\setcounter{equation}{0}

\subsection{First-order Hartree diagrams}
\label{first_order}

We consider a diffusive system of fermions in two dimensions which interact with a gauge field described by the transverse propagator
\be
U_{\alpha\beta}(\bk,\epsilon)=
\frac{1}{\chi_0 k^2-i\sigma(k)\epsilon}
\left[
\delta_{\alpha\beta}-\frac{k_\alpha k_\beta}{k^2}
\right],
\label{propfull}
\ee
where $\sigma(k)$ is the electric conductivity at finite wavevector $\bk$ and $\chi_0$ is the magnetic susceptibility of the electrons.
At small ``diffusive'' momenta $k\ll 1/l$ (where $l$ is the elastic mean free path), the propagator takes the form
\be
U_{\alpha\beta}(\bk,\epsilon)\simeq
\frac{1}{\sigma_0}\:
\frac{1}{k^2l^2T_0-i\epsilon}
\:\delta_{\alpha\beta}^\perp\:\:, \quad kl\ll 1\:\:.
\label{bare_gaugefield}
\ee
Here \mbox{$\sigma_0=e^2\nu D$} is the Drude conductivity, $D$ is the diffusion constant,  \mbox{$\nu=m/2\pi$} is the density of states per spin (we do not account for the spin degree of freedom, as appropriate for the fully polarized lowest Landau level), and we have introduced the short notation $$\delta_{\alpha\beta}^\perp\equiv\delta_{\alpha\beta}-k_\alpha k_\beta/k^2$$
for the transverse projector.
Equation~(\ref{bare_gaugefield}) has been written to display the characteristic temperature scale
\be
T_0=\frac{1}{12 g^2\tau}\:\:,
\label{T0}
\ee
where we have used the free-fermion susceptibility \mbox{$\chi_0=e^2/12\pi m$} resulting in \mbox{$e^2D/\chi_0=12\pi g$}, and \mbox{$g=2\pi\sigma_0/e^2=E_F\tau=k_Fl/2$} is the dimensionless conductance. For ease of notation we also define
\be
T_n\equiv g^nT_0\:\:.
\ee
The propagator (\ref{bare_gaugefield}) corresponds to a short-range interaction (Coulomb interaction screened by, say, an external gate) of the electrons before the transformation into composite fermions. At the end of the paper, in Section~\ref{Coulomb}, we will investigate the case of unscreened Coulomb interaction.

\begin{figure}

\hspace*{5mm}
\includegraphics[width=0.4\linewidth]{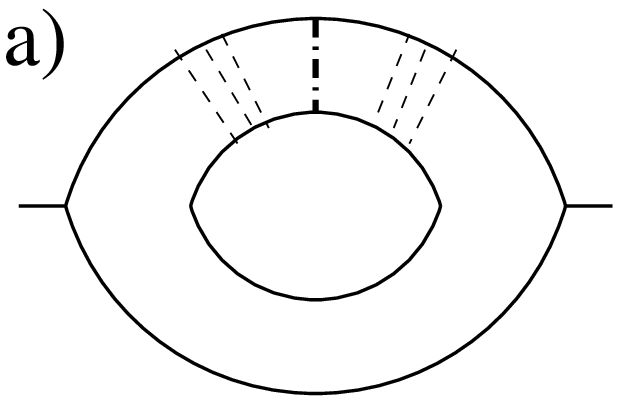}
\hfill
\includegraphics[width=0.4\linewidth]{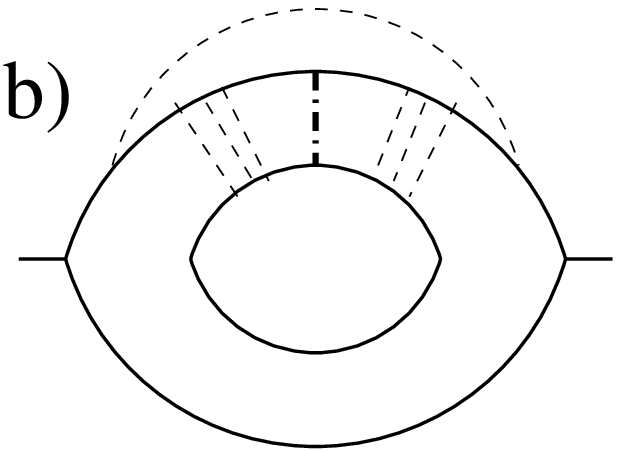}
\hspace*{5mm}

\vspace*{5mm}

\includegraphics[width=0.8\linewidth]{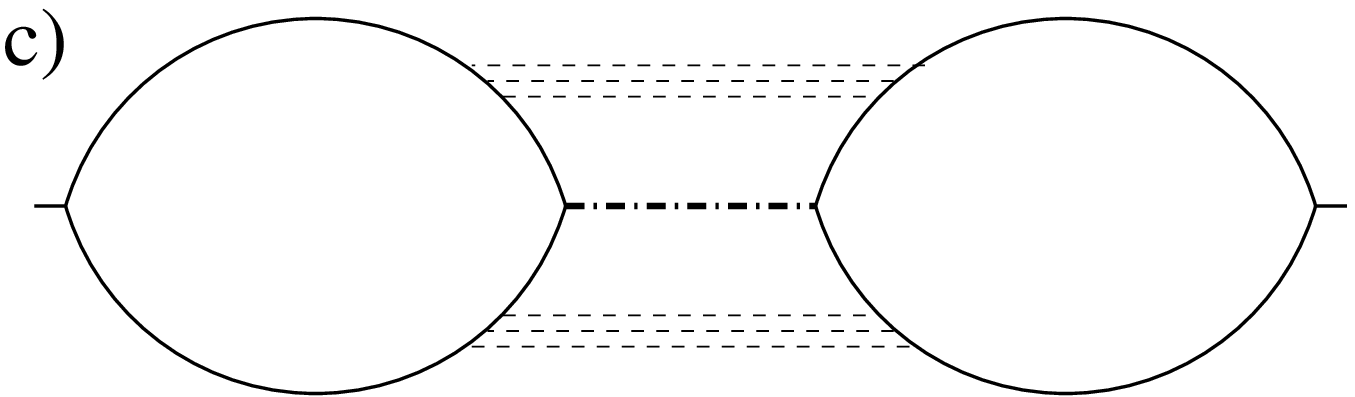}

\vspace*{5mm}

\includegraphics[width=0.4\linewidth]{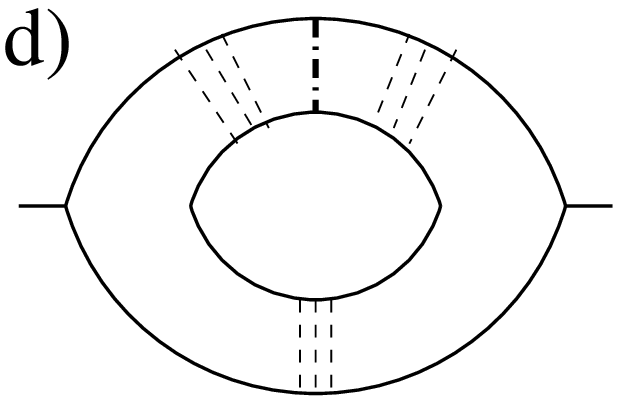}\\

\vspace*{5mm}

\includegraphics[width=0.8\linewidth]{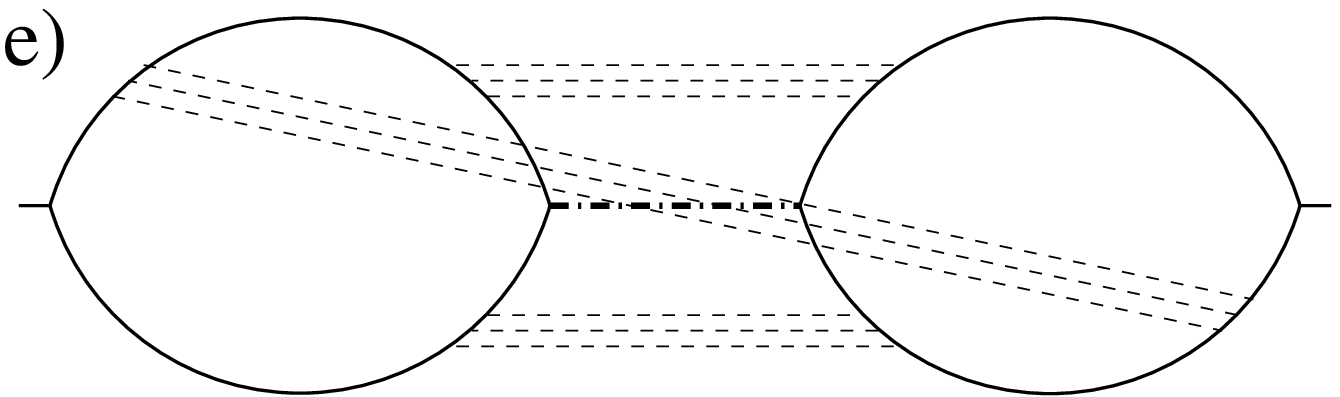}
\caption{\label{Hartree-diags} Diagrams contributing to the Hartree part of the conductivity correction.
Dashed lines represent impurities; the dot-dashed line denotes the bare gauge field propagator $U_{\alpha\beta}$ given by Eq.~(\ref{propfull})
Additionally, there is the possibility of diffusons crossing the gauge field line, as shown in Fig.~\ref{boxes} and discussed in the text.
}
\end{figure}

The vertices coupling the gauge field to the fermions carry factors \mbox{$e^*\bv$}, where at first we allow the coupling constant $e^*$ to be different from the electron charge $e$. This allows us to construct a well-controlled perturbation theory with a small parameter \mbox{$\lambda=e^*/e$}, although the results are not small in the usual parameter $1/g$.
We will set the parameter $\lambda$ to unity in Sections~\ref{low_T} and \ref{high_T}.

\begin{figure*}
\begin{LARGE} ${\displaystyle \tilde{U}(q)\:=\:\:}$\end{LARGE}
\parbox{0.25\linewidth}{\includegraphics[width=1.0\linewidth]{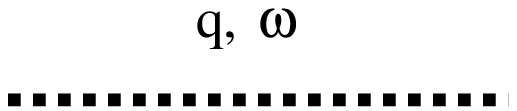}
}
\hfill
\\
\vspace*{5mm}
\begin{center}
\begin{LARGE}$=$   \end{LARGE}
\hfill
\parbox{0.25\linewidth}{
\includegraphics[width=1.0\linewidth]{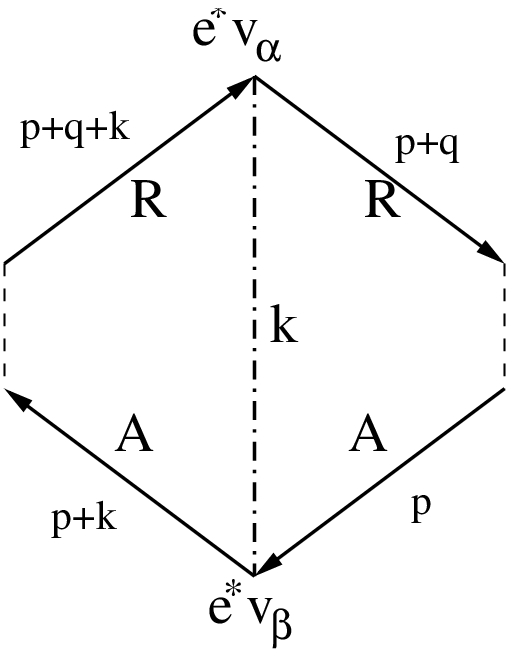}
}
\hfill
\begin{LARGE}$+$\end{LARGE}
\hfill
\parbox{0.25\linewidth}{
\includegraphics[width=1.0\linewidth]{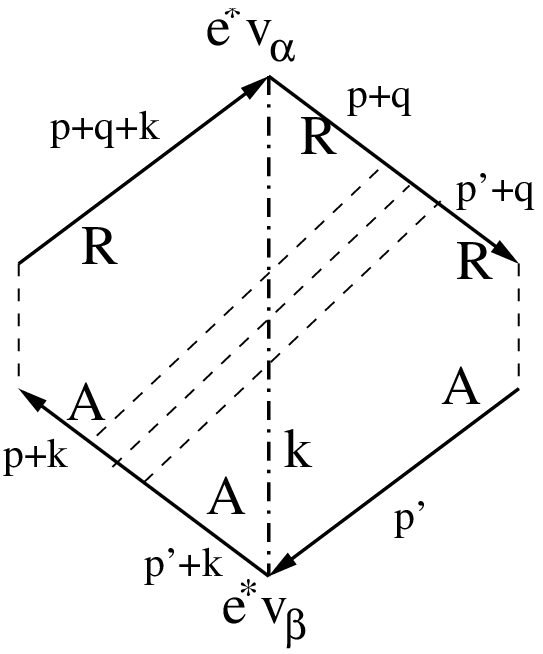}
}
\hfill
\begin{LARGE}$+$\end{LARGE}
\hfill
\parbox{0.25\linewidth}{
\includegraphics[width=1.0\linewidth]{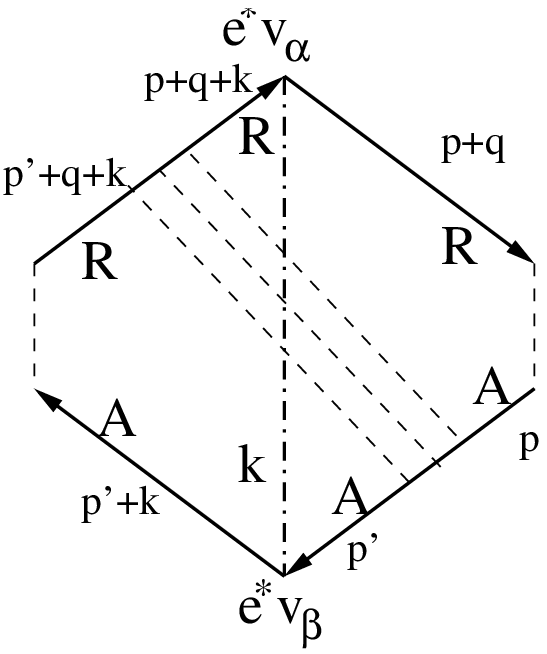}
}
\end{center}
\caption{\label{boxes} Diagrams contributing to the effective interaction $\tilde{U}(q)$. The dot-dashed line denotes the bare gauge field propagator $U_{\alpha\beta}$. The two possibilities with the diffuson crossing the interaction line provide a natural low-$k$ cutoff.
}
\end{figure*}

\begin{figure}

\includegraphics[width=0.4\linewidth]{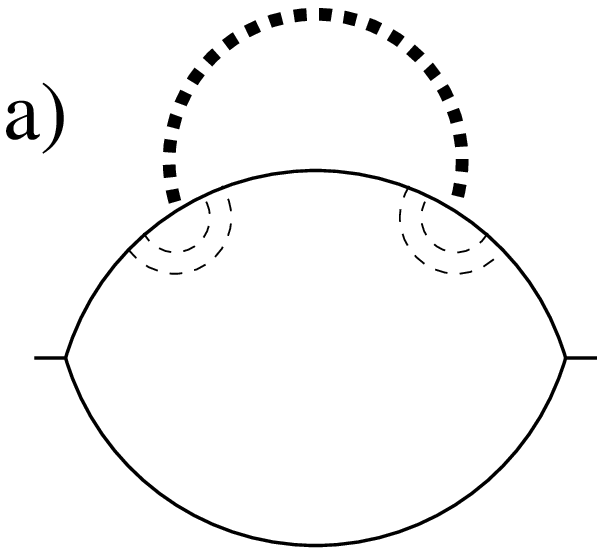}
\hfill
\includegraphics[width=0.4\linewidth]{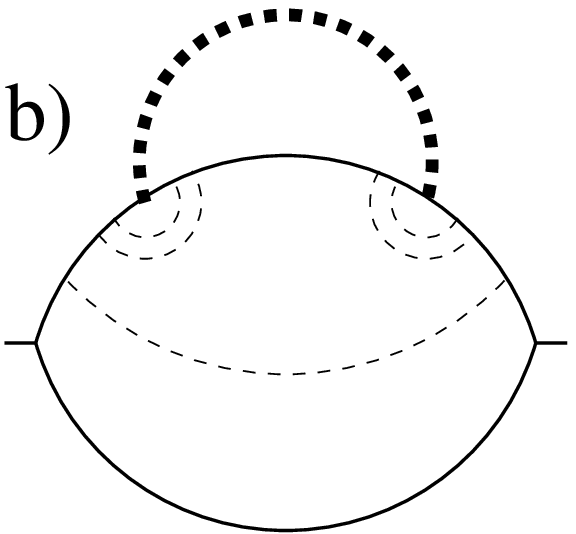}

\includegraphics[width=0.4\linewidth]{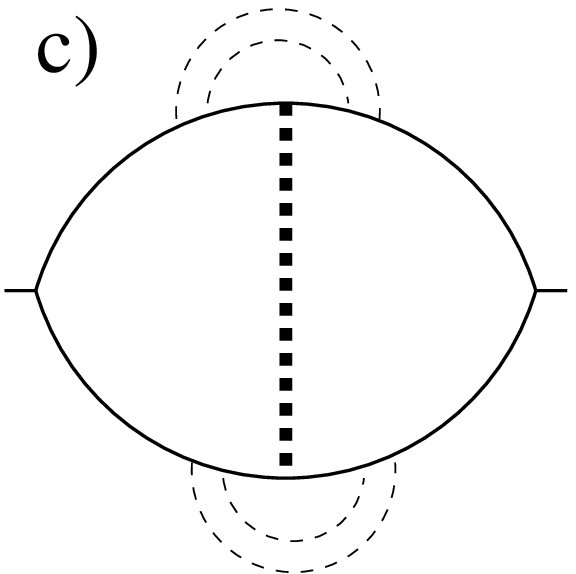}

\hfill
\includegraphics[width=0.4\linewidth]{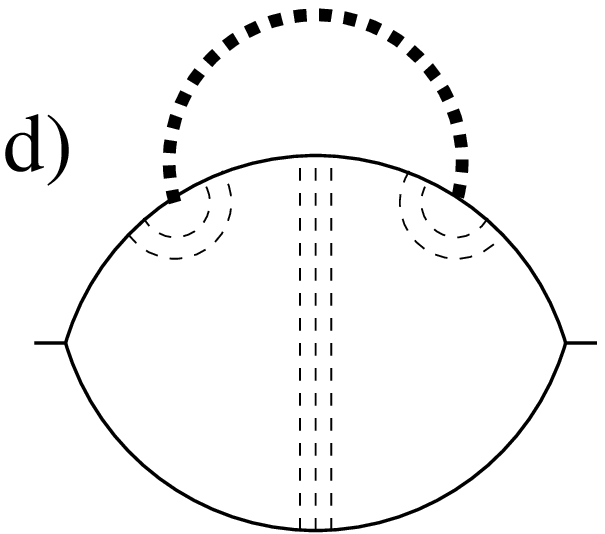}
\hfill
\includegraphics[width=0.4\linewidth]{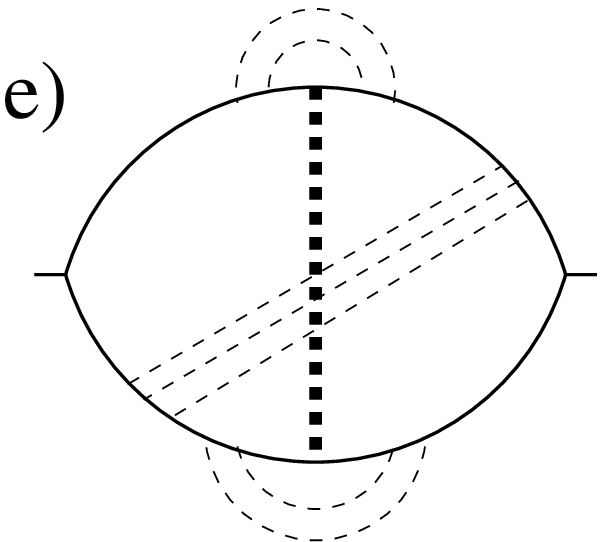}
\hfill
\caption{\label{AA-diags} Using the effective interaction $\tilde{U}$ as defined in Fig.~\ref{boxes} (thick dotted line), the Hartree diagrams of Fig.~\ref{Hartree-diags} can be mapped onto the standard exchange diagrams of Ref.~\onlinecite{Altshuler_Aronov_85}.
}
\end{figure}

We will concentrate on systems with broken time-reversal symmetry. This is in particular the case for the half-filled lowest Landau level, where the external magnetic field and random potential of impurities induce, after the Chern-Simons transformation, a random magnetic field as the dominant disorder for composite fermions.

We are interested in the Hartree contribution to the conductivity correction, which to leading order in the fermion-gauge field coupling is given by the diagrams shown in Fig.~\ref{Hartree-diags}. We consider the diffusive regime \mbox{$T\tau\ll 1$} and do not treat the details of the diffusive-ballistic crossover, which involve some extra diagrams\cite{ZNA,Gornyi_Mirlin_04}.

Defining the effective interaction $\tilde{U}$ as shown in Fig~\ref{boxes} (see Appendix~\ref{app-eff-int}), the Hartree diagrams with respect to the bare interaction $U_{\alpha\beta}$ can be written as exchange diagrams\cite{Galitski_05} with respect to $\tilde{U}$, see Figs.~\ref{Hartree-diags} and \ref{AA-diags}. The two diagrams of Fig.~\ref{boxes} with the diffuson crossing the interaction line cancel the bare box at \mbox{$k\ll q$} and are negligible at \mbox{$k\gg q$}. They thus provide a natural lower cutoff for the gauge field momenta $k$, which has been missed in Ref.~\onlinecite{Galitski_05}.
In Appendix~\ref{app-extra-diags} we derive the diagrams for the Hartree conductivity correction using a generating functional and
show that all relevant diagrams involve the effective interaction block $\tilde{U}$ as given by Fig.~\ref{boxes}.
It is also shown there that to the leading order
the Hartree conductivity correction can be equivalently represented either as a sum of diagrams \ref{Hartree-diags}$a)$+\ref{Hartree-diags}$b)$
or  a sum of diagrams \ref{Hartree-diags}$d)$+\ref{Hartree-diags}$e)$, as in the case of conventional Coulomb interaction~\cite{Altshuler_Aronov_85}.

The effective interaction $\tilde{U}$ is evaluated in Appendix~\ref{app-eff-int}, with the result
\be
\tilde{U}(q)
\equiv\tilde{U}(q,\epsilon=0)
=
\frac{3 g \lambda^2}{\pi \nu} \ln{\frac{1}{q^2l^2}}\:\:.
\label{tilde_U}
\ee
It is instructive to compare the effective interaction block (\ref{tilde_U}) with the dynamically screened Coulomb interaction $U_{C}(q,\omega)$
in the conventional interaction correction~\cite{Altshuler_Aronov_85}:
\be
U_{C}(q,\omega)=\frac{1}{2\nu} \frac{Dq^2-i\omega}{Dq^2}.
\label{CoulRPA}
\ee
The main difference is the appearance of the parameter \mbox{$g\lambda^2$} in the prefactor, which makes the effective gauge-field interaction block \mbox{$\nu \tilde{U}(q)\sim g\gg 1$} very strong in the realistic case \mbox{$\lambda=1$}. Furthermore, for characteristic values of diffuson momenta and frequencies $Dq^2\sim\omega$, the interaction $\tilde{U}(q)$ diverges logarithmically with decreasing $q$, while the screened Coulomb interaction can be replaced by a constant. Note that the gauge invariance ensures that the $q^{-2}$ singularity of $U_{C}(q,\omega)$ at fixed $\omega$ does not lead to anomalies in the gauge-invariant quantities like the conductivity, see Refs.~\onlinecite{Finkelstein,Kamenev,ZNA,Gornyi_Mirlin_04,Adamov} for discussion.

Using the standard expression for the first-order exchange diagrams\cite{Altshuler_Aronov_85}
\ref{AA-diags}$d$)+\ref{AA-diags}$e$)
\bea
\delta\sigma^H
&=&
2\sigma_0
\int\limits\frac{d\omega}{2\pi}\:
\frac{\partial}{\partial\omega}\left[\omega\,{\rm coth}\frac{\omega}{2T}\right]\nonumber\\
& &\times\:
\int(dq)\:{\rm Im}\left\{\tilde{U}(q)\frac{Dq^2}{\left(Dq^2-i\omega\right)^3}\right\}
\label{delta_sigma_raw}
\eea
(we use the compact notation \mbox{$\int(dq)\equiv\int d^2q/(2\pi)^2$}) and the effective interaction $\tilde{U}(q)$ given by Eq.~(\ref{tilde_U}), we find the positive conductivity correction
\bea
\delta\sigma^H
&\simeq&
\frac{3g\left(e^*\right)^2}{4\pi^3}\:
\int\limits_0^{1/\tau}\frac{d\omega}{\omega}\:
\frac{\partial}{\partial\omega}\left[\omega\,{\rm coth}\frac{\omega}{2T}\right]
\ln{\frac{1}{\omega\tau}}\nonumber\\
&=&
\frac{3}{4\pi^2}\:\lambda^2\:\sigma_0\:{\rm ln}^2\frac{1}{T\tau}\:\:.
\label{delta_sigma}
\eea
The correction is proportional to the parameter \mbox{$\lambda^2 g$}, as expected from the comparison of Eqs.~(\ref{tilde_U}) and (\ref{CoulRPA}). The stronger ($\ln^2 T$) temperature dependence as compared to the standard Altshuler-Aronov interaction correction\cite{Altshuler_Aronov_85} (which is proportional to $\ln T$) arises due to the logarithmic infrared singularity of the effective interaction block (\ref{tilde_U}).
The overall sign is the result of including an additional minus sign relative to the standard formula for the exchange correction, due to the closed fermionic loop of the Hartree diagram.
In Section~\ref{Hartree_by_current_corr} we will present another derivation reproducing Eq.~(\ref{delta_sigma}).

The velocity factors at the interaction vertices introduce the factor $g$ in the effective interaction (\ref{tilde_U}) which compensates the usual factor $1/g$, so that Eq.~(\ref{delta_sigma}) is small compared to $\sigma_0$ only through the parameter $\lambda$. In the course of this paper, we will therefore develop more careful treatments beyond first order in the interaction, in order to calculate the Hartree conductivity correction in the situation \mbox{$\lambda=1$} relevant for the half-filled lowest Landau level.

Even though the relative Hartree conductivity correction $\delta\sigma^H/\sigma_0$ is not small in $1/g$, Eq.~(\ref{delta_sigma}) does not diverge with the system size, at variance with the results found in Ref.~\onlinecite{Galitski_05}. This is because small gauge field momenta \mbox{$k\ll q$} are cancelled (see Fig.~\ref{boxes} and Appendix~\ref{app-eff-int} for details), so that a static uniform gauge field does not contribute to the correct effective interaction and gauge invariance requirements~\cite{Aronov_Mirlin_Woelfle_93,Aronov_Altshuler_Mirlin_Woelfle_EPL_95,Aronov_Altshuler_Mirlin_Woelfle_PRB_95, Gornyi_Mirlin_Woelfle_01} are satisfied.

\subsection{Alternative derivation of the Hartree correction from mesoscopic current fluctuations}
\label{Hartree_by_current_corr}

\begin{figure}
\includegraphics[width=0.8\linewidth]{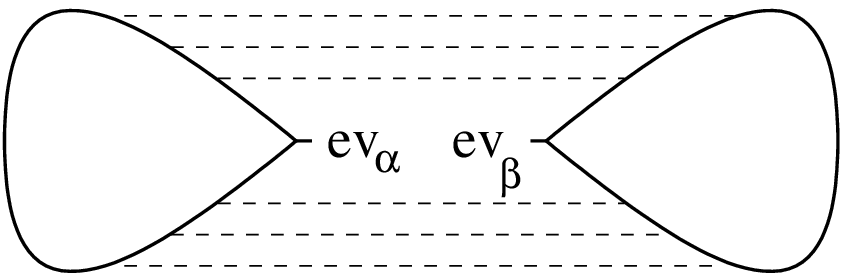}
\vspace*{5mm}

\includegraphics[width=0.8\linewidth]{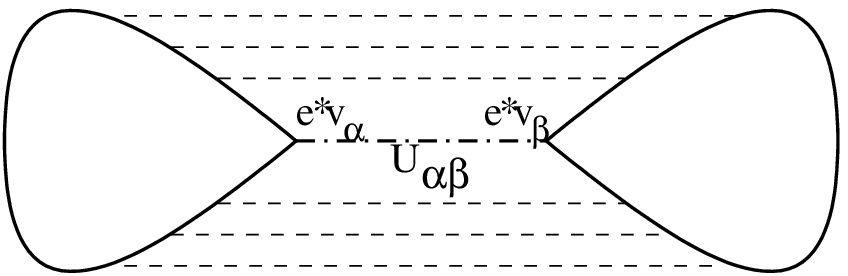}
\caption{\label{current-corr-fig} Upper part: example of a contribution to the current correlator, Eq.~(\ref{current-corr}).
Lower part: The corresponding contribution to the current correlator can be obtained from the DOS correlator by removing the interaction line, keeping the velocity vertices.
}
\end{figure}

\begin{figure}
\includegraphics[width=1.0\linewidth]{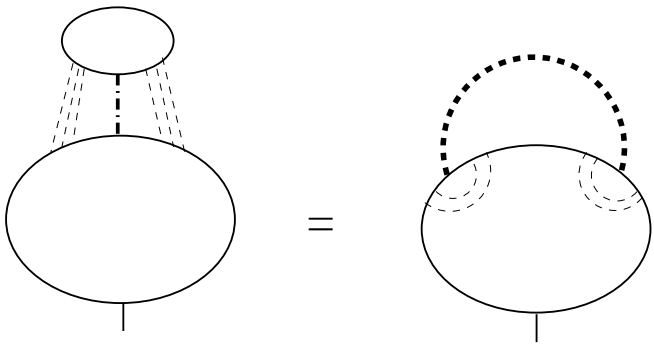}
\caption{\label{nu1fig} The first-order correction to the TDoS, Eq.~(\ref{nu1}).}
\end{figure}

In order to confirm Eq.~(\ref{delta_sigma}) and shed more light on the underlying physics, we now provide an alternative derivation, based on an existing result for the equilibrium current fluctuations in a disordered system without time-reversal symmetry.
In Ref.~\onlinecite{Gornyi_Mirlin_Woelfle_01}, the following result for the correlation function of local mesoscopic currents has been derived for the relevant range \mbox{$L_\omega^{-1}\ll k\ll l^{-1}$},
\be
\Bigl\langle j_\alpha(E+\omega)\,j_\beta(E)\Bigr\rangle_k
=
\frac{e^2}{2\pi^3}\ln{(kL_\omega)}\:
\delta_{\alpha\beta}^\perp
\label{current-corr}
\ee
with \mbox{$L_\omega=\left(D/\omega\right)^{1/2}$}.
In the limit \mbox{$k\to 0$} the current-current correlator vanishes as $(kL_\omega)^{2}$ as a result of the gauge invariance, which is closely related to the infrared regularization of the effective interaction (\ref{tilde_U}). Equation (\ref{current-corr}) describes the upper part of Fig.~\ref{current-corr-fig} and similar diagrams. It can be identified as the Hartree correction to the tunnelling density of states (TDoS), $\delta\nu^H$ (Fig.~\ref{nu1fig}, which is generated by insertion of a scalar vertex into the lower part of Fig.~\ref{current-corr-fig} and similar diagrams with three diffusons) with the interaction line $U_{\alpha\beta}(k)$ removed.

In analogy to Ref.~\onlinecite{Altshuler_Aronov_85}, we consider the Hartree contribution to the energy shift of a state with unperturbed energy $E_m$ above the Fermi sea (due to interactions with levels below the Fermi energy),
\bea
\Sigma_m^H
&=&
\sum\limits_{E_n<0}\int d\br\,d\br'\:
U_{\alpha\beta}(\br-\br')\nonumber\\
& &
\times\:
\psi_m^*(\br)\:e^*\hat{v}_\alpha\:\psi_m(\br)\:
\psi_n^*(\br')\:e^*\hat{v}_\beta\:\psi_n(\br')\:\:.\nonumber\\
\eea
Averaging over all states with this energy, the mean energy shift is
\bea
\Sigma_\epsilon^H
&=&
\frac{\left(e^*\right)^2}{\nu V}\sum\limits_m
\Bigl\langle
\delta(\epsilon-E_m)\:\Sigma_m^H
\Bigr\rangle\nonumber \\
&=&
\frac{\left(e^*\right)^2}{\nu V}\sum\limits_m
\Bigl\langle
\delta(\epsilon-E_m)
\sum\limits_{E_n<0}\int d\br\,d\br'\nonumber\\
& &
\times\:U_{\alpha\beta}(\br-\br')
\psi_m^*(\br)\hat{v}_\alpha\psi_m(\br)\:
\psi_n^*(\br')\hat{v}_\beta\psi_n(\br')
\Bigr\rangle\nonumber\\
&=&
\frac{\left(e^*\right)^2}{\nu V}\sum\limits_m
\int\limits_\epsilon^\infty d\omega\:
\biggl\langle
\delta(\epsilon-E_m)
\sum\limits_{E_n<0}
\delta(\epsilon-\omega-E_n)\nonumber\\
& &
\times\:\int d\br\,d\br'\:
U_{\alpha\beta}(\br-\br')\nonumber\\
& &
\times\:\psi_m^*(\br)\hat{v}_\alpha\psi_m(\br)\:
\psi_n^*(\br')\hat{v}_\beta\psi_n(\br')
\biggr\rangle\nonumber \\
&=&
\frac{\lambda^2}{\nu}
\int\limits_\epsilon^\infty d\omega\:
\int d(\br-\br')\:
U_{\alpha\beta}(\br-\br')\:
\bigl\langle
j_\alpha\,j_\beta
\bigr\rangle_{\br-\br',\omega}\:,\nonumber
\\
& &\label{Hartree-self-energyAA}
\eea
which after the Fourier transformation to momentum representation
reads
\be
\Sigma_\epsilon^H
=
\frac{\lambda^2}{\nu}
\int\limits_\epsilon^\infty d\omega\:
\int (dk)\:
U_{\alpha\beta}(\bk)\:
\Bigl\langle
j_\alpha\,j_\beta
\Bigr\rangle_{\bk,\omega}\:,
\ee
with $U_{\alpha\beta}(\bk)$ given by Eq.~(\ref{propfull}).
Using the formula
\be
\frac{\delta\nu^H}{\nu}=-\frac{\partial\Sigma_\epsilon^H}{\partial\epsilon}
\ee
and the fact that
\be
\frac{\delta\sigma^H}{\sigma_0}
=
\frac{\delta\nu^H(\epsilon\sim T)}{\nu}
\label{DoSsigma}
\ee
(which can be directly shown by inserting velocity vertices into the diagrams for the TDoS correction),
we find
\bea
\frac{\delta\sigma^H}{\sigma_0}
&=&
\frac{\lambda^2}{\nu}
\int (dk)\:
U_{\alpha\beta}(\bk)\:
\bigl\langle
j_\alpha\,j_\beta
\bigr\rangle_{\bk,T}\nonumber \\
&=&
\frac{\lambda^2}{\nu}
\int (dk)\:
\frac{1}{\chi_0 k^2}
\frac{e^2}{2\pi^3}\:
\ln{\left(kL_T\right)}\nonumber \\
&=&
\frac{\lambda^2e^2}{4\pi^4\nu\chi_0}
\int\limits_0^{\ln{(L_T/l)}}d\ln{(kL_T)}\:
\ln{\left(kL_T\right)}\nonumber \\
&=&
\frac{\lambda^2e^2}{32\pi^4\nu\chi_0}\:
{\rm ln}^2\frac{1}{T\tau}\:\:,
\label{delta_sigma_again}
\eea
which is identical with Eq.~(\ref{delta_sigma}) since we have used the free-fermion relation \mbox{$\nu\chi_0=e^2/24\pi^2$}.

The calculation in this subsection helps to clarify the physical origin of the Hartree correction (\ref{delta_sigma}) and (\ref{delta_sigma_again}):
this contribution to the conductivity is induced by scattering off static mesoscopic fluctuations of local currents, whose correlation function is given by Eq.~(\ref{current-corr}).

It is also instructive to explicitly calculate the first-order perturbative correction to TDoS (see Fig.~\ref{nu1fig}),
\be
\delta\nu^H(E)
=
-\frac{1}{\pi}\:{\rm Im}\int(dp)\:\delta G^R(E,p)\:\:.
\label{nu1}
\ee
Here $\delta G^R$ is the interaction-induced correction to the retarded
Green's function of a fermion.
For simplicity, we restrict ourselves to the zero-$T$ case.
Calculating the ``exchange'' correction using the effective interaction propagator $\tilde{U}$ given by Eq.~(\ref{tilde_U}), we need to include an additional minus sign due to the closed fermionic loop. After averaging over disorder, the first-order correction to the TDoS reads:
\bea
\delta\nu^H(E)
&\simeq&
\frac{1}{\pi}\,\frac{3 \lambda^2 g}{\pi \nu}
\:{\rm Re}\int(dp)\left[G^R(E,p)\right]^2 G^A(E,p)\nonumber\\
&\times&
\int(dq)\int\limits_E^{1/\tau}\frac{d\omega}{2\pi}
\left[
\frac{1}{\tau(Dq^2-i\omega)}
\right]^2
\ln {\frac{1}{q^2l^2}}\nonumber \\
&=&
\nu \lambda^2 \frac{3}{4\pi^2} \ln^2(E\tau)\:\:.
\eea
Thus the Hartree correction to the conductivity and to the TDoS indeed have the same form Eq.~(\ref{DoSsigma}), as expected.

Our aim now is to proceed on to the case where the coupling of the fermions to the gauge field (or, more precisely, the product \mbox{$\lambda^2g$}) is not small, \mbox{$\lambda^2 g \gg 1$}, including the realistic case \mbox{$\lambda=1$}.
Since Eq.~(\ref{delta_sigma_again}) is small only in the parameter \mbox{$\lambda^2g$}, it is necessary to take higher-order interaction effects into account. These involve dephasing (by real processes) and renormalization (by virtual processes). For a general review of these effects, we refer the reader to Refs.~\onlinecite{AAG_99} and \onlinecite{GMP07}. Since dephasing effects in the present case\cite{Aronov_Woelfle_PRL_94,Aronov_Woelfle_94,Lee_Mucciolo_Smith_96,Woelfle_2000,TL_diss_06} are much stronger than in the standard situation\cite{AAK_82,AAG_99,Aleiner_Blanter_2002} even in the weak coupling regime, we first study dephasing (at \mbox{$\lambda^2g\ll 1$}) in detail in Section~\ref{dephasing}. We return to the discussion of the Hartree correction to conductivity in Section~\ref{low_T}, where we will find that we may deal exclusively with renormalization effects at low temperatures. In Section~\ref{high_T}, we will then study the intermediate-temperature situation, where dephasing and renormalization compete. Finally, in Section~\ref{dephasing-strong-coupling}, we discuss the renormalization effects on the dephasing rate at strong coupling.

\section{Dephasing due to weak gauge field fluctuations}
\label{dephasing}
\setcounter{equation}{0}

Even in a normal disordered metal, electrons are subject to transverse gauge field fluctuations\cite{Rukhadze_Silin_61,AAK_82,Altshuler_Aronov_85,Stern_Aharonov_Imry_90}, however the transverse part of the electromagnetic fluctuations is in that case small in \mbox{$v_F/c$} compared to the longitudinal one, and may usually be neglected. In the composite-fermion model of the half-filled lowest Landau level\cite{Jain_89,Lopez_Fradkin_91}, a situation occurs with a similar gauge field propagator at the random-phase approximation (RPA) level\cite{HLR_93}, but a fermion-gauge field coupling of order unity. Therefore the effects of the gauge field interaction may greatly exceed those of the Coulomb interaction.

The correlator of gauge field fluctuations can be obtained according to the fluctuation-dissipation theorem from Eq.~(\ref{propfull}),
\be
\left\langle a_\alpha a_\beta\right\rangle_{\bk,\epsilon}
=
{\rm coth}\frac{\epsilon}{2T}\:
{\rm Im}\:U_{\alpha\beta}(\bk,\epsilon)\:\:.
\label{FDT}
\ee
In this Section, we are interested in classical thermal fluctuations with \mbox{$\epsilon\ll T$}. The characteristic energy scale in Eq.~(\ref{propfull}) is \mbox{$\epsilon\lesssim T_0$} while the corresponding transferred momenta $\bk$ fall into the diffusive range \mbox{$k\ll 1/l$}. This allows us to use Eq.~(\ref{bare_gaugefield}) for $U_{\alpha\beta}(\bk,\epsilon)$, yielding
\be
\left\langle a_\alpha a_\beta\right\rangle_{\bk,\epsilon}
\approx
\frac{2T}{\sigma_0}\:
\frac{\delta_{\alpha\beta}^\perp}{\left(k^2l^2T_0\right)^2+\epsilon^2}\:\:,\quad T\gg\epsilon\:\:.
\label{gaugefield_fluct}
\ee
Since the correlator is sharply peaked as a function of the transferred energy \mbox{$\epsilon\lesssim T_0$}, for many situations the static approximation is appropriate, which collects all the weight in a $\delta$-function in energy space,
\be
\left\langle a_\alpha a_\beta\right\rangle_{\bk,\epsilon}
=
\frac{T}{\chi_0 k^2}\:
\delta_{\alpha\beta}^\perp\:
2\pi\delta(\epsilon)\:\:.
\label{static_correlator}
\ee 
It is convenient to introduce the correlation function of the static vector potential
\be 
\left\langle a_\alpha a_\beta\right\rangle_{\bk}
\equiv
\frac{T}{\chi_0 k^2}\:
\delta_{\alpha\beta}^\perp\:\:.
\label{static_correlator_def}
\ee
This corresponds to a static random magnetic field (RMF) with the correlator
\be
\left\langle h(\br)\,h(\br')\right\rangle
=
\frac{T}{\chi_0}\:\delta(\br-\br')\:\:.
\label{rmf-corr}
\ee
Below we use the static approximation (\ref{static_correlator}) whenever appropriate and return to the full dynamical form (\ref{gaugefield_fluct}) if necessary, see discussion around Eqs.~(\ref{tauphi-coop-3}) and (\ref{wl_cf_highT}). It should be emphasized that the dephasing effects arising within the static approximation are purely geometric effects due to the phases associated with encircling magnetic flux and do not involve any energy transfer.
The phase-space available for inelastic processes only appears through the magnitude of the correlation function of the RMF, Eq.~(\ref{rmf-corr}).

We will first discuss dephasing effects for Cooperons and conductance-fluctuation diffusons (which are the usual manifestations of the dephasing), and then refine the approach for a treatment of the (finite-delay time) diffusons appearing in the Hartree diagrams.

\subsection{Cooperon dephasing}
\label{Cooperon_dephasing}

In this section, the Cooperon amplitude and the weak localization correction are calculated for a system of fermions weakly coupled to a fluctuating gauge field (\mbox{$\lambda^2g\ll 1$}).
At variance with the rest of the paper where we concentrate on systems with broken time-reversal invariance (having composite fermions in mind as a particularly important example), here we consider the case of usual scalar-potential disorder which preserves the time-reversal symmetry.
As discussed above, in the half-filled lowest Landau level, the time-reversal symmetry is broken by the static disorder acquiring a vector component via screening effects. Therefore, for the composite-fermion problem, the true Cooperon is completely suppressed by disorder-induced RMF. The analysis in this Section should be then considered as an auxiliary calculation that helps to understand dephasing effects showing up in mesoscopic conductance fluctuations, Section~\ref{ucf_dephasing}.

The weak localization correction to conductivity\cite{Gorkov_Larkin_Khmelnitskii_79,Altshuler_Khmelnitskii_Larkin_Lee_80,Chakravarty_Schmid_86} is given by
\be
\delta\sigma_{\rm WL}
=
-\frac{2e^2D}{\pi}
\int\limits_\tau^\infty dt\:
\left\langle {\cal C}^{t_0}(0,0;t,-t)\right\rangle,
\label{deltaWL-C}
\ee
where the Cooperon ${\cal C}^{t_0}(\br,\br';t,t')$ in the presence of a random gauge field satisfies
\begin{widetext}
\be
\left\{
\partial_t
+D\Bigl[
-i\nabla
-\lambda e{\bf a}\left(\br,t_0+t/2\right)
-\lambda e{\bf a}\left(\br,t_0+t'/2\right)
\Bigr]^2
\right\}{\cal C}^{t_0}(\br,\br';t,t')=\delta(\br-\br')\:\delta(t-t')\:\:.
\label{Cooperon}
\ee
\end{widetext}
and the average \mbox{$\langle\cdots\rangle$} is over the configurations of the random gauge field.
The Cooperon ${\cal C}^{t_0}(\br,\br';t,-t)$ determining the weak localization correction (\ref{deltaWL-C}) describes coherent propagation of a particle from $\br'$ to $\br$ and of a hole along the backward path from  $\br$ to $\br'$, both processes starting at \mbox{$t_0-t/2$} and ending at \mbox{$t_0+t/2$}. The averaged Cooperon does not depend on $t_0$ in view of translational invariance in time.

We write the Cooperon ${\cal C}^{t_0}(0,0;t,-t)$ as a path integral,
\be
{\cal C}^{t_0}(0,0;t,-t)=
\int\limits_{\br(-t)=0}^{\br(t)=0}{\cal D}[\br(t')]\:
{\rm exp}\bigl\{{}-S_0+iS_1\bigr\}
\ee
with the kinematic part of the action
\be
S_0=\int\limits_{-t}^{t}dt'\:\frac{\dot{\br}^2(t')}{4D}
\ee
describing the diffusive dynamics,
and
\bea
S_1
&=&
-\lambda e\int\limits_{-t}^{t}dt'\:\dot{\br}(t')\cdot {\bf a}[\br(t_0+t'/2)]\nonumber\\
& &-\lambda e\int\limits_{-t}^{t}dt'\:\dot{\br}(t')\cdot {\bf a}[\br(t_0-t'/2)]\:\:.
\eea
Averaging over the gauge field configurations with Gaussian weight, we have
\be
\left\langle{\cal C}^{t_0}(0,0;t,-t)\right\rangle=
\int\limits_{\br(-t)=0}^{\br(t)=0}{\cal D}[\br(t')]\:
{\rm exp}\bigl\{{}-S_0-\Delta S\bigr\}
\ee
with [we now drop the ``mute'' variable $t_0$ and denote ${\cal C}^{t_0}(0,0;t,-t)\equiv {\cal C}(t)$]
\begin{widetext}
\bea
\Delta S(t)
&=&
\frac{1}{2}\,\lambda^2e^2\int\limits_{-t}^{t}dt_1\int\limits_{-t}^{t}dt_2\:
\dot{r}_\alpha(t_1)\,\dot{r}_\beta(t_2)\nonumber\\
& &\times\biggl\{
\Bigl\langle a_\alpha[\br_1,t_1/2]\,a_\beta[\br_2,t_2/2]\Bigr\rangle
+\Bigl\langle a_\alpha[\br_1,t_1/2]\,a_\beta[\br_2,-t_2/2]\Bigr\rangle\nonumber\\
& &\quad{}
+\Bigl\langle a_\alpha[\br_1,-t_1/2]\,a_\beta[\br_2,t_2/2]\Bigr\rangle
+\Bigl\langle a_\alpha[\br_1,-t_1/2]\,a_\beta[\br_2,-t_2/2]\Bigr\rangle\biggr\}\:\:,
\label{DeltaS-dyn}
\eea
where \mbox{$\br_i\equiv\br(t_i)$}.
Within the static gauge field approximation as described by Eq.~(\ref{static_correlator}), Eq.~(\ref{DeltaS-dyn}) reduces to
\be
\Delta S=
2\lambda^2 e^2\int\limits_{-t}^{t}dt_1\int\limits_{-t}^{t}dt_2\:
\dot{r}_\alpha(t_1)
\,\Bigl\langle a_\alpha[\br(t_1)]\,a_\beta[\br(t_2)]\Bigr\rangle\,
\dot{r}_\beta(t_2)\:\:.
\ee
\end{widetext}
As discussed in the end of this subsection, the static approximation is sufficient for the present problem, except for very low temperatures, where some refinement will be needed. The time dependence of the gauge fields will, however, become crucial in Section~\ref{diffuson_dephasing} where dephasing of ``delayed diffusons'' relevant to the Hartree correction will be analyzed.

It is convenient to define an effective action \mbox{$\Delta S^{\rm eff}(t)$} with the property \cite{Woelfle_2000}
\bea
\left\langle{\cal C}(t)\right\rangle&=&
{\rm exp}\bigl\{- \Delta S^{\rm eff}(t) \bigr\}
\int\limits_{\br(-t)=0}^{\br(t)=0}{\cal D}[\br(t')]\:
{\rm exp}\bigl\{-S_0\bigr\}\nonumber\\
&=&
{\rm exp}\bigl\{-\Delta S^{\rm eff}(t)\bigr\}\:
{\cal C}^{(0)}(t)\:\:,
\eea
where \mbox{${\cal C}^{(0)}(t)\equiv{\cal C}^{(0)}(0,0,t)=\left(4\pi Dt\right)^{-1}$} is the unperturbed Cooperon in two dimensions.

To second order in the coupling constant \mbox{$\lambda e$}, \mbox{$\Delta S^{\rm eff}(t)$} can be evaluated as the average of \mbox{$\Delta S$} weighted with the unperturbed Cooperon,
\be
\Delta S^{\rm eff}(t)\simeq
\frac{1}{{\cal C}^{(0)}(t)}
\!\!\int\limits_{\br(-t)=0}^{\br(t)=0}\!\!\!\!{\cal D}[\br(t')]\:
{\rm exp}\bigl\{-S_0\bigr\}\:\Delta S\bigl[\br(t'),t\bigr]\:.
\label{cooperon_expansion}
\ee
The integral can be identified as the term of second order in \mbox{$\lambda e$}
of an expansion of the Cooperon
\mbox{${\cal C}=\left(-D\nabla^2\right)^{-1}=\left(D{\bf \hat{q}}^2\right)^{-1}$} after coupling to the gauge field by the substitution \mbox{$-i\nabla \to (-i\nabla-2\lambda  e{\bf a})\,$},
\mbox{${\bf \hat{q}} \to ({\bf \hat{q}}-2\lambda e{\bf a})\,$},
\bea
{\cal C}&=&
{\cal C}^{(0)}
+2\lambda eD{\cal C}^{(0)}\left\{a_\alpha,\hat{q}_\alpha\right\}{\cal C}^{(0)}
-4\lambda^2e^2D{\cal C}^{(0)}a_\alpha a_\alpha{\cal C}^{(0)}\nonumber\\
& &{}
+4\lambda^2e^2D^2{\cal C}^{(0)}\left\{a_\alpha,\hat{q}_\alpha\right\}
   {\cal C}^{(0)}\left\{a_\beta,\hat{q}_\beta\right\}{\cal C}^{(0)}\:,\nonumber \\
\label{Cooperon_expansion}
\eea
where $\{\cdot,\cdot\}$ is the anticommutator and summation over $\alpha$ and $\beta$ is implied.
In the static approximation the two gauge field terms in Eq.~(\ref{Cooperon}) simply add, so that the Cooperon couples with the charge \mbox{$2\lambda e$} to the static gauge field.

Performing the average over the gauge field fluctuations, we get
\bea
\left\langle{\cal C}\right\rangle
&=&
{\cal C}^{(0)}
-4\lambda^2e^2D\,{\cal C}^{(0)}\left\langle a_\alpha a_\alpha\right\rangle{\cal C}^{(0)}\nonumber\\
& &{}
+4\lambda^2e^2D^2{\cal C}^{(0)}
  \left\langle
    \left\{a_\alpha,\hat{q}_\alpha\right\}
    {\cal C}^{(0)}
    \left\{a_\beta,\hat{q}_\beta\right\}
  \right\rangle
  {\cal C}^{(0)}\:.\qquad \nonumber\\
\label{Cooperon_expansion_av}
\eea
It is worth stressing that, within the approach based on Eq.~(\ref{Cooperon_expansion}), it is not necessary to distinguish to which fermionic line of the Cooperon the ends of the gauge field line are connected. The reason is that within the static approximation (\ref{static_correlator}) for the gauge field propagator, fermionic self-energy and vertex parts contribute equally, as can be seen from Eq.~(\ref{Cooperon}). This is at variance with the conventional case of scalar density-density interaction, where, both in Cooperon and diffuson, the fermionic self-energy and vertex interaction parts have opposite signs. In the present case of current-current interaction, the vertex interaction line in a Cooperon acquires an additional minus sign due to reversing the velocity in one of the fermionic lines constituting the Cooperon.

\begin{figure}
\includegraphics[width=1.0\linewidth]{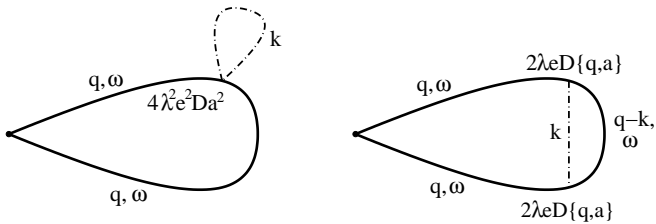}
\caption{\label{2nd_order_processes} Illustration of the processes contributing to Eqs.~(\ref{deltaS-raw}) and (\ref{deltaS-2d}). The left and right diagram give the first and second term in brackets, respectively. }
\end{figure}

Employing the static approximation (\ref{static_correlator_def}), the averaged action \mbox{$\langle\Delta S\rangle$} can thus be written as
\begin{widetext}
\bea
\Delta S^{\rm eff}(t)&=&
\frac{1}{{\cal C}^{(0)}(t)}
\:
\int\frac{d\omega}{2\pi}\:{\rm exp}\{i\omega t\}
\int(dq)
\int(dk)\:
\frac{1}{\left(D\bq^2-i\omega\right)^2}\,4\lambda^2e^2D\nonumber\\
& &
\times\:\left[
{}-\bigl\langle a_\alpha a_\alpha\bigr\rangle_\bk
+\frac{4D}{D\left({\bf q-k}\right)^2-i\omega}
 \left(q-\frac{k}{2}\right)_\alpha
   \bigl\langle a_\alpha a_\beta\bigr\rangle_\bk
 \left(q-\frac{k}{2}\right)_\beta
\right]\:\:.
\label{deltaS-raw}
\eea
 With the gauge field correlator (\ref{gaugefield_fluct}) we find
\bea
\Delta S^{\rm eff}(t)&=&
4\pi Dt\:
\frac{4\lambda^2e^2DT}{\chi_0}
\int\frac{d\omega}{2\pi}\:{\rm exp}\{i\omega t\}
\int\limits_0^{l^{-1}}\frac{q\,dq}{2\pi}
\int\limits_0^{l^{-1}}\frac{k\,dk}{2\pi}
\int\limits_0^{2\pi}\frac{d\phi}{2\pi}\:
\frac{1}{\left(Dq^2-i\omega\right)^2}\nonumber\\
& &\times\:
\frac{1}{k^2}
\left[
 {}-1
 +\frac{4D\:q^2\,{\rm sin}^2\phi}{Dq^2-2Dqk\,{\rm cos}\phi+Dk^2-i\omega}
\right]\:\:,
\label{deltaS-2d}
\eea
\end{widetext}
where $\phi$ is the angle between the directions of the momenta $\bq$ and $\bk$. An illustration of this equation is given in Fig.~\ref{2nd_order_processes}.

Equation~(\ref{deltaS-2d}) bears a close similarity to Eq.~(21) of Ref.~\onlinecite{Woelfle_2000}, which was derived in a different way (explicitly including ballistic propagation). Here the first term $(-1)$ in brackets takes the role of the ballistic term of Ref.~\onlinecite{Woelfle_2000}. The cancellation of the two terms in brackets at \mbox{$\omega=0$} and \mbox{$k\to 0$} is a result of gauge invariance:
A static uniform gauge field should not affect observables. This cancellation is the path-integral counterpart of the cancellation in the effective interaction box $\tilde U$ (Fig.~\ref{boxes}) in the Hartree correction, Section~\ref{Hartree_wo_dephasing}.

Inspection of Eq.~(\ref{deltaS-2d}) shows that the  $k$-integral is logarithmic in the range $q<k<l^{-1}$, with the second term in the brackets providing the low-$k$ cutoff at \mbox{$k\sim q$}. Evaluating then $q$-integrals in Eq.~(\ref{deltaS-2d}), we find
\be
\Delta S^{\rm eff}(t)
=
\frac{24}{\pi}\,\lambda^2gTt\int\limits_0^\infty d\omega\:
\frac{{\rm sin}\,\omega t}{\omega}\ln{\frac{L_\omega}{l}}\:\:.
\label{deltaS-calc}
\ee
Setting $\omega\sim 1/t$ under the logarithm, we  evaluate (\ref{deltaS-calc}) with logarithmic accuracy, arriving at
\be
\Delta S^{\rm eff}(t)
=
6\,\lambda^2gTt\ln{\frac{t}{\tau}}\:\:,
\label{deltaS-result-2d}
\ee
in agreement with Ref.~\onlinecite{Woelfle_2000}. Let us note that in Ref.~\onlinecite{Woelfle_2000} this result was obtained by treating explicitly the ballistic dynamics in a particular model of isotropic scattering. On the other hand, we have derived Eq.~(\ref{deltaS-result-2d}) within a purely {\it diffusive} calculation without the need of taking details of ballistic propagation into account. This could be expected since the relevant physics happens on the large length scales of diffusive propagation and does not depend on the details of microscopic scattering processes.

The physical picture behind Eq.~(\ref{deltaS-result-2d}) is the following: For a typical closed diffusive path, the {\it geometrical} area covered by it will be proportional to its duration $t$, and so will be the average absolute value of the flux through this area, suggesting \mbox{$\langle \Delta S \rangle (t) \propto t$}.
However, the path may encircle some areas more than once. Since the gauge field configuration does not change appreciably during the time in between, the phases picked up from that area will add up coherently, so that the quantity relevant for dephasing is the {\it non-oriented} (Amperean) area enclosed, with the result\cite{Aronov_Woelfle_PRL_94,Aronov_Woelfle_94}
\be
\Delta S=\frac{2\lambda^2e^2T}{\chi_0}\sum\limits_i n_i^2\,A_i\:\:,
\label{sum_nonoriented}
\ee
where \mbox{$n_i\in{\mathbb Z}$} is the number of times the area $A_i$ is encircled. Ref.~\onlinecite{Aronov_Woelfle_94} then proceeded by setting all \mbox{$n_i=1$}, approximating the non-oriented area \mbox{$\sum_i n_i^2A_i$} by the geometrical area \mbox{$A=\sum_i A_i\sim Dt$} to obtain a linear-in-$T$ dephasing rate \mbox{$1/\tau_\varphi\sim \lambda^2gT$}. The logarithmic correction in Eq.~(\ref{deltaS-result-2d}) is thus due to diffusive paths forming multiple loops.

The dephasing rate $1/\tau_\varphi$ and corresponding dephasing length \mbox{$L_\varphi=\left(D\tau_\varphi\right)^{1/2}$} (note that this relation between $\tau_\varphi$ and $L_\varphi$ only holds for weak coupling $\lambda^2g\ll 1$, see Section~\ref{dephasing-strong-coupling}) can be defined using Eq.~(\ref{deltaS-result-2d}) and the condition \mbox{$\Delta S^{\rm eff}(t=\tau_\varphi)=1$},
\be
\frac{1}{\tau_\varphi}
=
\displaystyle 6\,\lambda^2gT
\ln{\frac{T_1}{\lambda^2T}}\:,\quad T_0/\lambda^2 g\ll T\ll  T_1/\lambda^2\:\:.
\label{tauphi-coop}
\ee
The weak localization correction is now easily calculated,
\bea
\delta\sigma_{\rm WL}&=&
-\frac{2e^2D}{\pi}
\int\limits_\tau^\infty dt\:
\left\langle {\cal C}(0,0,t)\right\rangle\nonumber\\
&=&
-\frac{2e^2D}{\pi}
\int\limits_\tau^\infty dt\:
{\cal C}^{(0)}(0,0,t)\:
{\rm exp}\bigl\{-\Delta S^{\rm eff}(t)\bigr\}\nonumber\\
&=&
-\frac{e^2}{2\pi^2}
\ln{\frac{\tau_\varphi(T)}{\tau}}\:\:.
\label{WL}
\eea
At higher temperatures, \mbox{$T\gtrsim T_1/\lambda^2$}, the weak localization amplitude is dominated by very short Cooperon paths of duration \mbox{$t\lesssim\tau$}, so that the present calculations for the diffusive regime do not apply. We do not attempt an analysis of the diffusive-ballistic crossover and of the ballistic regime in this paper.

At sufficiently low temperatures, \mbox{$T\ll T_0/\lambda^2 g$}, the fully static approximation is no longer valid, since the characteristic times \mbox{$t\sim \tau_\varphi$} in the Cooperon propagator become longer than $1/T_0$ [see also the discussion below Eq.~(\ref{effective_correlator_for_diffuson})].
At such long times, the correlations between the forward and backward interfering paths disappear due to the slow dynamics of the gauge-fields: only two out of four terms in Eq.~(\ref{DeltaS-dyn}) survive, related to the correlations within the same (forward or backward) path. For those remaining correlations, the static approximation still applies, as long as \mbox{$T>T_0$}.
As a result, at \mbox{$T\ll T_0/\lambda^2 g$} the dephasing action becomes smaller than Eq.~(\ref{deltaS-result-2d}) by a factor of 2,
yielding
\be
\frac{1}{\tau_\varphi}
=
\displaystyle 3\,\lambda^2gT
\ln{\frac{T_1}{\lambda^2T}}\:\:,\quad T_0 \ll T\ll T_0/\lambda^2g\:\:.
\label{tauphi-coop-3}
\ee
Note that this intermediate regime disappears in the strong-coupling regime \mbox{$\lambda^2 g\gg 1$}.

At the lowest temperatures \mbox{$T<T_0$}, the result is further modified by the fact that the allowed phase space for the inelastic energy transfers, \mbox{$|\epsilon|\lesssim T$}, does not cover the whole peak of the correlator (\ref{gaugefield_fluct}). This results in the appearance of the ratio $T/T_0$ under the logarithm in the dephasing action
\be
\Delta S^{\rm eff}(t)
=
3\,\lambda^2gTt\ln{\frac{T t}{T_0 \tau}}\:\:,\quad T\ll T_0\:\:,
\label{deltaS-result-lowT}
\ee
which in turn makes the logarithmic factor in the dephasing rate
$T$-independent:
\be
\frac{1}{\tau_\varphi}
=
\displaystyle 3\,\lambda^2gT
\ln{\frac{g}{\lambda^2}}\:\:,\quad   T\ll T_0\:\:.
\label{tauphi-coop-lowT}
\ee

The above results for the dephasing rate are only valid in the regime of weak coupling \mbox{$\lambda^2 g\ll 1$}. For stronger coupling, including the realistic case \mbox{$\lambda=1$}, one should take into account the renormalization of the interfering paths by virtual processes, which are reflected in the strong interaction-induced $Z$-factor in the Cooperon propagators. This situation is discussed in Section~\ref{dephasing-strong-coupling} below.

Since the logarithmic correction in Eq.~(\ref{deltaS-result-2d}) is the result of multiple return processes, it is instructive to make a short digression and to inspect \mbox{$\Delta S^{\rm eff}$} for a quasi-onedimensional wire of width \mbox{$w\gg l$}. In that situation the Cooperon dephasing rate is\cite{TL_diss_06}
\be
\frac{1}{\tau_\varphi}
=
24\lambda^2g_\square T\ln{\frac{w}{l}}\:\:,\quad T\ll T_1/\lambda^2\:\:
\label{tauphi_q1d}
\ee
($g_\square$ is the conductance per square),
without any infrared anomalies. Equation~(\ref{tauphi_q1d}) results from a linear-in-$t$ behavior of \mbox{$\Delta S^{\rm eff}$}, consisting of a factor $t^{1/2}$ from the normalization of the unperturbed quasi-onedimensional Cooperon and an {\it algebraic} correction factor $t^{1/2}$ due to the enhancement of the non-oriented area over the geometric one.
In view of the absence of infrared divergences, the different behavior of dephasing rates associated with Aharonov-Bohm oscillations in quasi-1D rings\cite{Ludwig_Mirlin_2004} and weak localization in quasi-1D wires, due to different low-momentum cutoff conditions, does not occur for the dephasing by gauge field fluctuations\cite{TL_diss_06}.

\subsection{Diffuson dephasing: mesoscopic conductance fluctuations}
\label{ucf_dephasing}

We now turn to the dephasing applicable to mesoscopic conductance fluctuations\cite{Altshuler_85,Stone_85,Lee_Stone_85,Altshuler_Shklovskii_86,Lee_Stone_Fukuyama_87,Kane_Serota_Lee_88}.
Later on, we will relate it to the treatment of Cooperon dephasing within the context of weak localization (Section~\ref{Cooperon_dephasing}) and of the ``delayed diffuson'' dephasing (Section~\ref{diffuson_dephasing}) relevant for Hartree conductivity correction.
The variance of the conductance can be written as
\bea
\left\langle\delta g^2\right\rangle
&=&
\frac{16\pi D^2}{3TL^4}
\int {\bf dr_1dr_2} \int dt\,dt'\:
\tilde{\delta}(t-t')\nonumber\\
& &
\times
\left\langle
{\cal D}^{12}(\br_1,\br_2,t)\,
{\cal D}^{21}(\br_2,\br_1,t^\prime)
\right\rangle\:,
\label{dgdg}
\eea
where $L$ is the system size, the function $\tilde{\delta}(t-t')$ describes the thermal smearing of the two Fermi distribution functions,
\bea
\tilde{\delta}(t-t')
&=&
12\pi T\int\frac{d\epsilon_1}{2\pi}\,\frac{d\epsilon_2}{2\pi}\:
f'(\epsilon_1)\,f'(\epsilon_2)\nonumber\\
& &
\times\:
{\rm exp}\left\{i\left(\epsilon_1-\epsilon_2\right)\left(t-t'\right)\right\}\nonumber \\
&=&
3\pi T^3\left(t-t'\right)^2{\rm sinh}^{-2}\left[\pi T(t-t')\right]\nonumber \\
\label{tilde-delta}
\eea
[which for \mbox{$T(t-t')\gg 1$} may be replaced by a true delta function], and  ${\cal D}^{12}$ is a diffuson satisfying
\begin{widetext}
\be
\left\{
\partial_t
+D\Bigl[
-i\nabla
-\lambda e{\bf a}_1\left(\br,t\right)
+\lambda e{\bf a}_2\left(\br,t\right)
\Bigr]^2
\right\}{\cal D}^{12}(\br,\br',t)=\delta(\br-\br')\:\delta(t)\:\:.
\label{CF_diffuson}
\ee
\end{widetext}
Here the two measurements denoted by $1,2$ see uncorrelated gauge field configurations. In the absence of interaction-induced dephasing [${\bf a}_i(\br,t)=0$], Eqs.~(\ref{dgdg})-(\ref{CF_diffuson}) lead to \mbox{$\left\langle\delta g^2\right\rangle\sim 1$} independent of the system size (only dependent on its dimensionality and shape)
-- the famous universal conductance fluctuations (UCF). Dephasing manifests itself in a suppression of conductance fluctuations as compared to the fully coherent UCF-regime.

Since the ``UCF diffuson'' ${\cal D}^{12}$ involves two separate measurements, it is not subject to particle number conservation\cite{Aleiner_Blanter_2002}. In other words, the cancellation between self-energy and vertex corrections known from the ``true'' diffuson does not occur since vertex corrections, with an interaction line connecting the fermionic lines, are absent: A truly static random magnetic field would indeed drop out of Eq.~(\ref{CF_diffuson}); however a slowly varying random gauge field does not drop out when the diffuson is formed by two Green functions related to two separate measurements.

The essential ingredient of the calculations is the following assumption about the timescales involved: Characteristic frequencies of the gauge field fluctuations are much smaller than those of the electron diffusion, so that to a good approximation a fermion experiences a static random gauge field while diffusing through the sample. The duration of a measurement, in turn, is assumed to be much longer than the timescale set by the gauge field fluctuations, so that a measurement samples many electrons and performs a complete ensemble average over realizations of the random gauge field. Finally, different measurements will see no correlation between their respective gauge field configurations. Technically, this means that the static gauge field correlator (\ref{static_correlator}) is used for correlations experienced by any individual fermion, and correlators between separate measurements are completely dropped.

The aim of this section is to establish a formal relation between weak localization and mesoscopic conductance fluctuations 
when 
the dephasing is governed by a fluctuating gauge field. A relation of this kind has been demonstrated for the case of the usual screened Coulomb interaction by Aleiner and Blanter\cite{Aleiner_Blanter_2002}. It has been formulated in a more general way in Ref.~\onlinecite{Ludwig_Mirlin_2004} (and more recently also been confirmed independently\cite{Texier_Montambaux_2005}), where it has been shown that a manipulation of the path-integral expressions can transform these quantities one into another without actually evaluating the path integrals.
When applied to a ring geometry, this general relation links $h/e$ (mesoscopic) and $h/2e$ (weak-localization) Aharonov-Bohm oscillations\cite{Ludwig_Mirlin_2004,TL_diss_06}.

We now present a similar calculation for the case of a fluctuating gauge field. This situation differs in the following points from the case of the screened Coulomb interaction: First, the characteristic time scales of the gauge field are much longer than the time scales of diffusion. This is important because, as will be discussed below, it results in an ensemble averaging effect which suppresses mesoscopic conductance fluctuations but not weak localization. On the other hand, weak localization is suppressed by the time-reversal symmetry breaking due to the fluctuating gauge field, while mesoscopic conductance fluctuations are insensitive to time-reversal breaking once it has resulted in the transition from the orthogonal to the unitary symmetry class.

\begin{figure*}[t]
\includegraphics[width=0.6\linewidth]{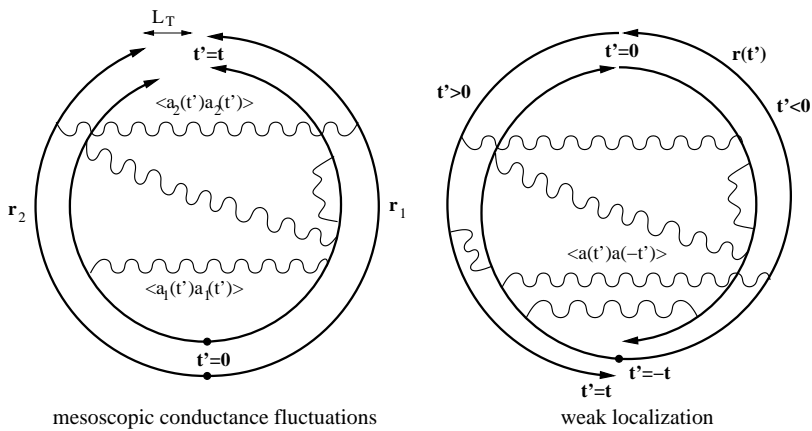}
\caption{\label{wl-ucf} Illustration of the path integral transformation which related weak localization to mesoscopic conductance fluctuations. The detailed presentation of the transformation can be found in Appendix~\ref{path_integral_trafo}. As indicated in the left part of this figure, the two paths only need to end within one thermal length $L_T$ of each other, changing the short-scale cutoff in Eq.~(\ref{wl_cf_highT}) (see Ref.~\onlinecite{Aleiner_Blanter_2002}).
For mesoscopic conductance fluctuations, vertex corrections (interaction lines connecting the two copies of the same path) are absent. For the Cooperon (right part of the figure), the vertex corrections are present and {\it add} to the self-energy terms.}
\end{figure*}

The detailed path-integral transformation is performed in Appendix~\ref{path_integral_trafo}, and illustrated in Fig.~\ref{wl-ucf}. The results have slightly different forms depending on temperature ranges.
For temperatures \mbox{$T>T_0/\lambda^2 g$} (the range of validity of the static approximation for the Cooperon dephasing), the thermal prefactor can be written for \mbox{$T\gg D/L^2$},
\bea
\left\langle\delta g^2\right\rangle(T)&=&
\frac{2\pi D}{3TL^2}\,
\biggl|\delta g_{\rm WL}\biggr|_{ T\to T/2, \ l\to L_T}\:\:\nonumber \\
&=& \frac{2D}{3 T L^2} \ln[T\tau_\varphi(T/2)],
\label{wl_cf_highT}
\eea
where the dimensionless conductance correction $\delta g_{\rm WL}=2\pi\delta\sigma_{\rm WL}/e^2$ is given by Eq.~(\ref{WL}) and the dephasing rate $1/\tau_\varphi(T)$ is given by Eq.~(\ref{tauphi-coop}).
Equation~(\ref{wl_cf_highT}) is the equivalent of Eqs.~(38) and (50) of Ref.~\onlinecite{Aleiner_Blanter_2002} which were derived for the standard screened Coulomb interaction.

The essential new feature of Eq.~(\ref{wl_cf_highT}) is that the interaction via a static gauge field has given rise to a relative factor of $1/2$ in the temperature argument of the dephasing time. The reason is that for the case of conductance fluctuations half of the possible gauge field correlators are between the two different measurements, so that only the other half of them remains.
This issue does not appear in the standard situation because the correlator of the screened Coulomb interaction is not slow but local in time. At lower temperatures, \mbox{$T<T_0/\lambda^2 g$} (when \mbox{$\tau_\varphi(T)\gg 1/T_0$}), the fully static approximation is no longer valid. The Cooperon dephasing rate, Eqs.~(\ref{tauphi-coop-3}) and (\ref{tauphi-coop-lowT}), becomes then twice smaller and the relative factor of $1/2$ in Eq.~(\ref{wl_cf_highT}) disappears.

The replacement \mbox{$l\to L_T$} is due to the different short-scale cutoffs of the two quantities involved: Mesoscopic conductance fluctuations are given by pairs of paths which end within a distance of \mbox{$L_T\equiv\left(D/T\right)^{1/2}$} of each other, while combining them to one closed loop, as needed for weak localization, requires them to end within one mean free path of each other.
This modification has no effect in \mbox{$d=1$} and only enters logarithmically in \mbox{$d=2$}. In Ref.~\onlinecite{Aleiner_Blanter_2002}, where the case of Coulomb interaction has been considered, this has been formulated using differences of the quantities on both sides, taken at different fields, so that the logarithmic cutoff drops out.

Equation (\ref{wl_cf_highT}) states that, also in the presence of a fluctuating gauge field, there is a deep relation between weak localization and mesoscopic conductance fluctuations, similar to the ones found in Refs.~\onlinecite{Aleiner_Blanter_2002,TL_diss_06}, and that these two quantities feature essentially the same dephasing rates. The time-reversal breaking effect of the gauge field on weak localization is mapped onto the ensemble-averaging effect of slowly varying gauge field configurations on mesoscopic conductance fluctuations. For many conceptual purposes, it is therefore convenient to study whichever of these two quantities is more accessible.

A generalization of these formulas involving the correlation between conductances at two magnetic field values and diffuson and Cooperon contributions with the difference and the sum of the magnetic field arguments appearing is straightforward in analogy to Ref.~\onlinecite{Aleiner_Blanter_2002}.

Since the connection between mesoscopic conductance fluctuations and weak localization has been made on the path-integral level, the corresponding generalizations to $h/e$ and $h/2e$ Aharonov-Bohm oscillations are straightforward, as in the standard case\cite{Aleiner_Blanter_2002,Ludwig_Mirlin_2004,TL_diss_06}.

\vspace*{0.5cm}
\subsection{Delayed-diffuson dephasing}
\label{diffuson_dephasing}
\vspace*{-0.5cm}
While the Cooperon and the UCF diffuson are dephased by all gauge field fluctuations, it is well known that a ``true'' diffuson is subject to particle number conservation. More precisely, while the total number of particles is conserved, the number of particles with given energy is not conserved if inelastic processes are taken into account. Then the diffuson at fixed particle energy acquires a dephasing rate, cutting off the infrared singularity. In position-time representation, the energy-dependence of the diffuson transforms into a dependence on the delay time $\eta$ between the particle and hole propagators.
In the limit of \mbox{$\eta\to 0$}, corresponding to integration over all energies, the full diffusion pole is restored.

The diffusons in the Hartree interaction diagrams have this intermediate character: Since they connect two different fermionic bubbles, and the gauge field has much slower dynamics than the diffusion processes, these diffusons allow for a delay time between the fermionic lines. A related situation has been investigated in Ref.~\onlinecite{Polyakov_Samokhin_98} in the context  of the second-loop weak localization and its dephasing due to Coulomb interaction.

In the presence of a fluctuating gauge field, the delayed diffuson satisfies the equation
\begin{widetext}
\be
\left\{
\partial_t
+D\Bigl[
-i\nabla
-\lambda e{\bf a}\left(\br,t+\eta/2\right)
+\lambda e{\bf a}\left(\br,t-\eta/2\right)
\Bigr]^2
\right\}{\cal D}_{\eta}(\br,\br',t)=\delta(\br-\br')\:\delta(t)\:\:.
\label{delayed_diffuson}
\ee
\end{widetext}
Clearly, fluctuations which are static on the scale of the delay time do not contribute to dephasing, so that in the limit of zero delay time there is no dephasing (this corresponds to the conservation of the total particle number),
\be
{\cal D}_{\eta=0}(\omega,q)=\frac{1}{Dq^2-i\omega}\:\:,
\ee
and in the limit of infinite delay time the result for the UCF diffuson is recovered,
\be
{\cal D}_{\eta=\infty}={\cal D}^{12}\:.
\ee
Unlike for the conductance-fluctuation diffuson (\ref{CF_diffuson}), the static approximation (\ref{static_correlator}) cannot be used directly, since it assumes that the dynamics of the gauge field is the slowest scale of the system (only the time separation between two independent measurements is longer).

To account for {\it finite} delay times $\eta$, we refine the static approximation by inspecting the diffuson analog of Eq.~(\ref{DeltaS-dyn}),
\begin{widetext}
\bea
\Delta S_\eta(t)
&=&
\frac{1}{2}\,\lambda^2e^2\int\limits_{-t}^{t}dt_1\int\limits_{-t}^{t}dt_2\:
\dot{r}_\alpha(t_1)\,\dot{r}_\beta(t_2)\nonumber\\
& &\times\biggl\{
\Bigl\langle a_\alpha[\br_1,(t_1+\eta/2)]\,a_\beta[\br_2,(t_2+\eta/2)]\Bigr\rangle
-\Bigl\langle a_\alpha[\br_1,(t_1+\eta/2)]\,a_\beta[\br_2,(t_2-\eta/2)]\Bigr\rangle\nonumber\\
& &\quad{}
-\Bigl\langle a_\alpha[\br_1,(t_1-\eta/2)]\,a_\beta[\br_2,(t_2+\eta/2)]\Bigr\rangle
+\Bigl\langle a_\alpha[\br_1,(t_1-\eta/2)]\,a_\beta[\br_2,(t_2-\eta/2)]\Bigr\rangle\biggr\}\:\:.
\label{DeltaS-diff}
\eea
\end{widetext}
The action $\Delta S_\eta$ of the delayed diffuson can be obtained from the action of the Cooperon, $\Delta S$, by defining the effective correlator
\be
\left\langle a_\alpha a_\beta\right\rangle_{\bk,\epsilon}^{\rm diff}(\eta)
=
\left\langle a_\alpha a_\beta\right\rangle_{\bk,\epsilon}
\left[1-{\rm cos}\,\epsilon\eta\right]
\label{effective_correlator_for_diffuson}
\ee
where the factor in the brackets\cite{Polyakov_Samokhin_98} arises from the combinations of the time arguments in Eq.~(\ref{DeltaS-diff}).
For the Cooperon in the static approximation the factor corresponding to the brackets in Eq.~(\ref{effective_correlator_for_diffuson}) is simply $2$ because in that case self-energy and vertex contributions {\it add} equally. In this context it is important to note that applying the static approximation (\ref{static_correlator}) and taking the limit of infinite $\eta$ in Eq.~(\ref{DeltaS-diff}) do not commute. For the UCF diffuson, the correct procedure used in Section~\ref{ucf_dephasing}, is to first send \mbox{$\eta\to\infty$}, which results in the vertex contributions dropping out.
We also remind that for the same reason, at very long times \mbox{$t\gg 1/T_0$} the static approximation is not applicable, and the Cooperon dephasing rate reduces to {\it half} the value given by Eq.~(\ref{tauphi-coop}), see Eqs.~(\ref{tauphi-coop-3}) and (\ref{tauphi-coop-lowT}).

For the delayed diffuson, the cosine term in Eq.~(\ref{effective_correlator_for_diffuson}) effectively removes the fluctuations which are slow on the time scale $\eta$ from the calculations for the conductance-fluctuation diffuson (Section~\ref{ucf_dephasing}): For short delay times \mbox{$\eta\ll 1/T$} it results in an extra suppression factor \mbox{$\frac{1}{2}\epsilon^2\eta^2$}. This suppression factor is due to the cancellation of self-energy and vertex terms imposed by particle number conservation.\cite{Polyakov_Samokhin_98} For long delay times \mbox{$\eta\gg 1/T$} the oscillating contribution drops out, resulting in the long-$\eta$ dephasing rate of the delayed diffuson being half the dephasing rate of the Cooperon in a {\it static} RMF, Eq.~(\ref{tauphi-coop}).

The contribution of thermal gauge field fluctuations relevant for dephasing of a diffuson with finite delay time $\eta$ can thus be written in analogy to Eq.~(\ref{static_correlator_def}) as
\be
\left\langle a_\alpha a_\beta\right\rangle_{\bk}^{\rm diff}(\eta)
=
\int\limits_{-T}^{T}\frac{d\epsilon}{2\pi}\:
\left\langle a_\alpha a_\beta\right\rangle_{\bk,\epsilon}^{\rm diff}(\eta)\:\:.
\label{eff_static_corr}
\ee
For \mbox{$\eta\gg 1/T$}, Eq.~(\ref{eff_static_corr}) can be approximated as
\begin{widetext}
\bea
\left\langle a_\alpha a_\beta\right\rangle_{\bk}^{\rm diff}(\eta)
&=&
\int\limits_{|\epsilon|>1/|\eta|}^{|\epsilon|<T}\frac{d\epsilon}{2\pi}\:
\left\langle a_\alpha a_\beta\right\rangle_{\bk,\epsilon}
=
\frac{2T}{\pi\chi_0 k^2}\:
\delta_{\alpha\beta}^\perp
\left[
{\rm arctan}\,T_k|\eta|
-{\rm arctan}\,\frac{T_k}{T}
\right]\:\:
\nonumber\\
&\approx&
\frac{2T}{\pi\chi_0 k^2}\:
\delta_{\alpha\beta}^\perp\:
{\rm arctan}\,T_k|\eta|\:\:,\quad T\gg T_0\:,\:\:|\eta|\gg 1/T\:\:,
\label{fast_modes}
\eea
where \mbox{$T_k\equiv\chi_0 k^2/\sigma_0=k^2l^2T_0$}.
For short delay times \mbox{$|\eta|\ll 1/T$}, we find
\bea
\left\langle a_\alpha a_\beta\right\rangle_{\bk}^{\rm diff}(\eta)
&\approx&
\int\limits_{-T}^T\frac{d\epsilon}{2\pi}\:
\left\langle a_\alpha a_\beta\right\rangle_{\bk,\epsilon}
\frac{1}{2}\,\epsilon^2\eta^2
=
\frac{T\eta^2}{\pi\sigma_0}
\left[
T-T_k\:{\rm arctan}\displaystyle\frac{T}{T_k}
\right]
\delta_{\alpha\beta}^\perp\nonumber \\
&\approx&
\left\{
\begin{array}{ll}
\displaystyle\frac{T^2\eta^2}{\pi\sigma_0}\:\delta_{\alpha\beta}^\perp\:\:,&T\gg T_k\:,\:\:|\eta|\ll 1/T\\[0.3cm]
\displaystyle\frac{T^4\eta^2}{3\pi\sigma_0T_k^2}\:\delta_{\alpha\beta}^\perp\:\:,&T\ll T_k\:,\:\:|\eta|\ll 1/T\:\:.
\end{array}
\right.\nonumber\\
& &\label{fast_modes_short_eta}
\eea
\end{widetext}
Using the correlator (\ref{fast_modes}) or (\ref{fast_modes_short_eta}) instead of (\ref{static_correlator}) in Eq.~(\ref{deltaS-raw}), we can derive a dephasing action for the delayed diffuson which corresponds to the dephasing action (\ref{deltaS-result-2d}) for the Cooperon.
In contrast to the Cooperon dephasing in a static RMF only gauge field fluctuations which are {\it fast} on the scale $\eta$ contribute. These fluctuations add up incoherently (instead of coherently for the static ones). The dephasing rate of the infinitely-delayed diffuson is therefore {\it half} the dephasing rate of the Cooperon in a truly static RMF, and the same as that of a Cooperon for times \mbox{$t\gg 1/T_0$}, when the gauge field cannot be regarded as static.

Compared to the situation in Section~\ref{Cooperon_dephasing}, finite delay times modify the low-$k$ cutoff $L_\omega^{-1}$ in Eq.~(\ref{deltaS-calc}) in the following way: \mbox{$L_\eta\equiv\left(\chi\eta/\sigma_0\right)^{1/2}=\left(T_0\eta\right)^{1/2}l$} replaces $L_\omega$ if it is shorter, modifying the cutoff of the logarithm. If $\eta$ is so short that \mbox{$L_\eta\approx l$}, the logarithm collapses and the otherwise subleading term becomes the dominant one. We find the following dephasing action for \mbox{$T_0\ll T\ll T_1/\lambda^2$},
\be
\Delta S^{\rm eff}_\eta(t)
\approx
\left\{
\begin{array}{ll}
\displaystyle 3\lambda^2gTt\ln{\displaystyle\frac{t}{\tau}}\:\:,&|\eta|\gg g^2t\\[0.2cm]
\displaystyle 3\lambda^2gTt\ln{\left(T_0|\eta|\right)}\:\:,&1/T_0\ll|\eta|\ll g^2t\\[0.2cm]
\displaystyle\frac{6}{\pi}\lambda^2gTtT_0|\eta|\:\:,&1/T\ll|\eta|\ll 1/T_0\\[0.3cm]
\displaystyle\frac{3}{\pi}\lambda^2gT^2tT_0\eta^2\:\:,&|\eta|\ll 1/T\:\:.
\end{array}
\right.
\label{DeltaS_delayed}
\ee
The dephasing rates are defined by the condition \mbox{$\Delta S^{\rm eff}_\eta(t=\tau_\varphi)=1$} and read
for \mbox{$T_0\ll T\ll T_1/\lambda^2$}
\be
\frac{1}{\tau_\varphi(\eta)}
\approx
\left\{
\begin{array}{ll}
\displaystyle 3\lambda^2gT\ln{(T_0\eta^*)}\:\:,&1/T\ll 1/T_0\ll|\eta|\\[0.2cm]
\displaystyle \frac{6}{\pi}\lambda^2gTT_0|\eta|\:\:,&1/T\ll|\eta|\ll 1/T_0\\[0.3cm]
\displaystyle\frac{3}{\pi}\lambda^2gT^2T_0\eta^2\:\:,&|\eta|\ll 1/T\ll 1/T_0\:\:,
\end{array}
\right.
\label{tauphi_fastmodes}
\ee
where
\be
\eta^*={\rm min}\left\{|\eta|,\:g/\lambda^2T\right\}\:\:.
\label{eta-star}
\ee

For later reference (to ensure that we may neglect dephasing in Section~\ref{low_T}), we also estimate the dephasing action for the case \mbox{$T\ll T_0$}. Then,  in addition to the low-$k$ cutoff, a high-$k$ cutoff appears, such that \mbox{$k\lesssim l_T^{-1}\equiv\left(T/T_0\right)^{1/2}l^{-1}$}. This condition is stronger than the cutoff by the elastic mean free path at low temperatures, \mbox{$T\lesssim T_0$}, see Eq.~(\ref{deltaS-result-lowT}). As a result, we find the dephasing action
\be
\Delta S^{\rm eff}_\eta(t)
\approx
\left\{
\begin{array}{ll}
\displaystyle 3\lambda^2gTt\ln{\displaystyle\frac{t}{\tau^*}}\:\:,&|\eta|\gg g^2t\\[0.2cm]
\displaystyle 3\lambda^2gTt\ln{\left(T|\eta|\right)}\:\:,&1/T\ll|\eta|\ll g^2t\\[0.2cm]
\displaystyle\frac{4}{\pi}\lambda^2gT^3\eta^2t\:\:,&|\eta|\ll 1/T
\end{array}
\right.
\label{DeltaS_delayed_lowT}
\ee
with \mbox{$\tau^*=\left(T_0/T\right)\tau$}.
Equation~(\ref{DeltaS_delayed_lowT}) gives the following dephasing rates for \mbox{$T\ll T_0$},
\be
\frac{1}{\tau_\varphi(\eta)}
\approx
\left\{
\begin{array}{ll}
\displaystyle 3\lambda^2gT\ln{(T\eta^*)}\:\:,&1/T_0\ll 1/T\ll|\eta|\\[0.2cm]
\displaystyle\frac{4}{\pi}\lambda^2gT^3\eta^2\:\:,&|\eta|\ll 1/T\:\:.
\end{array}
\right.
\label{tauphi_fastmodes_lowT}
\ee

It should be noted that in the long-time limit \mbox{$t\to\infty$} the quasiclassical approximation (employed in the path-integral calculation) breaks down. This occurs at the time scale \mbox{$t\gg t^*\sim E_F\eta\tau_\varphi(\eta)$}, which in particular satisfies \mbox{$t^*\gg\tau_\varphi(\eta)$} for any $\eta$.
The reason for the breakdown is that the two quasiclassical trajectories can eliminate the delay (and therefore further suppression) by quantum ``tunneling'' (assuming non-classical velocities during some time). This happens at the cost of an extra phase difference which is, however, smaller than the one for the delayed paths with classical velocities. As a result, for \mbox{$t\gg t^*$} the diffuson is no longer decaying with increasing time $t$; particle number conservation and the corresponding diffusion pole for small frequencies are restored in the long-time limit. However the weight of the diffusion pole is exponentially small due to the suppression factor associated with the non-classical pieces of the trajectories. The results of this paper are not affected by the breakdown of the quasiclassical approximation since all relevant time scales are shorter than $t^*$.

It is also worth mentioning that the virtual interaction processes renormalize the quasiclassical trajectories (velocity, diffuson constant). For the case of weak coupling considered above, these effects are negligible. However, they become important in the strong coupling limit, see Section~\ref{strong}.

\subsection{Diffuson dephasing: two-loop localization correction}
\label{2-loop-dephasing}

Since we have understood in the previous sections that dephasing of Cooperons and diffusons in the presence of a slowly fluctuating gauge field are mostly time-reversal breaking and ensemble-averaging effects, respectively, rather than true rates of loss of phase memory of the fermions, it is natural to ask if it is possible to avoid these effects and access the ``true'' dephasing rates. The diffuson contribution to the two-loop (second order in $g$) weak localization is sensitive to neither time-reversal breaking (since it contains no Cooperons) nor ensemble-averaging (since weak localization already is an ensemble-averaged quantity, corresponding to the ``same measurement'', in contrast to the UCF diffusons).
Therefore the diffuson two-loop weak localization is not expected to be subject to the very high dephasing rates applicable to the Cooperons and the UCF diffusons. (It should be noted, however, that this two-loop correction to the conductivity is very hard to study experimentally, since it is insensitive to magnetic fields and much smaller than interaction corrections.)

The diffuson contribution to two-loop weak localization correction $\delta\sigma_{\rm WL}^{\rm D}$ includes the structure\cite{Gorkov_Larkin_Khmelnitskii_79,Polyakov_Samokhin_98}
\be
A_2=
\int\limits_0^t dt'
\left\langle
{\cal D}_{t'-t}(\br,\br,t')\:
{\cal D}_{t'}(\br,\br,t-t')
\right\rangle
\label{two-loop-WL}
\ee
with two delayed diffusons of the type (\ref{delayed_diffuson}) along with a similar structure $A_3$ consisting of three delayed diffusons.
As already seen in Eq.~(\ref{delayed_diffuson}), a gauge field which does not change on the scale of the delay time drops out\cite{Polyakov_Samokhin_98,TL_diss_06}. Here the delay times are given by the duration of the respective other path. It should be noted that anomalous sets of paths with one very small loop, which are not suppressed strongly because the short loop gives a short delay time of the other loop, drop out because of a cancellation of the contribution (\ref{two-loop-WL}) with a three-diffuson contribution.\cite{Polyakov_Samokhin_98} As a result, relevant $t$ and \mbox{$t-t'$} in Eq.~(\ref{two-loop-WL}) are of the same order.
To estimate the dephasing time associated with the two-loop correction, we may therefore self-consistently set
\mbox{$1/\tau_\varphi^{\rm D}(\eta=\tau_\varphi^{\rm D})=1/\eta$} in the respective delayed-diffuson dephasing rates (\ref{tauphi_fastmodes}) and (\ref{tauphi_fastmodes_lowT}).
In the weak coupling regime, $\lambda^2g\ll 1$, the results are
\be
\frac{1}{\tau^{\rm D}_\varphi}
\sim
\left\{
\begin{array}{lll}
\lambda^2gT\:\:, &T\ll  T_0/\lambda^2 g \:,\\
\lambda\left(TT_1\right)^{1/2}\:\:,& T_0/\lambda^2 g\ll T\ll  T_1/\lambda^2\:.
\end{array}
\right.
\label{tauphi-2loop}
\ee
As expected, $1/\tau^{\rm D}_\varphi$ is smaller than the Cooperon dephasing rate $1/\tau_\varphi$ as given by Eqs. (\ref{tauphi-coop}), (\ref{tauphi-coop-3}) and (\ref{tauphi-coop-lowT}).
This is at variance with the conventional Coulomb interaction, for which the diffuson dephasing rate in the second-loop weak localization correction is of the same order\cite{Polyakov_Samokhin_98} as the Cooperon dephasing rate.

In the strong coupling regime \mbox{$\lambda^2 g \gg 1$}, the second-loop diffuson dephasing rate becomes of the order of the temperature, owing to the interaction-induced renormalization of the paths contributing to the weak-localization correction, see the discussion in Section~\ref{dephasing-strong-coupling}.

As mentioned in Section~\ref{Cooperon_dephasing}, for the half-filled lowest Landau level the time-reversal symmetry is broken because of the strong magnetic field. In the context of composite fermions this manifests itself in the way that also the static scalar impurities acquire a vector component due to screening. Therefore weak localization is absent in the first order in $1/g$, and the two-diffuson contributions discussed here give the leading localization correction. This correction, along with the interaction correction at low temperatures which we will discuss in Section~\ref{low_T}, strongly indicates that the system of composite fermions interacting via a Chern-Simons gauge field is localized in the limit \mbox{$T\to 0$}.

\section{Strong coupling: Hartree correction and dephasing}
\label{strong}
\setcounter{equation}{0}

\subsection{Diffuson self-energies}
\label{diff_self_energy}

Let us now discuss the situation of strong coupling, \mbox{$\lambda^2g \gg 1$}. Since the first-order result (\ref{delta_sigma}) is not a small correction then, it is necessary to take the interaction into account to all orders.
We have to include both virtual (renormalization) processes, which are determined by the real part of the interaction propagator, and real (dephasing) processes, which  are determined by the imaginary part of the interaction propagator. For the conceptual framework of treatment of interaction effects in disordered systems, we refer the reader to Refs.~\onlinecite{AAG_99,GMP07,vonDelft_etal_05}.

The virtual processes are taken into account by inserting the self-energies calculated diagrammatically in Appendix~\ref{app_selfenergies} into the delayed diffusons. The treatment of the dephasing processes by using the path-integral method as discussed in Section~\ref{diffuson_dephasing} is complemented by the diagrammatic calculation of the dephasing-induced self-energy of the delayed diffusons.

We first calculate the self-energies in the first order in the effective interaction for \mbox{$\lambda\ll 1$}. As we are going to show below, neither Hartree conductivity correction nor the dephasing rate depend on $\lambda$ already for \mbox{$\lambda^2 \gg 1/g$}. This allows us to evaluate these quantities in the relevant case of $\lambda=1$ using the results derived for \mbox{$1/g\ll \lambda^2 \ll 1$}, up to numerical prefactors (stemming from the contribution of higher-order interaction terms in the interaction blocks).
The situation is somewhat similar to the conventional Coulomb interaction case, where the Hartree ladder in the triplet channel depends on Fermi-liquid constants~\cite{ZNA}, which makes it impossible to calculate analytically the numerical coefficients in the conductivity corrections for \mbox{$r_s\gtrsim 1$} (where $r_s$, an analog of $\lambda$ here, is the standard gas parameter) starting from the microscopic theory.

It is important that higher-order-in-$\lambda$ contributions do not lead to any further singularities, in contrast to the clean situation~\cite{gauge}.
The expansion of the effective interaction in $\lambda$ is regular in the present disordered case, since disorder cures the infrared singularities arising in the clean~\cite{gauge} theory. 
The two new (as compared to the clean situation) characteristic energy scales are introduced by disorder: $T_0$ and $1/\tau$. 
On scales longer than the mean free path the dynamics of the system is diffusive.
In the first-order interaction terms, the infrared singularity at low momenta $\bk$ transferred through the gauge-interaction lines is cut off by additional diagrams involving impurity ladders (see Fig.~\ref{boxes}), as discussed in Section~\ref{Hartree_wo_dephasing}.
Furthermore, a resummation of the higher-order interaction terms in the fermionic self energies is not needed in the diffusive regime, i.e. as long as the first-order self-energy does not exceed $1/\tau$.

The peculiarity of the interaction block $\tilde{U}(q)$ (thick dotted line in Fig.~\ref{boxes}) is that its magnitude is very large \mbox{($\propto g$)} in a narrow window of small momentum transfers \mbox{$q\lesssim 1/l$}, see Eq.~(\ref{tilde_U}). A similar situation takes place in a normal metal with Coulomb interaction, where the bare interaction is singular \mbox{($\propto 1/q$)} and hence can be arbitrarily strong in the limit of small $q$. The screening of Coulomb interaction in normal metals is described by the RPA and results in a much less singular effective interaction, Eq.~(\ref{CoulRPA}). Therefore, in analogy to the standard situation, we first consider the resummation of an infinite number of higher-order virtual processes. This can be described by an integral equation for the diffuson which sums up blocks of the (real part of the) effective interaction in an RPA-inspired way, yielding a renormalized diffuson. The dephasing will be included later on in Section~\ref{disconnected-diff}.

\begin{figure}
\parbox{0.23\linewidth}{\includegraphics[height=0.9\linewidth]{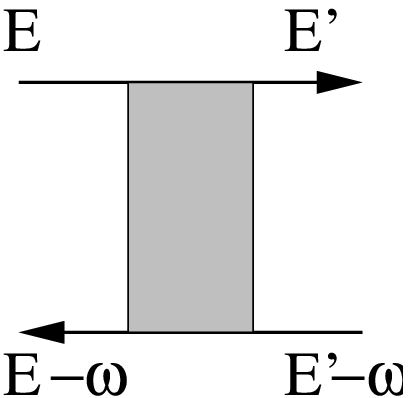}}
$=$
\parbox{0.22\linewidth}{\includegraphics[height=0.7\linewidth]{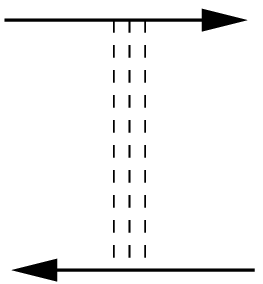}}
$+$
\parbox{0.44\linewidth}{\includegraphics[height=0.35\linewidth]{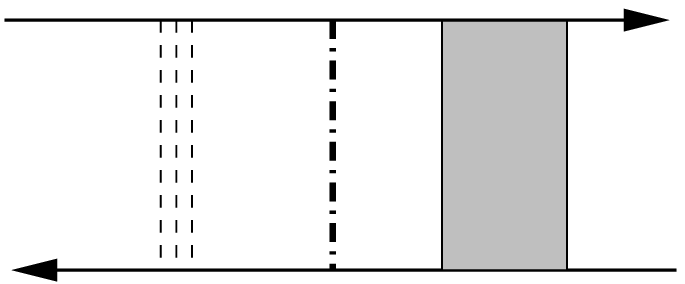}}\\[0.5cm]
$=$
\parbox{0.22\linewidth}{\includegraphics[height=0.7\linewidth]{bare_diffuson.eps}}
$+$
\parbox{0.44\linewidth}{\includegraphics[height=0.35\linewidth]{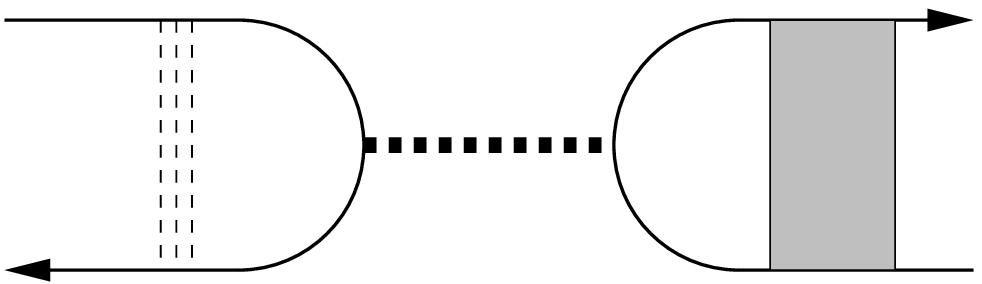}}
\caption{\label{diffuson-ladder-fig}
The interaction-dressed diffuson described by Eq.~(\ref{diffuson-vertex}).
The impurity ladders here denote the ``disconnected diffusons'' $\tilde{\cal D}_\infty$ with only self-energy interaction lines included;
the vertex interaction (dash-dotted) line enters along with the additional diagrams from Fig.~\ref{boxes} which form the effective interaction $\tilde U$ (thick dotted line).}
\end{figure}

As a result of the resummation, we will find the renormalized one-loop Hartree correction to the conductivity. It still contains only one Hartree fermionic bubble, as the perturbative correction in Section~\ref{Hartree_wo_dephasing}, so that all diffusons form a single loop. Another source of higher-order Hartree corrections is provided by higher-loops diagrams  with many Hartree fermionic bubbles attached to the main ``conductivity bubble'' (before disorder averaging they correspond to diagrams with many tadpoles). These diagrams are relevant in the situation when the one-loop result exceeds the Drude conductivity.
An efficient way of resummation of such diagrams (a certain type of self-consistent approximation) has been proposed in Ref.~\onlinecite{Kee_Aleiner_Altshuler_98} in the context of the tunneling density of states in superconductors. However, as we are going to show, in the present problem, the renormalization of diffusons prevents the one-loop Hartree correction from being larger than the Drude conductivity. Therefore, the additional resummation~\cite{Kee_Aleiner_Altshuler_98} of higher-loop diagrams is not needed.

\begin{figure}
\vskip1cm
\includegraphics[width=1.0\linewidth]{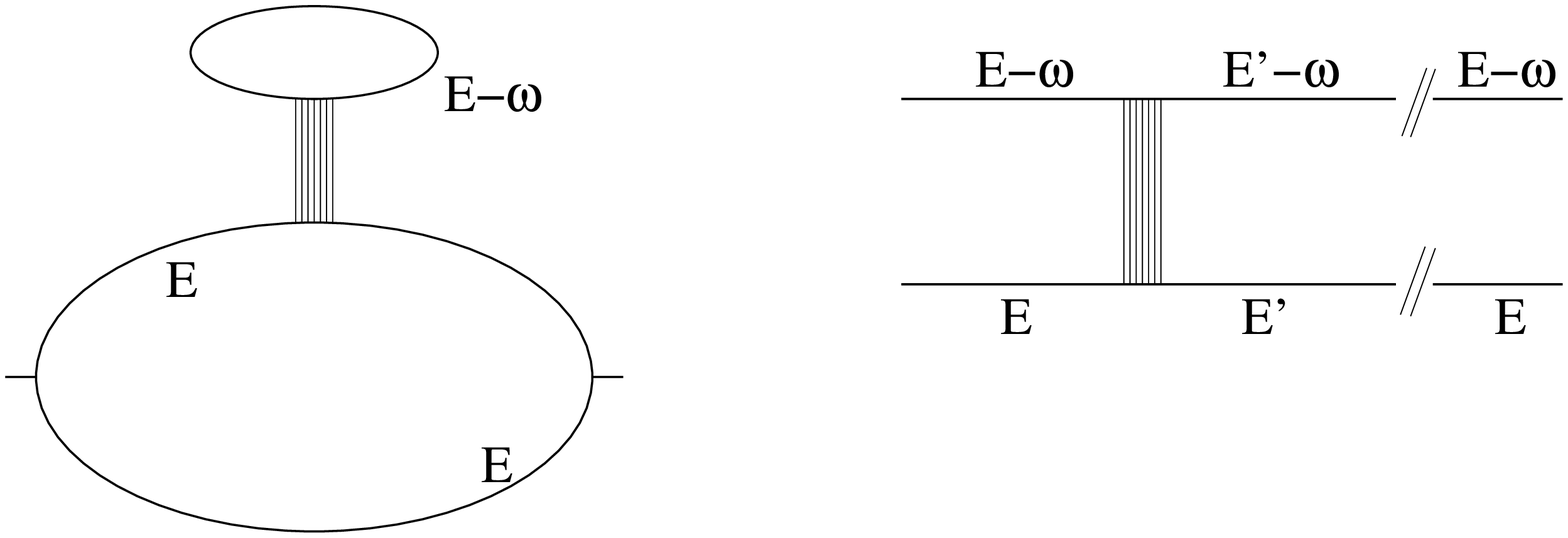}
\caption{\label{diffuson_energies}
The energies of the renormalized diffusons in the Hartree diagrams satisfy
\mbox{$E,E'\lesssim T$} and \mbox{$\omega\gtrsim T$}.
}
\end{figure}

\begin{figure}
[t]
\includegraphics[width=0.4\linewidth]{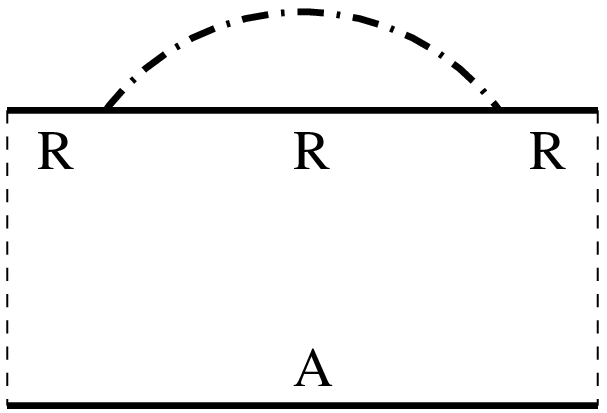}
\hfill
\includegraphics[width=0.4\linewidth]{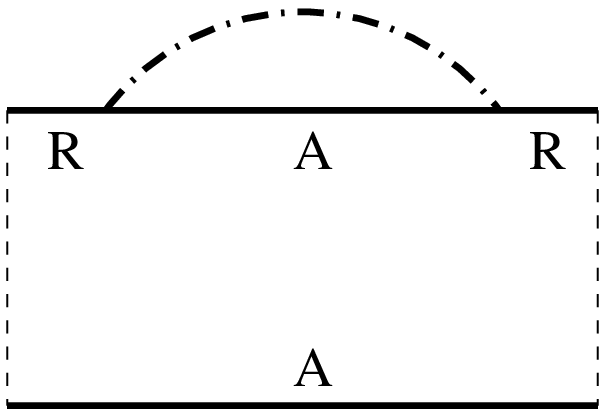}
\bigskip

\includegraphics[width=0.4\linewidth]{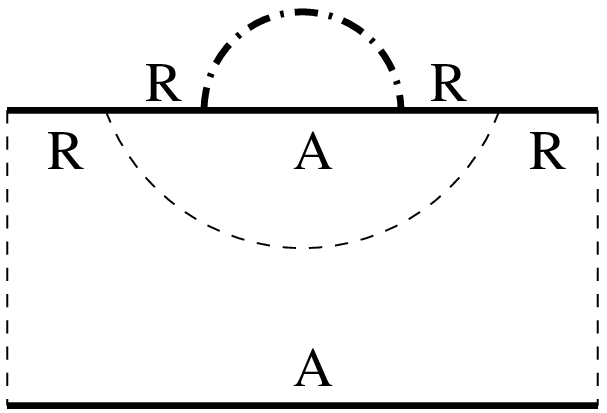}
\hfill
\includegraphics[width=0.4\linewidth]{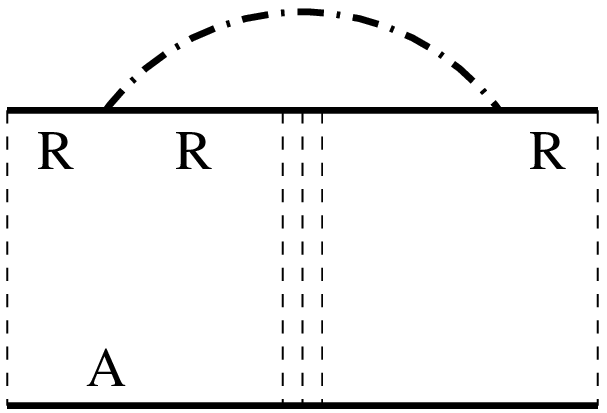}
\caption{\label{selfenergy-boxes}
Self-energy counterparts to the vertex interaction (Fig.~\ref{boxes}), resulting in the self-energy (\ref{disconnected-selfenergy-main}). In addition to the diagrams shown, equivalent possibilities to insert self-energy lines into the advanced Green's function exist. The detailed calculation can be found in Appendix~\ref{app_selfenergies}.
}
\end{figure}

In the presence of interaction, the energies of the (retarded and advances) Green's functions forming the diffuson may change due to the vertex interaction lines transferring finite energy from the upper to the lower fermionic line. Therefore, the interaction-dressed diffuson not only depends on the difference $\omega$ between the energies of two (retarded and advances) Green's functions but also on the incoming and outcoming energies, $E$ and $E'$ (for definiteness, these are the energies of the retarded Green's functions). The diffuson $\tilde{{\cal D}}(E,E';\omega,q)$  dressed by vertex and self-energy gauge-field interaction lines satisfies the equation (see Figs.~\ref{diffuson-ladder-fig} and \ref{diffuson_energies})
\begin{widetext}
\be
\tilde{{\cal D}}(E,E';\omega,q)
=
\delta(E-E')\,{\cal D}_0(\omega,q)
+2\pi \nu {\cal D}_0(\omega,q) \int\limits_{E-\omega}^E \frac{d\epsilon}{2\pi}\:
(-i){\rm Re}\,\tilde{U}(\epsilon)
\left[\tilde{\cal D}(E-\epsilon,E';\omega,q)-\tilde{\cal D}(E,E';\omega,q)\right]\:\:.
\label{diffuson_ladder}
\ee
\end{widetext}
Here we set \mbox{$T=0$} for simplicity. At finite temperature a standard combination of thermal factors \mbox{$\tanh[(E-\epsilon)/2T]-\tanh[(E-\omega-\epsilon)/2T]$} appears which in effect softens the limits of $\epsilon$-integration on the scale of $T$ (for $\omega\gg T$ this effect is inessential). The first term in brackets under the integral in Eq.~(\ref{diffuson_ladder}) is the vertex interaction part (see Fig.~\ref{boxes}), the second one is the self-energy part (see Fig.~\ref{selfenergy-boxes}), and
\be
{\cal D}_0=\frac{1}{Dq^2-i\omega}
\label{D0barediff}
\ee
is the bare (noninteracting) diffuson. The effective interaction $\tilde{U}(\epsilon)$ corresponds to the interaction block in Fig.~\ref{boxes} in the vertex part; in the self-energy part, the same structure arises from diagrams in Fig.~\ref{selfenergy-boxes}. The real part of $\tilde{U}(\epsilon)$ involved in the renormalized diffuson (\ref{diffuson_ladder}) reads (see Appendix~\ref{app_selfenergies}):
\be
{\rm Re}\,\tilde{U}(\epsilon)\simeq
\frac{3\lambda^2 g}{\pi \nu} 
\left\{
\begin{array}{ll}
\displaystyle \ln{\frac{1}{q^2l^2}}\:,&  |\epsilon|\ll T_0 q^2l^2\\[0.5cm]
\displaystyle     \ln{\frac{T_0}{|\epsilon|}}\:,&  T_0 q^2l^2\ll|\epsilon|\ll T_0\\[0.5cm]
\displaystyle \frac{4\pi}{3} \left(\frac{T_0}{2|\epsilon|}\right)^{1/3}\!\!,& |\epsilon|\gg T_0, \end{array}
\right.
\label{Re-tilde-U-epsilon}
\ee

It is worth recalling that here we are dealing with spinless (spin-polarized) fermions (which is in particular the case for the composite fermions at the lowest Landau level).
At variance with the singlet channel in the standard case, the Hartree gauge-field ladder is not affected by the possible insertion of gauge-field ``exchange" parts. The reason is that, due to the vector character of the vertices, no disorder lines (and hence extra diffusons) may separate the exchange interaction line from the adjacent Hartree interaction lines. Therefore, inserting an exchange interaction line gives a result which is smaller by a factor of the order of \mbox{$q^2l^2\ll 1$}.

The range of energy integration for the single self-energy block  is \mbox{$\left[E-\omega,E\right]$} and hence depends on the total energy of the diffuson.
The full ladder is part of a Hartree diagram, which features \mbox{$E\lesssim T$} (the energy in the ``conductivity bubble'' is restricted by the thermal factors from the Kubo formula), but \mbox{$E-\omega$} (energies of the attached bubble) are unbounded from below and characterized by \mbox{$\omega\gtrsim T$}, owing to the standard thermal factor $(\partial/\partial\omega)[\omega \coth(\omega/2T)]$ in the interaction correction to the conductivity, see Fig.~\ref{diffuson_energies}.
It is convenient to introduce a ``disconnected'' diffuson $\tilde{\cal D}_\infty \delta(E-E')$ dressed only by self-energy lines,
\bea
\tilde{{\cal D}}_\infty
&=&
\frac{\,{\cal D}_0}{1+2\pi\nu\,{\cal D}_0\int\limits_{E-\omega}^{E}\frac{d\epsilon}{2\pi}\:(-i){\rm Re}\,\tilde{U}(\epsilon)}\:\:\nonumber\\
&=&
\frac{1}{Dq^2-i\omega- i\nu\int\limits_{E-\omega}^{E}d\epsilon\:{\rm Re}\,\tilde{U}(\epsilon)}\:\:.\nonumber \\
\label{disconnected_E}
\eea
For the disconnected diffuson, the fact that we are interested in $\omega\gtrsim T \gtrsim E$ allows us to neglect $E$ in the limits of the $\epsilon$-integration.

In the full diffuson (\ref{diffuson_ladder}) including the vertex lines, the energy is no longer conserved along the fermionic Green functions, so that the $E$-dependence of the integration limits becomes important. Although we are still interested in \mbox{$\omega\gg E$}, the integral equation for the full diffuson involves propagators with \mbox{$E\sim \omega$} at the intermediate steps of the ladder, thus making the exact analytical solution of Eq.~(\ref{diffuson_ladder}) impossible.
In what follows we will simplify the equation for the diffuson, neglecting the energy dependence of the integration limits. 

This approximation, which is closely related to that of Ref.~\onlinecite{vonDelft_etal_05}, while giving the correct $T$-dependence of the Hartree conductivity correction, does not allow us to find the exact numerical prefactor at low temperatures.
This prefactor, however, is not too important since the low-$T$ dependence of the total conductivity correction will be dominated by the exchange contribution, as we will show below.
In this regime (\mbox{$T\ll T_0$}), the relevant delay times are short, \mbox{$\eta\lesssim 1/\omega$}, and fluctuate within the window $\omega^{-1}$ from one step of the diffuson ladder to another.
At higher temperatures, when the conductivity correction is dominated by long \mbox{$\eta\gg 1/\omega$}, the diffuson delay time is well defined and our approximation is controlled by the parameter \mbox{$\omega/T\gg 1$} [see, e.g., Eq.~(\ref{111}) below, which is governed by $T<\omega<T_0$].

Within the approximation described above, Eq.~(\ref{diffuson_ladder}) is replaced by the equation for the diffuson which now depends only on the difference of the two energies \mbox{$E-E'$}, corresponding to a fixed delay time $\eta$ in the time domain (to simplify notation we do not write the diffuson frequency $\omega$ and momentum $q$; we also assume $T=0$ here):
\begin{widetext}
\be
\tilde{{\cal D}}(E-E')
=
{\cal D}_0 \, \delta(E-E')
+2\pi \nu {\cal D}_0\int\limits_{-\omega}^0 \frac{d\epsilon}{2\pi}\:
(-i){\rm Re}\,\tilde{U}(\epsilon)
\left[\tilde{\cal D}(E-\epsilon-E')-\tilde{\cal D}(E-E')\right]\:\:.
\label{diffuson_ladder_approx}
\ee
\end{widetext}
It can be solved by Fourier transformation to the (delay) time domain with respect to \mbox{$E-E'$}:
\be
\tilde{\cal D}_\eta(q,\omega)=\frac{1}{Dq^2-i\omega-i \Sigma_\eta^Z}\:\:,
\ee
with
\be
\Sigma_\eta^Z=2\pi \nu\int\limits_{-\omega}^0 \frac{d\epsilon}{2\pi}\:
{\rm Re}\,\tilde{U}(\epsilon)\left[1-\cos\epsilon\eta\right]\:\:.
\label{Sigma-U-eta}
\ee

Equation (\ref{diffuson_ladder_approx}) for the fully renormalized diffuson can be rewritten
as
\begin{widetext}
\be
\tilde{\cal D}(E-E')
=
\tilde{\cal D}_\infty \delta(E-E')
+2\pi \nu\tilde{\cal D}_\infty\int\limits_{-\omega}^0 \frac{d\epsilon}{2\pi}\:
(-i){\rm Re}\,\tilde{U}(\epsilon)\:\tilde{\cal D}(E+\epsilon-E')\:\:,
\label{diffuson-vertex}
\ee
\end{widetext}
in terms of the disconnected diffuson \mbox{$\tilde{\cal D}_\infty$} given by Eq.~(\ref{disconnected_E}), with $E$ neglected in the limits of the $\epsilon$-integration.

Let us discuss how the delayed diffusons $\tilde{\cal D}_\eta$ enter the Hartree correction. For the lowest-order Hartree diagrams in Fig.~\ref{Hartree-diags}, the condition of zero energy transfer through the single interaction line can be rewritten by the substitution
\be
(-i){\rm Re}\,\tilde{U}(q,\epsilon=0)
\cdot
{\cal B}
\to
\int d\eta\:
(-i)\: u(\eta)\:
{\cal B}(\eta)\,,
\label{Fourier_substitution}
\ee
where $u(\eta)$ is the Fourier transform of the real part ${\rm Re}\,\tilde{U}(\epsilon)$ [Eqs.~(\ref{Re-tilde-U-epsilon-k}) and (\ref{Re-tilde-U-epsilon})] of the effective interaction $\tilde{U}(\epsilon)$,
\bea
 u(\eta)
\approx
\left\{
\begin{array}{ll}
\displaystyle \frac{3 \lambda^2 g}{2\pi \nu}\, \gamma_1 T_0^{1/3}\eta^{-2/3}\!,\quad & \quad \displaystyle 0<|\eta|\ll\frac{1}{T_0} \\
&\\
\displaystyle\frac{3 \lambda^2 g}{2\pi \nu |\eta|}\:,& \displaystyle\frac{1}{T_0}\ll|\eta|\ll \frac{1}{q^2l^2T_0}\:,\end{array}
\right.
\nonumber \\
\label{Re-tilde-U-eta}
\eea
[with $\gamma_1=\Gamma(5/3)/2^{1/3}$, where  $\Gamma(x)$ is the gamma-function]
and ${\cal B}$ is the fermionic part of the diagram.

Using the diagrams Fig.~\ref{Hartree-diags}$a$) and \ref{Hartree-diags}$b$), we can rewrite the first-order Hartree correction
in terms of the delayed interaction $u(\eta)$, 
\bea
\delta\sigma^H
&\approx&
\frac{\sigma_0}{2} \; {\rm Re}
\int\limits_{-\infty}^\infty\frac{d\omega}{2\pi}\:
\frac{\partial}{\partial\omega}\left[\omega\,{\rm coth}\frac{\omega}{2T}\right]
\nonumber \\
&\times&\int\limits_{-\infty}^\infty d\eta
\int(dq)\:
{\cal D}_0(\omega,q)(-i)u(\eta){\cal D}_0(\omega,q).
\nonumber \\
\label{perturbative-correction}
\eea
In this first-order correction, the fermionic part ${\cal B}$ does not depend on the delay time.
The correction (\ref{perturbative-correction}) is dominated by long delay times $\eta\gg 1/T_0$,
the integration over $\eta$ yielding the logarithm in (\ref{tilde_U}), which leads to the $\ln^2 T$ temperature dependence
(\ref{delta_sigma}) found in Section~\ref{first_order}.

For a more refined treatment including higher-order interaction terms, it is necessary to keep track of the energy arguments $E$ and $E'$ of the diffusons.
Then the Hartree correction to the conductivity can be evaluated using the diagrams which now should contain diffusons renormalized by additional self-energy and vertex interaction lines. This implies that the fermionic part ${\cal B}$ of the diagrams becomes $\eta$-dependent, since there is a 
time difference between the Hartree bubble and the main ``conductivity fermionic loop'', owing to the slow dynamics of gauge fields.

\begin{figure}[t]
\includegraphics[width=0.8\linewidth]{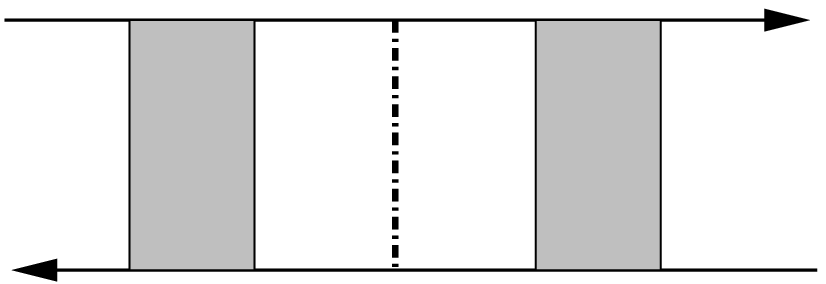}\\
$=\quad$\parbox{0.8\linewidth}{\includegraphics[width=1.0\linewidth]{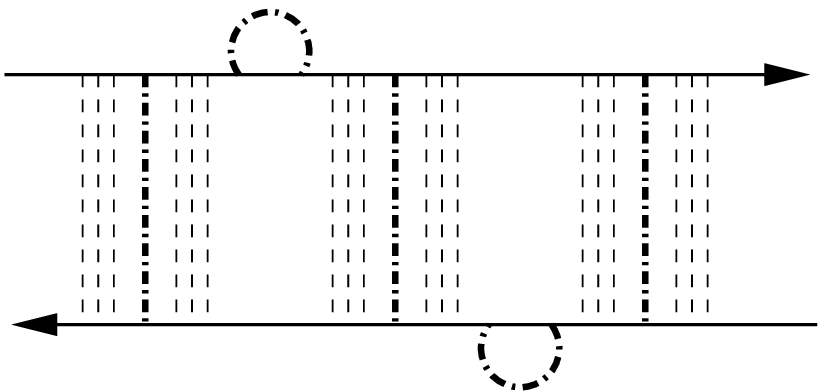}}\\
$-\quad$\parbox{0.8\linewidth}{\includegraphics[width=1.0\linewidth]{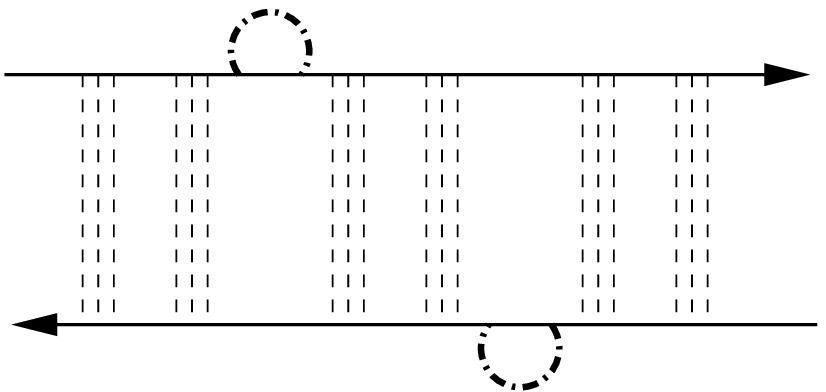}}
\caption{\label{diffuson-difference}
In the presence of self-energy and vertex interactions dressing the diffusons, the effective interaction block $\tilde{U}$ together with the two adjacent dressed diffusons can be more conveniently considered as one diffuson renormalized by the self-energy and vertex interaction lines, and the disconnected part subtracted to ensure the Hartree structure of at least one interaction line connecting the fermionic bubbles.
}
\end{figure}

In technical terms, in the first-order interaction diagrams in Fig.~\ref{AA-diags}, each effective interaction line ${\tilde U}$ is
dressed from both sides by impurity ladders, see, e.g., Eq.~(\ref{perturbative-correction}).
In higher-order diagrams, the product of the effective interaction $\tilde{U}$ with the two adjacent bare diffusons is replaced by a single fully dressed diffuson $\tilde{{\cal D}}_\eta$, minus the completely disconnected contribution $\tilde{{\cal D}}_\infty$ (with only self-energy interaction lines but no vertex interaction line), see Fig.~\ref{diffuson-difference}:
\be
\left[{\cal D}_0\,(-i){\rm Re}\{\tilde{U}\}\,{\cal D}_0\right]_\eta
\to
\frac{1}{2\pi \nu}\left[\tilde{{\cal D}}_\eta-\tilde{{\cal D}}_\infty\right]\:\:.
\ee
The subtraction of the disconnected part ensures the Hartree structure of the contribution under consideration: at least one vertex interaction line connects the two fermionic bubbles. This formula will be used in Section~\ref{low_T} for \mbox{$T\ll T_0$}, when dephasing can be neglected.

Transforming the fully-dressed diffuson (\ref{diffuson-vertex}) to delay time space and subtracting the disconnected part $\tilde{{\cal D}}_\infty$, we find
\be
\frac{1}{2\pi \nu}\left[
\tilde{{\cal D}}_\eta-\tilde{{\cal D}}_\infty
\right]
=
\tilde{{\cal D}}_\infty\:(-i)u(\eta)\:\tilde{{\cal D}}_\eta\:\:.
\label{D-D=DUD}
\ee
Since \mbox{${\rm Im}\,\tilde{U}$} enters the dephasing self-energy \mbox{$\Sigma^\varphi_\eta$}, only \mbox{${\rm Re}\,\tilde{U}$} should be kept for the renormalization.
Equation (\ref{D-D=DUD}) explicitly singles out a Hartree (renormalization) interaction line. The dephasing is then included into the diffusons on the right-hand side of (\ref{D-D=DUD}). This expression will be used in Section~\ref{high_T} for calculation of the Hartree conductivity correction at intermediate temperatures \mbox{$T_0\ll T \ll T_1$}, when dephasing may be strong compared to the renormalization effects.

\subsubsection{Disconnected diffuson}
\label{disconnected-diff}

Now we include the dephasing into the delayed diffuson. We start with the simpler case of a disconnected diffuson (infinite delay time), which has already appeared in Section~\ref{ucf_dephasing} as the UCF-diffuson. As discussed in Appendix~\ref{app_selfenergies}, the disconnected diffuson $\tilde{{\cal D}}_\infty$ has the form
\be
\tilde{{\cal D}}_\infty
=
\frac{1}{Dq^2-i\omega-i\Sigma^Z_\infty+\Sigma^\varphi_\infty}
\ee
with the renormalization part of the self-energy
\be
\Sigma^Z_\infty
=
\frac{3 \lambda^2}{2\pi}\,g\,\omega\cdot f_Z\left(\frac{T_0}{\omega}\right)
\label{disconnected-selfenergy-main}
\ee
where $f_Z(x)$ is a slowly varying function
\be
f_Z(x)
\approx
\left\{
\begin{array}{ll}
\ln{x}+1\:\:,&x\gg 1\\
\pi\left(4x\right)^{1/3}\:\:,&x\ll 1\:\:.
\end{array}
\right.
\ee
The high-$\omega$ behavior of Eq.~(\ref{disconnected-selfenergy-main}) is reminiscent of the ballistic behavior in a clean system.\cite{HLR_93}
The self energy $\Sigma^Z_\infty$ can be cast in the form
\be
\Sigma^Z_\infty=\omega [Z(\omega)-1],
\ee
where $Z(\omega)=1+(3 \lambda^2 g/2\pi)f_Z(T_0/\omega)$ represents an effective $Z$-factor renormalizing the frequency,
hence the superscript $Z$. 

The dephasing-induced part $\Sigma^\varphi_\infty$ of the self-energy of the disconnected diffuson is also given by diagrams in Fig.~\ref{selfenergy-boxes}, now with the imaginary part of the interaction propagator and the appropriate thermal factor [see Eq.~(\ref{coth-tanh})].
For \mbox{$\omega\ll T$} it can also be obtained by the path-integral calculation of Section~\ref{diffuson_dephasing}, using the classical (thermal) part of the interaction propagator for \mbox{$\epsilon\ll T$}, since for \mbox{$\omega,E\ll T$} the high-energy transfers with \mbox{$|\epsilon|>T$} are suppressed by the standard inelastic thermal factor $\coth(\epsilon/2T)-\tanh(\epsilon/2T)$.
Then $\Sigma^\varphi_\infty\propto \lambda^2 g T$ is given by Eqs.~(\ref{tauphi_fastmodes}) and (\ref{tauphi_fastmodes_lowT}) with \mbox{$\eta=\infty$} and the logarithmic factor (see Sec.~\ref{dephasing-strong-coupling} below) modified
by the renormalization processes.

For \mbox{$\omega\gtrsim T$} and \mbox{$E\lesssim T$}, which is the range relevant for the Hartree correction to the conductivity, the full inelastic thermal factor\cite{Altshuler_Aronov_85,AAG_99,Narozhny02,vonDelft_etal_05}
\bea
\coth\frac{\epsilon}{2T}&+&\frac{1}{2}\left[\tanh\frac{E-\epsilon}{2T} + \tanh\frac{E-\omega-\epsilon}{2T} \right]
\nonumber \\
&\simeq&
\coth\frac{\epsilon}{2T}-\frac{1}{2}\left[\tanh\frac{\epsilon}{2T} + \tanh\frac{\omega+\epsilon}{2T} \right]
\nonumber \\
&\simeq&
\left\{
\begin{array}{ll}
\displaystyle   0,&\epsilon\gg T \\[0.1cm]
\displaystyle   2T/\epsilon,& |\epsilon|<T \\[0.1cm]
\displaystyle  -1,& -\omega\ll\epsilon\ll -T\\[0.1cm]
\displaystyle   0,&\:\epsilon \ll -\omega\:\:
\end{array}
\right.
\label{coth-tanh}
\eea
allows also for real inelastic processes with energy transfers \mbox{$-\omega<\epsilon<-T$}. Thus, in addition to the standard range \mbox{$|\epsilon|<T$} where the thermal factor is classical, $2T/\epsilon$, we have a contribution described by the quantum factor $-1$ (the phase space available for inelastic scattering is then determined by $\omega$ rather than by temperature).
It is convenient to separate the thermal ($\Sigma^{\varphi,T}$) and frequency-induced ($\Sigma^{\varphi,\omega}$) contributions to the dephasing self-energy,
\be
\Sigma^\varphi_\infty=
\Sigma^{\varphi,T}_\infty+\Sigma^{\varphi,\omega}_\infty\:.
\label{sigma-varphi}
\ee

The thermal contribution can be still evaluated by the path-integral method and is given by Eqs.~(\ref{tauphi_fastmodes}) and (\ref{tauphi_fastmodes_lowT}) with the logarithmic factor determined by the diffuson momenta $q$ through the infrared cutoff established by the gauge-invariance:
\be
\Sigma^{\varphi,T}_\infty
=
3 \lambda^2 g\,T \times
\left\{
\begin{array}{ll}
\displaystyle
\ln\left(\frac{1}{q^2l^2}\right),&T\gg T_0\\[0.5cm]
\displaystyle
\ln\left(\frac{T}{T_0 q^2l^2}\right),&T\ll T_0\:\:.
\end{array}
\right.
\label{disconnected-selfenergy-dephT}
\ee
The same result is obtained diagrammatically along the lines described in Appendix~\ref{app_selfenergies}.

One sees that the thermal contribution to the dephasing is proportional to the conductance and temperature, \mbox{$\Sigma^{\varphi,T}_\infty\propto \lambda^2 g T$}. In the strong coupling regime \mbox{$\lambda^2g\gg 1$}, the thermal-dephasing part of the diffuson self-energy exceeds $T$. This result for the disconnected diffuson self-energy agrees with the path-integral calculation of the Cooperon and UCF-diffuson dephasing rate, Eqs.~(\ref{tauphi-coop}), (\ref{tauphi-coop-3}) and (\ref{tauphi-coop-lowT}): the path integral automatically chooses characteristic values of $q$ in Eq.~(\ref{disconnected-selfenergy-dephT}).

The frequency-induced dephasing self-energy is similar to the renormalization part $\Sigma^Z_\infty$: at high frequencies \mbox{$\omega\gg T_0$} it is also determined by the ballistic energy/momenta transfers:
\be
\Sigma^{\varphi,\omega}_\infty
=
\lambda^2 g\,\omega\cdot f_\varphi\left(\frac{T_0}{\omega}\right)\:,
\label{disconnected-selfenergy-deph-lowT}
\ee
with
\be
f_\varphi(x)\sim
\left\{
\begin{array}{ll}
\displaystyle
1,&x\gg 1\\
x^{1/3},&x\ll 1\:\:.
\end{array}
\right.
\label{fomega}
\ee
The high-frequency ($\omega\gg T_0$) asymptotics of Eq.~(\ref{disconnected-selfenergy-deph-lowT}), \mbox{$\Sigma^{\varphi,\omega}_\infty\propto \omega^{2/3}$} is, in fact, determined by the imaginary part of the  self-energy of a single particle Green's function and agrees with the result of Ref.~\onlinecite{HLR_93}. However, in contrast to the clean case, there is no need in resummation of the higher-order interaction terms (for not too high \mbox{$\omega<T_{3/2}$}), since disorder generates a larger self-energy $i/2\tau$ in the Green's function.

\subsubsection{Fully dressed delayed diffuson}

The fully dressed diffuson $\tilde{{\cal D}}_\eta$ is characterized by a cancellation between self-energy and vertex lines, which is complete at zero delay time
\be
\tilde{{\cal D}}_{\eta=0}(\omega,q)={\cal D}_0(\omega,q)
\ee
and partial at finite delay time,
\be
\tilde{{\cal D}}_\eta
=
\frac{1}{Dq^2-i\omega-i\Sigma^Z_\eta+\Sigma^\varphi_\eta}\:.
\label{delayed-diffuson-1}
\ee
The renormalization part $\Sigma^Z_\eta$ of the self-energy (see Fig.~\ref{selfenergy-boxes}) is at low frequencies \mbox{$\omega\ll T_0$} given by
\begin{widetext}
\be
\Sigma^Z_\eta
=
\displaystyle\frac{3\lambda^2}{2\pi}\,g\,\omega
\left[\ln{\frac{T_0}{\omega}}\left(1-\frac{{\rm sin}\,\omega\eta}{\omega\eta}\right)+1-\frac{{\rm Si}(\omega\eta)}{\omega\eta}
\right]\:\:,\qquad \omega\ll T_0\:\:,
\label{full-selfenergy-small-omega}
\ee
where ${\rm Si}(x)$ is the integral sine function.
At high frequencies \mbox{$\omega\gg T_0$}, the renormalization part of the self-energy reads
\be
\Sigma^Z_\eta
=
\left\{
\begin{array}{ll}
\displaystyle
{\cal O}(1)\cdot \lambda^2 g\,\omega\left(\omega\eta\right)^2\left(\displaystyle\frac{T_0}{\omega}\right)^{1/3}\:\:,
\qquad &\eta\ll 1/\omega\ll 1/T_0\\[0.3cm]
\displaystyle
\Sigma^Z_\infty-3\lambda^2 gT_02^{-1/3}\Gamma(5/3)\left(T_0\eta\right)^{-2/3}\:\:,\qquad &1/\omega\ll\eta\ll 1/T_0\\[0.2cm]
\displaystyle
\Sigma^Z_\infty-3 \lambda^2 g/\eta\:\:,\qquad &\:1/\omega\ll 1/T_0\ll\eta\:\:.
\end{array}
\right.
\label{full-selfenergy-main}
\ee
\end{widetext}

The dephasing part of the self-energy can be evaluated in a similar way.
Below we present only the leading terms in the relevant range of frequencies \mbox{$\omega\gg T$}. At low temperatures \mbox{$T\ll T_0$}, the dephasing part $\Sigma^\varphi_\eta$ of the self-energy is given by the frequency-induced part
\be
\Sigma^\varphi_\eta
\sim \left\{
\begin{array}{ll}
\displaystyle \lambda^2 g \omega (\omega\eta)^2 \:\:,&\eta\ll 1/\omega\\[0.2cm]
\displaystyle \lambda^2 g \omega \:\:,&\eta\gg 1/\omega
\end{array}
\right.
\label{deph-selfenergy-lowT-low-omega}
\ee
for \mbox{$T\ll\omega\ll T_0$}.
For higher frequencies \mbox{$T\ll T_0\ll\omega$} we get
\be
\Sigma^\varphi_\eta
\sim 
\left\{
\begin{array}{ll}
\displaystyle \lambda^2 g \omega^{2/3} T_0^{1/3} (\omega\eta)^2 \:\:,&\eta\ll 1/\omega,\\[0.2cm]
\displaystyle \Sigma_\infty^{\varphi,\omega} \:\:,&1/\omega\ll \eta \ll 1/T,\\[0.2cm]
\displaystyle 3 \lambda^2 g T \ln(T\eta) +\Sigma_\infty^{\varphi,\omega} \:\:,&1/T \ll \eta \ll 1/T_0 q^2l^2, \\[0.2cm]
\displaystyle \Sigma^{\varphi,T}_\infty+\Sigma^{\varphi,\omega}_\infty \:\:,&\eta \gg 1/T_0 q^2l^2,
\end{array}
\right.
\label{deph-selfenergy-lowT-high-omega}
\ee
which now features the competition between the thermal and frequency-induced contributions at long $\eta$. Finally, at temperatures \mbox{$T\gg T_0$}, we find a similar result
\be
\Sigma^\varphi_\eta
\sim
\left\{
\begin{array}{ll}
\displaystyle \lambda^2 g \omega^{2/3} T_0^{1/3} (\omega\eta)^2 \:\:,&\eta\ll 1/\omega\\[0.2cm]
\displaystyle \Sigma_\infty^{\varphi,\omega} \:\:,&1/\omega\ll \eta \ll 1/T_0\\[0.2cm]
\displaystyle 3 \lambda^2 g T \ln(T_0\eta) +\Sigma_\infty^{\varphi,\omega} \:\:,&1/T_0 \ll \eta \ll 1/T_0 q^2l^2 \\[0.2cm]
\displaystyle \Sigma^{\varphi,T}_\infty+\Sigma^{\varphi,\omega}_\infty \:\:,&\eta \gg 1/T_0 q^2l^2\:.
\end{array}
\right.
\label{deph-selfenergy-highT}
\ee
In the limit \mbox{$\omega\ll T$}, the dephasing part of the self-energy just reduces to the dephasing rate calculated using path integral in Section~\ref{diffuson_dephasing}.

In the strong coupling regime, when $\Sigma^Z_\infty \simeq \omega Z(\omega)\gg \omega$, the integration over the diffuson momentum $q$ in the Hartree conductivity correction is dominated by
\be
Dq^2\sim {\rm max}\left\{\Sigma^Z,\Sigma^{\varphi,\omega},\Sigma^{\varphi,T}\right\}\:.
\ee
One can see from Eqs.~(\ref{disconnected-selfenergy-main}), (\ref{disconnected-selfenergy-deph-lowT}) and (\ref{fomega}) that \mbox{$\Sigma^{\varphi,\omega}\lesssim\Sigma^Z$} for any frequency:
at \mbox{$\omega\ll T_0$} there is an extra logarithmic factor in $\Sigma^Z$, while at \mbox{$\omega\gg T_0$} the two quantities are of the same order.
Therefore, in what follows we neglect $\Sigma^{\varphi,\omega}$ for simplicity: its inclusion may only change the numerical coefficients in the results for the Hartree conductivity correction, which are anyway beyond the accuracy of the present approach.

On the other hand, the relation between $\Sigma^{\varphi,T}$ and $\Sigma^Z$ depends on the frequency and temperature.
At low temperatures inelastic scattering is suppressed. As a result, at \mbox{$T\ll T_0$} we have \mbox{$\Sigma^Z\gg \Sigma^{\varphi,T}$} for all relevant frequencies \mbox{$\omega\gg T$}.
For \mbox{$T\gg T_0$}, the thermal dephasing dominates in the range of sufficiently low frequencies, \mbox{$T\ll\omega\ll T(T/T_0)^{1/2}$}, while the $Z$-factor renormalization wins at higher frequencies, \mbox{$\omega\gg T(T/T_0)^{1/2}$}.
The self-energy part due to the thermal dephasing can always be evaluated using the path-integral approach developed in Section~\ref{diffuson_dephasing}.

The diffusion approximation breaks down at high frequencies, when the self-energy of the disconnected diffuson becomes comparable to the elastic scattering rate $1/\tau$. The dephasing part $\Sigma^{\varphi,T}$ restricts our subsequent consideration at \mbox{$\lambda=1$} to sufficiently low temperatures \mbox{$T\ll T_1$} since \mbox{$\Sigma^{\varphi,T}_\infty(T\sim T_1)\sim 1/\tau$}. The $Z$-factor poses the restriction on the frequency domain for diffusive dynamics, \mbox{$\omega<1/g^{1/2}\tau\equiv T_{3/2}$}, since \mbox{$\Sigma^Z_\infty(\omega\sim T_{3/2})\sim 1/\tau$}. We only consider the diffusive regime and hence restrict ourselves to the case \mbox{$T<T_1$}.
We are now in the position to calculate the Hartree conductivity correction, using the above results for the delayed diffuson self-energies.

\subsection{Low temperatures, $T\ll T_0$: strong renormalization by self-energies}
\label{low_T}

We start with the case of low temperatures, \mbox{$T\ll T_0$}. In this situation, the dephasing part of the self-energy $\Sigma^\varphi$ (\ref{sigma-varphi}) is smaller than the renormalization self-energy $\Sigma^Z$ (\ref{full-selfenergy-main}) for all $\omega$ and $\eta$ and may be neglected. In Section~\ref{high_T} we consider the situation \mbox{$T\gg T_0$}, where this is not the case and both renormalization and dephasing should be retained. In both temperature ranges we calculate the Hartree correction in the realistic case \mbox{$\lambda=1$}.

We use the generalization of diagrams $a$ and $b$ with fully dressed diffusons, as described by Eq.~(\ref{Fourier_substitution}). The result consists of a low-frequency contribution and a high-frequency contribution, with the low-frequency contribution given by
\bea
\delta\sigma^{H}_{\omega<T_0}
&=&
\frac{\sigma_0}{\pi \nu}
\int\limits_{-T_0}^{T_0}\frac{d\omega}{2\pi}\:\frac{\partial}{\partial\omega}
\left[\omega\:{\rm coth}\frac{\omega}{2T}\right]\nonumber \\
&\times& \int\limits_{-\infty}^\infty d\eta
\int(dq)\:
\left[
\tilde{{\cal D}}_\eta-\tilde{{\cal D}}_\infty
\right]\:\:.
\eea 
The details of the calculation of the low-temperature Hartree correction are presented in Appendix~\ref{app-lowT}. The results reads
\bea
\delta\sigma^{H}_{\omega<T_0}
=
\frac{e^2}{2\pi}
\left[
\frac{c_1}{\pi^2}\cdot\ln{\frac{T_0}{T}}
+\frac{1}{2\pi}\ln{g}\ln{\left[\ln{\frac{T_0}{T}}+1\right]}
\right]\:\:.\nonumber \\
\label{delta_sigma_lowfreq}
\eea
The value of the constant given by a dimensionless integral (\ref{integral-c1}) is \mbox{$c_1\approx 5$}. The strong renormalization of the frequency $\omega$ by a factor which is of order $g$ at \mbox{$T\ll T_0$} results in the prefactor of the $\ln(T_0/T)$-term in Eq.~(\ref{delta_sigma_lowfreq}) being of order unity rather than of order $g$. 

At the low temperatures \mbox{$T\ll T_0$} under consideration here, the contribution of high frequencies \mbox{$\omega\gg T_0$} saturates to a constant (again dephasing is unimportant compared to the self-energy),
\be
\delta\sigma^{H}_{\omega>T_0}
=
{\cal O}(1)\cdot e^2\,\left[g^{1/2}-\frac{1}{2}\ln{g}\right]\:\:,\quad T\ll T_0\:\:,
\label{delta_sigma_lowT_highfreq}
\ee
see also Eq.~(\ref{delta_sigma_highT}) below.
The coefficient depends on details of the high-frequency cutoff, which we do not attempt to calculate here.

Thus the temperature-dependent part of the Hartree correction at very low temperatures is given by Eq.~(\ref{delta_sigma_lowfreq})
\be
\delta\sigma^{H}(T)
=
\frac{e^2}{2\pi^3}\:
c_1\cdot\ln{\frac{T_0}{T}}\:\:,\quad T\ll T_{-1}\:\:.
\label{delta_sigma_lowfreq_simpler}
\ee
Due to the renormalization of the diffusons by the large self-energies, the coefficient of the low-temperature Hartree correction (\ref{delta_sigma_lowfreq_simpler}) is of order unity rather than of order $g$, in contrast to what could be expected from the perturbative result (\ref{delta_sigma}) taken at \mbox{$\lambda=1$}. As a result, the temperature dependence of the Hartree correction is overcompensated at low temperatures \mbox{$T<T_0$} by the negative exchange correction\cite{Mirlin_Woelfle_97}, which carries a coefficient of order $\ln{g}$, see Fig.~\ref{plot_delta_sigma}. 
It should be emphasized that in contrast to the Hartree contribution, the exchange correction involves only true diffusons ${\cal D}_0$ (zero delay time) which are affected neither by dephasing nor by renormalization.

At higher temperatures, \mbox{$T\gg T_0$}, dephasing and renormalization effects are of similar importance. This situation is considered in Section~\ref{high_T} below. We remind that at high \mbox{$T\gg T_1$} the diffusion approximation breaks down due to the strong dephasing ($L_\varphi\ll l$), so that we restrict ourselves to intermediate temperature range, \mbox{$T_0\ll T \ll T_1$}.

\subsection{Interplay of renormalization and dephasing at intermediate temperatures, $T_0\ll T \ll T_1$}
\label{high_T}

Using the results of Section~\ref{diffuson_dephasing} for self-energies of delayed diffusons, we will now construct a scheme to treat the interaction to all orders, while ensuring the Hartree structure of the calculated contributions also in the presence of dephasing by means of Eq.~(\ref{D-D=DUD}).
The Hartree correction at \mbox{$T_0\ll T\ll T_1$} can be then written as
\begin{widetext}
\be
\delta\sigma^H
=
\frac{\sigma_0}{\pi \nu}
\int\limits_{-T_{3/2}}^{T_{3/2}} \frac{d\omega}{2\pi}\:\frac{\partial}{\partial\omega}\left[\omega\:{\rm coth}\frac{\omega}{2T}\right]
\: {\rm Re}\:\int\limits_{-\infty}^\infty d\eta
\int(dq)\:
\tilde{{\cal D}}_\infty\cdot(-i)u(\eta)\:\tilde{{\cal D}}_\eta\:\:.
\label{DUD}
\ee

The dephasing self-energy $\Sigma^\varphi_\infty$ in one of the two diffusons ensures that the contribution from short delay times \mbox{$\eta\lesssim 1/T_0$} is subleading, so that the main contribution is given by
\bea
\delta\sigma^H
&\approx&
\frac{4\sigma_0}{\pi \nu}
\int\limits_0^{T_{3/2}}\frac{d\omega}{2\pi}\:\frac{\partial}{\partial\omega}\left[\omega\:{\rm coth}\frac{\omega}{2T}\right]
\: {\rm Re} \:
\int\limits_{1/T_0}^\infty d\eta
\int(dq)\:
\tilde{{\cal D}}_\infty\cdot(-i)u(\eta)\:\tilde{{\cal D}}_\eta\nonumber \\
&\sim &
e^2
\int\limits_T^{T_{3/2}} d\omega\:
\:{\rm Re}\:
\int\limits_{1/T_0}^{\eta_{\rm max}} d\eta\:
\frac{g}{\eta}\:
\frac{\ln{\displaystyle\frac{\Sigma^Z_\infty+i\Sigma^{\varphi,T}_\infty}
{\Sigma^Z_\eta+i\Sigma^{\varphi,T}_\eta}}}
{\left[\Sigma^Z_\infty-\Sigma^Z_\eta+i\Sigma^{\varphi,T}_\infty-i\Sigma^{\varphi,T}_\eta\right]}\:\:,
\label{110}
\eea
\end{widetext}
where $\eta_{\rm max}=g^2/{\rm max}\left\{\Sigma^Z_\infty,\Sigma^{\varphi,T}_\infty\right\}$ is related to the infrared momentum cutoff of the bare interaction propagator, which is established by the characteristic diffuson momenta $q$.

Using Eqs.~(\ref{full-selfenergy-main}) and (\ref{deph-selfenergy-highT}), we find for \mbox{$\eta\gg 1/T_0$}
\bea
\Sigma^Z_\infty-\Sigma^Z_\eta&\sim& g/\eta\nonumber \\
\Sigma^{\varphi,T}_\infty-\Sigma^{\varphi,T}_\eta&\sim& g T \ln(\eta_{\rm max}/\eta)\:.
\eea
For relevant values of \mbox{$1/T_0\lesssim\eta\lesssim\eta_{\rm max}$},
these differences arising due to the vertex interaction terms in the diffuson self-energies are small compared to $\Sigma^Z_\infty,\Sigma^{\varphi,T}_\infty$ which simplifies the expression for the conductivity correction:
\bea
\delta\sigma^H&\sim& e^2 \:{\rm Re}\: \int\limits_T^{g^{3/2} T_0} d\omega 
\int\limits_{1/T_0}^{\eta_{\rm max}} d\eta\:
\frac{g}{\eta} \:
\frac{1}{\Sigma^Z_\infty+i \Sigma^{\varphi,T}_\infty}\:.\nonumber \\
\label{111}
\eea

For higher frequencies, \mbox{$T^{3/2}/T_0^{1/2}\ll\omega\ll T_{3/2}$}, the $Z$-factor self-energy $\Sigma^Z$ is larger than $\Sigma^{\varphi,T}$, while for lower frequencies, \mbox{$T_0<T\lesssim\omega\lesssim T^{3/2}/T_0^{1/2}$}, the dephasing part dominates.
The result reads
\be
\delta\sigma^H
=
{\cal O}(1)\cdot e^2
\left\{g^{1/2}-\left(\frac{T}{T_0}\right)^{1/2}\left[1+\frac{1}{2}\ln{\frac{T_1}{T}}\right]
\right\},
\label{delta_sigma_highT}
\ee
where the temperature dependence arises because dephasing strongly suppresses the frequencies \mbox{$\omega\lesssim T^{3/2}/T_0^{1/2}$} in the Hartree correction.

\begin{figure}
\includegraphics[width=1.0\linewidth]{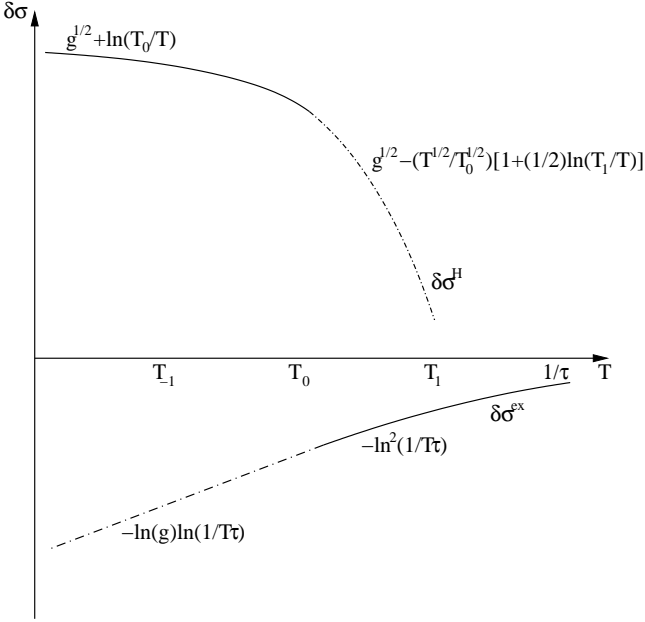}
\caption{\label{plot_delta_sigma} Schematic plot of the Hartree correction to the conductivity, $\delta\sigma^H(T)$ for $\lambda=1$, in the different ranges of temperatures (log-$T$ scale). The solid and dot-dashed lines of $\delta\sigma^H$ are described by the sum of Eqs.~(\ref{delta_sigma_lowfreq}) and (\ref{delta_sigma_lowT_highfreq}) and by Eq.~(\ref{delta_sigma_highT}), respectively. The exchange contribution\cite{Mirlin_Woelfle_97} $\delta\sigma^{\rm ex}$ is shown for comparison. Due to the high dephasing rates of delayed diffusons, the Hartree contribution can be calculated within the diffusive approximation only at temperatures below $T_1$. At low temperatures $T<T_0$, the exchange contribution overcompensates the $T$-dependent part of the
Hartree contribution, so that the total correction becomes localizing, $d\sigma/dT>0$.}
\end{figure}

A schematic overview of the result in the different temperature ranges in shown in Fig.~\ref{plot_delta_sigma}. 
In this plot, we also show the results for the exchange contribution\cite{Mirlin_Woelfle_97}
\be
\delta\sigma^{\rm ex}
= -\frac{e^2}{(2\pi)^2}\left\{
\begin{array}{ll}
\ln^2(T\tau)\:\:,& T\gg T_0\\[0.2cm]
4\ln g \ln (1/T\tau) \:\:,& T\ll T_0\:.
\end{array}
\right.
\label{exchange-point-MW97}
\ee
Most remarkably, while the positive Hartree contribution is larger than the negative exchange contribution over a wide range of temperatures, at low temperatures (\mbox{$T\ll T_0$}) the exchange contribution dominates the temperature dependence again.

\subsection{Dephasing at strong coupling: weak localization and mesoscopic conductance fluctuations}
\label{dephasing-strong-coupling}

Let us now turn to the dephasing rate of the Cooperon and UCF-diffuson.
The path-integral calculation of Section~\ref{Cooperon_dephasing} is modified by inclusion of the $Z$-factor in the Cooperon propagators in Eq.~(\ref{deltaS-2d}). In the presence of time-reversal symmetry, the relevant Cooperons are given by disconnected diffusons $\tilde{\cal D}_\infty$ which contain the renormalization part of the self-energy $\Sigma^Z_\infty$, given by Eq.~(\ref{disconnected-selfenergy-main}).

In this Section we will concentrate on the lowest-$T$ limit, \mbox{$T\ll T_0$}; the generalization onto the case of higher temperatures is straightforward.
For \mbox{$\omega \ll T_0$}, in the strong-coupling regime \mbox{$\lambda^2g\gg 1$} we have $Z(\omega)\simeq(3 \lambda^2 g/2\pi)\ln (T_0/\omega)\gg 1$, which implies that one should replace $\omega$ by $Z\omega\gg  \omega$ in all the Cooperon denominators in Eq.~(\ref{deltaS-2d}) [in the argument of the logarithm in $Z(\omega)$ the frequency can be replaced by its characteristic value, $\omega\sim 1/t$].
This amounts to the rescaling of frequency \mbox{$\omega\to Z\omega$} (inducing the prefactor $1/Z$ from the integration measure $d\omega$) and of the time variable \mbox{$t\to t/Z$}. The return probability is then given by
\be
{\tilde C}(t)=\frac{1}{Z} C^{(0)}(t/Z) = \frac{1}{Z} \frac{Z}{4\pi D t} = \frac{1}{4\pi D t}\:\:,
\ee
so that the corresponding prefactor in the dephasing action (\ref{deltaS-raw}) is not changed by the renormalization.

As a result of the rescaling, we get from Eq.~(\ref{deltaS-result-lowT})
\bea
\Delta {\tilde S}^{\rm eff}(t)
&\simeq&
3\,\lambda^2 g T \frac{t}{Z(t)}\ln{\left[\frac{T}{T_0}\frac{t}{Z(t)\tau}\right]}\:\nonumber \\
&\simeq&
\frac {3\,\lambda^2 g T t}{(3/2\pi) \lambda^2 g \ln(T_0 t)}\ln{\frac{T t}{ \lambda^2 g  T_0 \tau}}\:\nonumber\\
&\simeq&
2\pi T t\,\frac{\ln(g T t/\lambda^2)}{\ln(T_0 t)}\:\:.
\label{deltaS-result-renorm}
\eea
From the condition \mbox{$\Delta {\tilde S}^{\rm eff}(t\sim {\tilde \tau}_\varphi)\sim 1$}, we obtain the characteristic dephasing rate in the strong coupling regime
\be
\frac{1}{{\tilde \tau}_\varphi} \sim T\,\frac{\ln(g/\lambda^2)}{\ln(T_0/T)}\:\:.
\ee
We see that the conductance enters the dephasing rate only under the logarithm:
in the realistic case of \mbox{$\lambda=1$}, the Cooperon dephasing rate is given by temperature, up to logarithmic factors. This conclusion remains valid in the strong coupling regime also for the time-decay of the diffusons involved in the mesoscopic conductance fluctuations and in the second-loop weak localization correction.
It is also worth noticing that in the limit of \mbox{$t\to \infty$}, the dephasing action becomes
\be
\Delta{\tilde S}^{\rm eff}(t\to\infty) \simeq 2\pi T t\:.
\ee

On the other hand, the characteristic dephasing {\it length} $L_\varphi$ is not affected by the frequency renormalization. Indeed, the dephasing length is the ``static'' object, which is defined by the Cooperon at \mbox{$\omega=0$}. The term $Dq^2$ in the Cooperon denominator is not renormalized by interaction, in contrast to $-i\omega$, so that the renormalization effects related to the $Z$-factor are absent at \mbox{$\omega=0$}.
To extract the dephasing length $L_\varphi$, one has to compare $Dq^2$ at \mbox{$q\sim 1/L_\varphi$} with $\Sigma^\varphi$:
\be
L_\varphi\sim \sqrt{D/\Sigma_\infty^\varphi} \sim L_T/g^{1/2} \ll L_T\:,
\label{445}
\ee
where $L_T=(D/T)^{1/2}$ is the thermal length.
The dephasing length $L_\varphi$ is thus directly determined by the dephasing part of the Cooperon self-energy $\Sigma_\infty^\varphi$ and not by the dephasing rate \mbox{$1/{\tilde \tau}_\varphi\sim \Sigma_\infty^\varphi/Z$}.

Therefore, while the dephasing rate is moderate ($1/\tau_\varphi \sim T$) in the strong coupling regime, the dephasing length is anomalously short ($L_\varphi \ll L_T$). Note that the standard interference experiments (e.g., measuring the magnetoconductivity) usually probe directly the dephasing length rather than the dephasing rate.

The situation when $\tau_\varphi\sim 1/T$ but $L_\varphi$ and $L_T$ are parametrically different is an indicator of strong renormalization that may occur in strongly-correlated systems and in the vicinity of quantum critical points (in particular, at the Anderson transition in the presence of electron-electron interactions~\cite{Finkelstein}).

In the context of gauge-field models, a related physics has been encountered in Ref.~\onlinecite{Aronov_Woelfle_PRL_94} and \onlinecite{Aronov_Woelfle_94}, where the case of rapidly fluctuating gauge fields has been considered. This case may be realized in the gauge field formulation of the $t$-$J$-model of high temperature superconductors, with two species of pseudoparticles, holons and spinons. There the electric charge is carried by holons, while a fictitious gauge field effecting the projection onto the physical part of the Hilbert space is controlled by the spinons. The spinons are scattered much more weakly by impurities than the holons, allowing the anomalous skin effect regime with typical gauge field frequency \mbox{$\omega \propto q^3$}  to be reached at sufficiently high temperatures. In that regime the gauge field fluctuations lead to a spatially nonuniform diffusion coefficient. As a consequence, the phase breaking length varies with temperature as \mbox{$L_\varphi \propto T^{-1/6}$}, while the relevant time scale for phase breaking processes is still given by the inelastic scattering rate, \mbox{$1/\tau_\varphi \sim T$}. This behavior has been seen in experiment~\cite{jing91}.

\section{Unscreened Coulomb interaction}
\label{Coulomb}
\setcounter{equation}{0}

Finally, we contrast the results of the previous sections with a calculation for the case of a long-range electron-electron interaction. The aim of this section is to point out how the different (less singular) gauge field propagator modifies the quantities of interest. 
We use the susceptibility\cite{HLR_93}
\be
\chi(k)=\chi_0+\frac{e^2v(k)}{\left(2\pi\tilde{\phi}\right)^2}
\ee
with the 2D unscreened Coulomb interaction \mbox{$v(k)=2\pi e^2/k$}, the free-fermion susceptibility \mbox{$\chi_0=e^2/12\pi m$}, and the number of attached flux quanta $\tilde{\phi}$, which for the half-filled lowest Landau level is \mbox{$\tilde{\phi}=2$}.
While a short-range interaction, \mbox{$v(k)={\rm const}$}, would just renormalize the value of $\chi$, the long-range interaction leads to a less singular behavior of the gauge field propagator at small $k$,
\bea
U_{\alpha\beta}(\bk,\epsilon)
&=&
\frac{1}{\chi(k)\:k^2-i\sigma(k)\epsilon}\:\delta_{\alpha\beta}^\perp\nonumber \\
&=&
\frac{1}{\chi_0\kappa k-i\sigma(k)\epsilon}\:
\delta_{\alpha\beta}^\perp\:\:,\quad k\ll k_F\:,\nonumber \\
\label{less_singular_U}
\eea
and the corresponding correlator for thermal fluctuations in the static approximation
\bea
\left\langle a_\alpha a_\beta\right\rangle_{\bk,\epsilon}
&=&
\frac{T}{\chi(k)\,k^2}\:
\delta_{\alpha\beta}^\perp\:
2\pi\delta(\epsilon)\nonumber \\
&\approx&
\frac{T}{\chi_0k\kappa}\:\delta_{\alpha\beta}^\perp\:2\pi\delta(\epsilon)\:\:,
\label{less_singular_correlator}
\eea
with the inverse screening length
\be
\kappa=\frac{e^4}{2\pi\chi_0\tilde{\phi}^2}\:\:.
\ee
For the experimentally relevant case of composite fermions, $k_F$ and $\kappa$ are not independent, \mbox{$\kappa/k_F=3C_*/2$} with $C_*$ (see Ref.~\onlinecite{Mirlin_Woelfle_97}) a numerical constant, which is of order $10$ according to experiments.\cite{CF_data}
We will now consider this situation, \mbox{$\kappa\gtrsim k_F$}, to complement the results for the point-like interaction from the previous section with corresponding results for unscreened interaction.

\subsection{Cooperon dephasing}
\label{Coulomb_dephasing}

We will first consider the dephasing within the static approximation. As we discuss below, the latter is justified for not too low temperatures. For the situation that the Coulomb interaction is unscreened for all relevant momenta, \mbox{$\kappa\gg k_F$}, the less singular gauge field propagator (\ref{less_singular_correlator}) should be used instead of (\ref{static_correlator}) in Eq.~(\ref{deltaS-raw}).  Then the result is determined by characteristic momenta \mbox{$k\in\left[l^{-1},k_*\right]$}. Here $k_*\sim \sqrt{m T/C_*}$ is the highest momentum for which the static approximation is valid. Taking into account that the factor $D$ in the interaction vertex of the left diagram of Fig.~\ref{2nd_order_processes} acquires a $k$-dependence,
\be
D(k)\propto\sigma(k)
\simeq
\left\{
\begin{array}{ll}
\sigma_0\:\:,&kl\ll 1\\
2\sigma_0/kl\:\:,&kl\gg 1\:\:,
\end{array}
\right.
\ee
the equivalent of Eq.~(\ref{deltaS-2d}) (the dephasing action within the static approximation) reads
\begin{widetext}
\be
\Delta S^{\rm eff}(t)
=
4\pi Dt\:
\frac{4 \lambda^2 e^2DT}{\chi_0}
\int\frac{d\omega}{2\pi}\:{\rm exp}\{i\omega t\}
\int\limits_0^\infty\frac{q\,dq}{2\pi}
\int\limits_{l^{-1}}^{k_*}\frac{k\,dk}{2\pi}
\int\limits_0^{2\pi}\frac{d\phi}{2\pi}\:
\frac{1}{\left(Dq^2-i\omega\right)^2}\:
\frac{1}{\kappa k}\:\frac{-2}{kl}\:\:,
\label{deltaS-2d-longrange}
\ee
\end{widetext}
consistent with the interpolation formula derived in the microscopic calculation of Ref.~\onlinecite{Woelfle_2000}.
Compared to Eq.~(\ref{deltaS-2d}), we have dropped the second term in brackets, which is important only for the correct low-$k$ cutoff, since in the present case important $k$ are from the range \mbox{$k\in\left[l^{-1},k_*\right]$} rather than \mbox{$k\in\left[L_\omega^{-1},l^{-1}\right]$}. The condition \mbox{$k_*l\gg 1$}, which determines the range of validity of the static approximation, holds in the range of relatively high temperatures,
\be
T\gg C_*/g\tau\:\:.
\ee

Evaluating Eq.~(\ref{deltaS-2d-longrange}), we find
\be
\Delta S^{\rm eff}
=
\displaystyle\frac{16 \lambda^2}{C_*}\,Tt\ln{\frac{g T\tau}{C_*}}\:\:,\quad T\gg \frac{C_*}{g\tau}\:\:,
\ee
resulting in the dephasing rate
\be
\displaystyle\frac{1}{\tau_\varphi}
=
\frac{16 \lambda^2}{C_*}\,T\ln{\frac{g T\tau}{C_*}}\:\:,\quad T\gg \frac{C_*}{g\tau}\:\:.
\label{tauphi-longrange}
\ee
Equation~(\ref{tauphi-longrange}) is a more moderate dephasing rate than the corresponding result (\ref{tauphi-coop}) for the more singular gauge field propagator (\ref{static_correlator}) arising from a short-range interaction, since it does not carry the large parameter $g$ in the prefactor.

In the realistic case $\lambda=1$, we find at $T\gg C_*/g\tau$ a dephasing rate \mbox{$1/\tau_\varphi\sim (T/C_*) \ln(g T\tau/C_*)$}, similar to the result of Section~\ref{dephasing-strong-coupling}.
Using the conventional relation \mbox{$L_\varphi=(D\tau_\varphi)^{1/2}$} (which is now valid in view of the absence of strong renormalization effects), we get
\be
L_\varphi\sim (C_*/\ln g)^{1/2}L_T\:\:,
\label{510}
\ee
so that the dephasing length $L_\varphi$ is of the order of $L_T$ for realistic parameters.
This should contrasted to the case of short-range interaction, Section~\ref{dephasing-strong-coupling}, where the anomalously short dephasing length, \mbox{$L_\varphi\sim L_T/g^{1/2}\ll L_T$}, was obtained.

\begin{figure}
\includegraphics[width=1.0\linewidth]{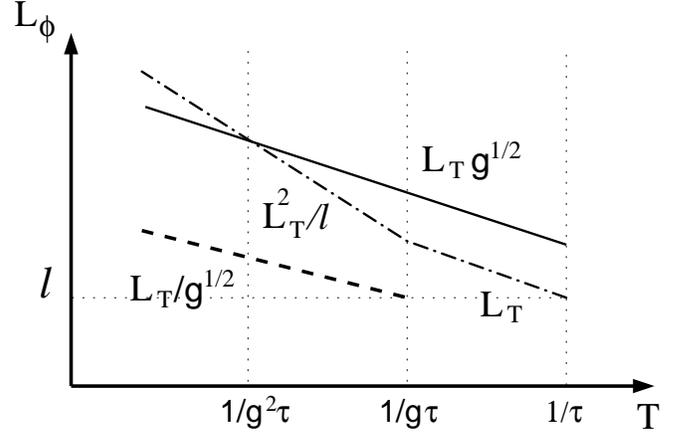}
\caption{
\label{dephasing_plot}
Schematic plot (log-log scale) of temperature dependence of dephasing length $L_\varphi$ due to gauge field interaction 
for the cases of screened (short-range) interaction [dashed line, Eq.~(\ref{445})]
and unscreened Coulomb interaction [dash-dotted line, Eqs.~(\ref{510}) and (\ref{511})].
The conventional dephasing length, Eq.~(\ref{512}), due to the scalar part of interaction 
(which dominates the lowest-$T$ total dephasing in the unscreened case) 
is shown by the solid line. For simplicity, the logarithmic factors and $C_*$ are suppressed.
}
\end{figure}

At lower temperatures, \mbox{$T\ll C_*/g\tau$}, the reduced thermal phase-space for inelastic scattering restricts the relevant transferred momenta
to the ``diffusive range'', $q\ll k \ll l^{-1}$.
The static approximation breaks down for such momenta: the fluctuations of the gauge-fields become fast on the scale of the dephasing time, yielding for \mbox{$\lambda=1$}
\be
\displaystyle\frac{1}{\tau_\varphi}
\sim 
\displaystyle \frac{T^2 g\tau }{C_*^2}\, \ln{\frac{C_*}{g T\tau}} \:\:,\quad T\ll C_*/g\tau\:\:,
\label{deph-intermediateT-Coul}
\ee
which corresponds to 
\be
L_\varphi\sim L_T^2/l \gg L_T.
\label{511}
\ee
Thus at \mbox{$T\ll C_*/g\tau$} we have a standard Fermi-liquid-type situation: \mbox{$1/\tau_\varphi\ll T$} and \mbox{$L_\varphi\gg L_T$}.
At the lowest temperatures \mbox{$T\ll C_*^2/g^2\tau$}, the dephasing is in fact governed by the scalar (density-density) part of the interaction and is the same as in the standard situation,
\be
\displaystyle\frac{1}{\tau_\varphi}
\sim 
\displaystyle \frac{T}{g}\ln g\:\:,\quad T\ll C_*^2/g^2\tau\:\:,
\label{deph-lowT-Coul}
\ee
with
\be 
L_\varphi\sim L_T g^{1/2} \gg L_T.
\label{512}
\ee 
Figure~\ref{dephasing_plot} illustrates of the behavior of the dephasing length in both the cases of screened (Section~\ref{dephasing-strong-coupling}) and
unscreened Coulomb interaction.

\subsection{First-order Hartree correction}
\label{Coulomb_Hartree}

We now turn to the evaluation of the effective interaction appearing in the Hartree correction. Since, as we will see below, the behavior of the quantum conductivity corrections are much less dramatic than for the short-range interaction, we set the coupling constant $\lambda$ to unity in this subsection.
Similar to the dephasing action \mbox{$\Delta S^{\rm eff}$}, the effective interaction box $\tilde{U}$ is now dominated by transferred momenta up to $k_F$. A calculation accounting for the momentum differences in the Green's functions is given in Appendix~\ref{App_interpolation}, with the following result to logarithmic accuracy,
\be
\tilde{U}(\epsilon=0)
=
\frac{4}{\pi\nu C_*}\ln{g}\:\:.
\label{tilde-U_for_long_range}
\ee
Eq.~(\ref{tilde-U_for_long_range}) results in the following first-order Hartree correction to conductivity,
\be
\frac{\delta\sigma^H(T)}{\sigma_0}
=
\frac{2}{\pi^2C_*g}\ln{g}\,\ln{\frac{1}{T\tau}}\:\:.
\label{delta_sigma_for_long_range}
\ee
Similar to the dephasing rate (\ref{tauphi-longrange}), the correction (\ref{delta_sigma_for_long_range}) is not as large as in the case of a short-range interaction, since the effective interaction (\ref{tilde-U_for_long_range}) is not too strong. For experimentally accessible parameters, the first-order result (\ref{delta_sigma_for_long_range}) is valid down to exponentially low temperatures.

\begin{figure}
\includegraphics[width=1.0\linewidth]{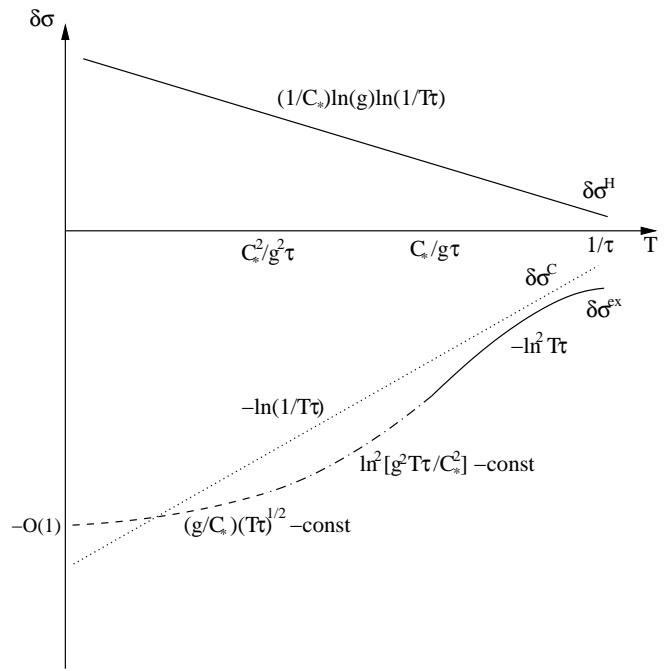}
\caption{\label{plot_delta_sigma_longrange}
Schematic plot (log-$T$ scale) of the contributions to the conductivity correction for unscreened Coulomb interaction: the Hartree contribution $\delta\sigma^H$ given by Eq.~(\ref{delta_sigma_for_long_range}), the exchange contribution $\delta\sigma^{\rm ex}$ given by Eq.~(19) of Ref.~\onlinecite{Mirlin_Woelfle_97}, and the standard Altshuler-Aronov contribution $\delta\sigma^C$ due to the scalar Coulomb exchange interaction\cite{Altshuler_Aronov_85}.
For realistically large conductances \mbox{$\ln{g}\ll C_*\ll g$}, $\delta\sigma^{ex}$ dominates in magnitude over $\delta\sigma^H$ in both magnitude and temperature-dependence at \mbox{$T\gtrsim C_*^2/g^2\tau$}, with the dependence in the vicinity of \mbox{$T\sim C_*/g\tau$} taking the form \mbox{$\delta\sigma^{\rm ex}\simeq -2\ln{(g/C_*)}\,\ln{(1/T\tau)}$}.
The temperature-dependence of $\delta\sigma^{\rm ex}$ saturates at low temperatures, so that at very low temperatures the standard contribution $\delta\sigma^C$ becomes dominant. The sum of all contributions is  negative for all $T$, in contrast to the situation of composite fermions with short-range interaction.
}
\end{figure}

It should be noted that in addition to the Hartree correction (\ref{delta_sigma_for_long_range}) and the exchange correction\cite{Mirlin_Woelfle_97}, 
\bea 
\delta\sigma^{\rm ex}
&=& - \frac{e^2}{(2\pi)^2} \nonumber \\
&\times& \left\{
\begin{array}{ll}
\displaystyle \ln^2(T\tau),&\quad\displaystyle \frac{C_*}{g\tau}\ll T\ll \frac{1}{\tau},\\[0.5cm]
\displaystyle {\cal A}-\ln^2\left[\frac{g^2 T \tau}{C_*^2}\right],&\quad \displaystyle \frac{C_*^2}{g^2\tau}\ll T\ll  \frac{C_*}{g\tau},\\[0.5cm]
\displaystyle {\cal A}-\frac{g}{C_*}(T\tau)^{1/2},&\quad\displaystyle  T\ll \frac{C_*^2}{g^2\tau},
\end{array}
\right.\nonumber \\
\label{exchange-Coul-MW97}
\eea
[with ${\cal A}=2\ln^2(g/C_*)$]
there is also the ``standard'' Altshuler-Aronov contribution 
\be
\delta\sigma^C=-\frac{e^2}{2 \pi^2} \ln(1/T\tau)
\label{AA-Coul-standard}
\ee 
from the scalar Coulomb interaction\cite{Altshuler_Aronov_85}. Using the experimental parameters of $g$ and $C_*$ for composite fermions in the lowest Landau level, $\delta\sigma^C$ dominates over the Hartree gauge field contribution over the whole range of temperatures. At very low temperatures, when the gauge field exchange contribution saturates, it is the ``standard'' contribution $\delta\sigma^C$ that determines the $T$-dependence of the conductivity (see Fig.~\ref{plot_delta_sigma_longrange}) for not too large conductance \mbox{$g<\exp[\pi C_*/4] \sim 10^3$}, resulting in a negative total correction to the conductivity. In the opposite (purely academic) limit of very high conductances, the effective coupling constant $\sim \ln(g)/C_*$ becomes larger than unity, leading to an effective $Z$-factor \mbox{$Z>1$}. In this situation, the resummation scheme of Section~\ref{strong} applies.

\section{Summary}
\label{summary}
\setcounter{equation}{0}

We have presented a systematic investigation of the quantum corrections to conductivity in a disordered system of fermions interacting via a fluctuating gauge field, concentrating on the Hartree-type contributions.
As a closely related effect, we also analyze the dephasing induced by the gauge-field interaction.
We have shown that the anomalously short dephasing length and strong Hartree interaction correction result from the infrared singularity of the gauge field propagator, arising from a short-range electron-electron interaction.

In general, an important outcome of our analysis is the absence of unphysical divergencies (all the singularities are cut off by disorder). In particular, the results of the paper prove that disorder stabilizes the ``mean-field'' composite-fermion model of the half-filled lowest Landau level, which in the clean situation suffers from infrared singularities.

For weak coupling \mbox{$\lambda^2 g \ll 1$} of the fermions to the gauge field discussed in Section~\ref{Hartree_wo_dephasing}, the first-order Hartree conductivity correction (\ref{delta_sigma}), \mbox{$\delta\sigma^H\sim\lambda^2g\,{\rm ln}^2T$}, exceeds the exchange contribution by a large factor $g$ (dimensionless conductance). The correction is found to be finite in the thermodynamic limit, at variance with Ref.~\onlinecite{Galitski_05}. This is intimately related to the gauge-invariance which requires summation of a certain set of the leading-order diagrams (Figs.~\ref{boxes} and \ref{AA-diags}). In Section~\ref{Hartree_by_current_corr} we elucidate the physical meaning of the obtained contribution and show that it is governed by scattering on static mesoscopic fluctuations of local currents. At exponentially low $T$, when the first-order correction becomes of the order of the Drude conductivity,
we expect that the resummation of higher-order interaction terms would effectively ``screen'' the $\ln^2 T$ contribution, similarly to the situation of strong coupling discussed in Section~\ref{strong}.

In Sections~\ref{Cooperon_dephasing} and \ref{ucf_dephasing}, we discuss dephasing of weak localization and mesoscopic conductance fluctuations at weak coupling. This, in particular, provides the background for the discussion of the dephasing effects on the Hartree correction to conductivity at strong coupling. Within the analysis of dephasing by thermal fluctuations of the gauge field, we have calculated the Cooperon dephasing rate, extending the result of Ref.~\onlinecite{Woelfle_2000} to a broader temperature range, see Eqs.~(\ref{tauphi-coop}), (\ref{tauphi-coop-3}), and (\ref{tauphi-coop-lowT}). We have also shown that, similarly to the case of Coulomb interaction\cite{Aleiner_Blanter_2002,Ludwig_Mirlin_2004}, first-order weak localization and mesoscopic conductance fluctuations are subject to essentially the same dephasing rate, see Eq.~(\ref{wl_cf_highT}). Since this demonstration has been performed on the path-integral level, it also applies to the counterparts of these phenomena in nontrivial geometries, e.g.~$h/e$ and $h/2e$ Aharonov-Bohm oscillations. We also have analyzed the dephasing rate applicable to the two-loop weak localization correction, Eq.~(\ref{tauphi-2loop}), which for composite fermions in the half-filled lowest Landau level is the leading one since the first-order Cooperon contribution is absent.

For stronger coupling \mbox{$\lambda^2 g\gg 1$}, an infinite summation of higher-order interaction terms is necessary, see Section~\ref{strong}. A surprisingly rich behavior is found in several distinct temperature regimes, owing to the interplay of the strong dephasing and the renormalization effects.   An important ingredient of the theory is the ``delayed diffuson'' characterized by large real and imaginary parts of the interaction-induced self-energy, see Eq.~(\ref{delayed-diffuson-1}). Virtual interaction processes manifest themselves in the delayed-diffuson frequency renormalization by the $Z$-factor in the self-energy: \mbox{$\Sigma^Z\sim g\omega\gg \omega$}. This renormalization effectively leads to the ``screening'' of the lowest-order contribution to the Hartree correction and dephasing rates.

We have identified two main temperature regimes, dominated by (i) strong frequency renormalization by the virtual processes (low temperatures, $T\ll T_0$, Section~\ref{low_T}) and by (ii) interplay of renormalization and dephasing (intermediate temperatures, $T_0\ll T \ll g T_0$, Section~\ref{high_T}). The temperature-dependent part of the Hartree conductivity correction is antilocalizing, \mbox{$d\delta\sigma^H/d T<0$}. At intermediate temperatures, the correction is given by Eq.~(\ref{delta_sigma_highT}) and is parametrically larger than the exchange correction~\cite{Mirlin_Woelfle_97}.
At lowest temperatures, the temperature-dependent part of the Hartree conductivity correction, Eqs.~(\ref{delta_sigma_lowfreq}) and (\ref{delta_sigma_lowT_highfreq}), is logarithmically divergent with a prefactor of order unity, \mbox{$\delta\sigma^H\sim\ln{(T_0/T)}$}. As a result, the negative exchange contribution~\cite{Mirlin_Woelfle_97} \mbox{$\delta\sigma^{\rm ex}\propto-\ln{g}\ln{(1/T)}$} becomes dominant, yielding localization in the limit of \mbox{$T\to 0$}.

Taking into account the influence of the renormalization processes on dephasing at strong coupling, we show in Section~\ref{dephasing-strong-coupling} that for $\lambda=1$ the dephasing rates are of the order of $T$: the renormalization of the frequency by virtual processes compensates the large factor of $g$ in the dephasing part of the self-energy. On the other hand, the dephasing length is anomalously short compared to the thermal length, \mbox{$L_\varphi\sim L_T/g^{1/2} \ll L_T$}.

Finally, in Section~\ref{Coulomb} we have considered composite fermions that, in addition to the gauge field, interact via unscreened Coulomb interaction which leads to a less singular gauge field propagator. As a result, the large parameter $g$ does not appear in the perturbative expressions for the dephasing rate as well as the first-order Hartree correction and the resummation of  higher-order gauge-field interaction terms is not needed. At not too low temperatures, the dephasing rate is of the order of the temperature and $L_\varphi\sim L_T$, while at $T\ll g T_0$, we find $L_\varphi\sim L_T^2/l \gg L_T$. 
For lowest temperatures $T\ll T_0$, the conventional dephasing due to scalar interaction becomes dominant.
The Hartree correction, Eq.~(\ref{delta_sigma_for_long_range}), takes the conventional form, \mbox{$\delta\sigma^H\propto\ln{(1/T)}$} (the prefactor is proportional to $\ln{g}$ but becomes large for unrealistically high conductances only).

On the experimental side, our results have an important implication for composite-fermion systems at half-filling of the lowest Landau level. In view of the parametrically different results for interaction corrections in systems with Coulomb and short-range interactions at intermediate temperatures, we expect a strong influence of an external gate (located sufficiently close to the 2D gas) on transport properties of the system. Also the dephasing lengths due to gauge field fluctuations arising from screened or unscreened electron-electron interaction differ parametrically. This should be important for the interpretation of experiments where interference of composite fermions might be observed.
In Ref.~\onlinecite{Rokhinson_Su_Goldman_95}, a logarithmic temperature-dependence of the conductivity has been reported in high-mobility samples. In those samples most likely the Coulomb interaction was unscreened, so that the Hartree correction was small compared to the exchange contribution.
Thus the interpretation of the experimental data in terms of the gauge-field exchange correction~\cite{Mirlin_Woelfle_97} retains its validity.

We close on a more general note. Low-temperature transport and quantum coherence phenomena in strongly-correlated systems have become a field of great research interest. The present work, where the interplay of disorder, strong renormalization, and dephasing effects was studied in a system with singular gauge-field interaction, demonstrates the complexity of physics emerging in this context. We expect that the ideas and methods developed here may be useful for the analysis of mesoscopic phenomena in a broad class of strongly-correlated systems.

\begin{acknowledgments}

We wish to thank A.~R. Akhmerov, V.~J. Goldman, J.~K. Jain, A. Kamenev, D.~V. Khveshchenko, D.~L. Maslov, A. Ossipov, P.~M. Ostrovsky, D.~G. Polyakov, and R.~A. Smith for useful discussions.
Additionally, we thank D.~G.~Polyakov for providing the unpublished version of Ref.~\onlinecite{Polyakov_Samokhin_98}.
T.L. was supported by the Dutch Science Foundation NWO/FOM.
The work of I.V.G., conducted as a part of the project ``Quantum Transport in Nanostructures'' made under the EUROHORCS/ESF EURYI Awards scheme, was supported by funds from the Participating Organizations of EURYI and the EC Sixth Framework Programme.
Part of this work was done when three of the authors (T.L., I.V.G., and A.D.M.) participated in the Workshop ``Interactions, excitations and broken symmetries in quantum Hall systems'' at the Max-Planck-Institut f\"ur Physik Komplexer Systeme in Dresden, Germany.
T.L. is grateful for the hospitality of the Institut f\"ur Nanotechnologie of the Forschungszentrum Karlsruhe, and of the Institut f\"ur Theorie der kondensierten Materie of the University of Karlsruhe during several visits while this work was done.

\end{acknowledgments}

\appendix

\section{Effective interaction}
\label{app-eff-int}
\renewcommand{\theequation}{A.\arabic{equation}}
\setcounter{equation}{0}

In this Appendix, we calculate the effective interaction box $\tilde{U}$ given by Eq.~(\ref{tilde_U}),
\be
\tilde{U}=\tilde{U}^{(0)}+\tilde{U}^{(1)}+\tilde{U}^{(2)}\:\:,
\ee
with the three contributions arising from the three diagrams shown in Fig.~\ref{boxes}. Since $\tilde{U}$ is part of the first-order Hartree diagrams, there is no energy transfer through the gauge field line. The bare box, shown in the left of Fig.~\ref{boxes}, is calculated as follows,
\begin{widetext}
\bea
\tilde{U}^{(0)}
&=&
\frac{1}{\left(2\pi\nu\tau\right)^2}\:
\int\limits^{l^{-1}}(dk)\:
U_{\alpha\beta}(\bk,\epsilon=0)\:\left(e^*\right)^2\:
\int(dp)\:v_\alpha v_\beta\:
G^R({\bf p})G^R({\bf p-k})G^A({\bf p})G^A({\bf p-k})\nonumber \\
&=&
\frac{1}{\left(2\pi\nu\tau\right)^2}\:
\int\limits^{l^{-1}}(dk)\:
\frac{\left(e^*\right)^2}{\chi_0k^2}
\int(dp)\:v_x^2{\rm sin}^2\phi\:\tau^2\left[G^R({\bf p})G^A({\bf p-k})+G^A({\bf p})G^R({\bf p-k})\right]\nonumber\\
&=&
\frac{1}{\left(2\pi\nu\tau\right)^2}\:
\int\limits^{l^{-1}}(dk)\:
\frac{\left(e^*\right)^2}{\chi_0k^2}\:
2\tau^2\:
2\pi i\nu v_F^2
\int\frac{d\phi}{2\pi}\:\frac{{\rm sin}^2\phi}{(p_Fk/m)\,{\rm cos}\phi+i/\tau}\nonumber\\
&=&
\frac{1}{\left(2\pi\nu\tau\right)^2}\:
\frac{\left(e^*\right)^2v_F^2}{\chi_0}\:
4\pi\nu\tau^3
\int(dk)\:
\frac{1}{k^2}\:\frac{1}{1+\sqrt{1+k^2l^2}}\nonumber\\
&=&
8\pi\left(e^*\right)^2\frac{3 g}{\pi \nu}
\int(dk)\:
\frac{1}{k^2}\:\frac{1}{1+\sqrt{1+k^2l^2}}
\eea
with $\phi$ the angle between $\bk$ and ${\bf p}$. This integral is determined by small momenta $k$,
\be
\tilde{U}^{(0)}
\approx
2\left(e^*\right)^2\frac{3 g}{\pi \nu}\int\limits^{l^{-1}}\frac{dk}{k}
\:\:.
\label{U0}
\ee
The boxes with the diffusons crossing the gauge field line are calculated by expanding both fermionic boxes to the leading order in \mbox{$ql\ll 1$} and \mbox{$kl\ll 1$}. We find
\bea
\tilde{U}^{(1)}
&=&
\frac{1}{\left(2\pi\nu\tau\right)^2}\:
\int\limits^{l^{-1}}(dk)\:U_{\alpha\beta}(\bk)\:
\frac{1}{(\bq-\bk)^2-i\omega/D}\:\frac{\left(e^*\right)^2v_F^2}{2}\:
2\pi\nu\tau^3
\biggl[{}
-4q_\alpha q_\beta+2q_\alpha k_\beta+2k_\alpha q_\beta-k_\alpha k_\beta
\biggr]\:\:,\\
\tilde{U}^{(2)}
&=&
\frac{1}{\left(2\pi\nu\tau\right)^2}\:
\int\limits^{l^{-1}}(dk)\:U_{\alpha\beta}(\bk)\:
\frac{1}{(\bq+\bk)^2-i\omega/D}\:\frac{\left(e^*\right)^2v_F^2}{2}\:
2\pi\nu\tau^3
\biggl[{}
-4q_\alpha q_\beta-2q_\alpha k_\beta-2k_\alpha q_\beta-k_\alpha k_\beta
\biggr]\:\:.
\eea
\end{widetext}
$\tilde{U}^{(1)}$ and $\tilde{U}^{(2)}$ do not contribute at \mbox{$k\gg q$}, while at \mbox{$k\ll q$} they cancel the low-$k$ divergence of the bare box $\tilde{U}^{(0)}$: Inserting the sum of the three contributions into the standard exchange diagrams of type d+e, the average of the entire diagram (consisting of the effective interaction $\tilde{U}$ times fermionic part ${\cal B}$) over the relative angle $\phi$ between $\bq$ and $\bk$ has the structure
\bea
\left\langle\tilde{U}\,{\cal B}_{d+e}\right\rangle_\phi
&\sim&
\left\langle\left[\delta_{\alpha\beta}-\frac{k_\alpha k_\beta}{k^2}\right]
\left[\delta_{\alpha\beta}-4\frac{q_\alpha q_\beta}{q^2}\right]\:q_xq_x\right\rangle_\phi\nonumber\\
&\sim&
1
-\left\langle{\rm cos}^2\phi\right\rangle
-4\left\langle{\rm cos}^2\phi\right\rangle
+4\left\langle{\rm cos}^4\phi\right\rangle\nonumber\\
&=&
0\:.
\label{angular_av}
\eea
For the logarithmic accuracy of the calculations of this paper, the details of the low-$k$ cutoff are not important. Since $\tilde{U}$ is always integrated against a fermionic part containing diffusion propagators, which set \mbox{$\omega\sim Dq^2$}, this also holds for the $\omega$-dependence. We may thus apply the low-$k$ cutoff $q$ to the integral in Eq.~(\ref{U0}) and find the effective interaction (\ref{tilde_U}).

\section{Gauge invariance and extra diagrams}
\label{app-extra-diags}
\renewcommand{\theequation}{B.\arabic{equation}}
\setcounter{equation}{0}

While calculating the Hartree conductivity correction for the gauge-field problem, one encounters certain cancellations between diagrams,
similarly to the case of conventional scalar interaction~\cite{Altshuler_Aronov_85}. Such cancellations greatly facilitate the evaluation of
$\delta \sigma^H$, as discussed in this Appendix.

\begin{figure}
\includegraphics[width=0.3\linewidth]{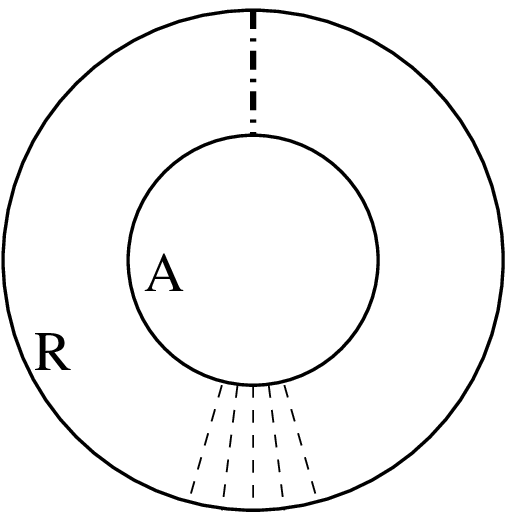}
\hfill
\includegraphics[width=0.3\linewidth]{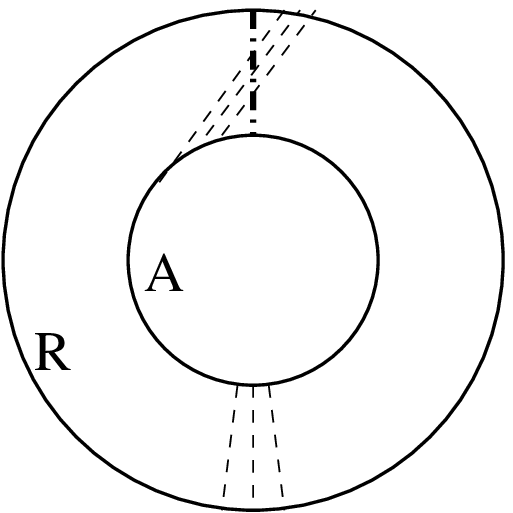}
\hfill
\includegraphics[width=0.3\linewidth]{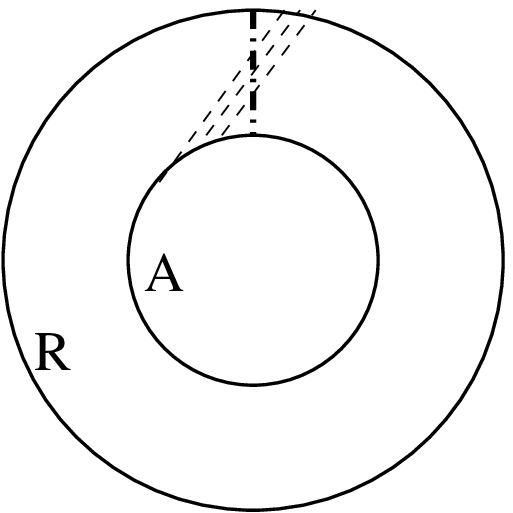}
\caption{\label{functionals} Generating functionals which lead to Hartree diagrams. Mirrored versions of the middle and right diagram are also possible.}
\end{figure}

There are two types of such cancellations. 
First, all diagrams arising from the variation of the generating functionals shown
in Fig.~\ref{functionals} can be combined to zero.
This follows from gauge-invariance arguments
in the limit of zero gauge field momentum (the variation of a gauge-invariant functional with respect to a static uniform gauge field must vanish),
and has been explicitly demonstrated in Ref.~\onlinecite{Altshuler_Aronov_Larkin_Khmelnitskii_81}.
The result is that
only diagrams where retarded Green's functions are changed into advanced ones at the external current vertices
[``retarded-advanced" (RA) diagrams of type $a$ and $b$] remain.  This kind of cancellation is used in the strong-coupling regime (Section~\ref{strong}), where diagrams of the RA-type ($a$ + $b$) with interaction-dressed diffusons were evaluated.

An alternative cancellation allows one to take into account only diagrams of type $d$ and $e$, using the argument of Ref.~\onlinecite{Altshuler_Aronov_85} that the sum of diagrams $a$,$b$, and $c$ is zero in the diffusion approximation.
In view of this, the result (\ref{delta_sigma}) has been calculated from diagrams $d$ and $e$ (see Fig.~\ref{AA-diags})
in terms of the effective interaction $\tilde{U}$, Fig.~\ref{boxes}.

Let us note that one should exercise certain caution employing this type of cancellation to the gauge-field problem.
The point is that additional diagrams with diffusons crossing the gauge field line which do not
contain the closed effective interaction box $\tilde{U}$ exist. We show below, however, that these ``non-standard''
diagrams cancel out to the leading order.
This justifies using the standard diagrams of type $d$ and $e$ with the effective interaction $\tilde{U}$
for calculating the Hartree correction.

\begin{figure}
\hspace*{3mm}
\includegraphics[width=0.4\linewidth]{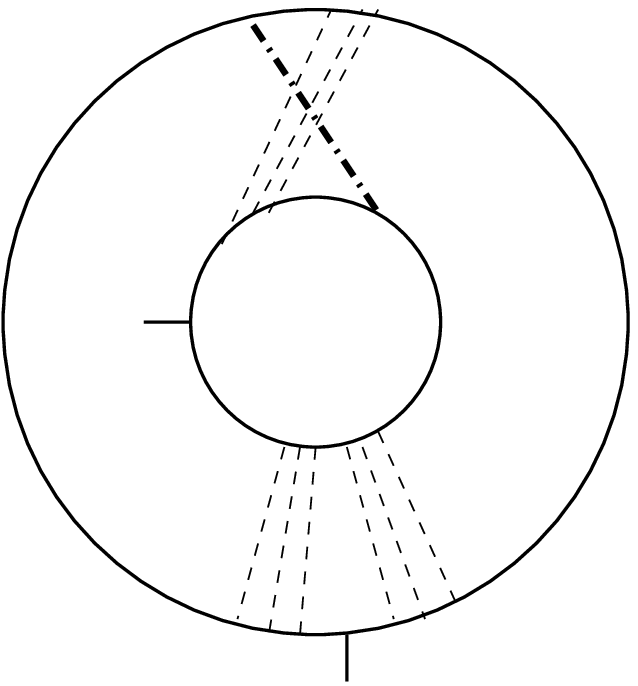}
\hfill
\includegraphics[width=0.4\linewidth]{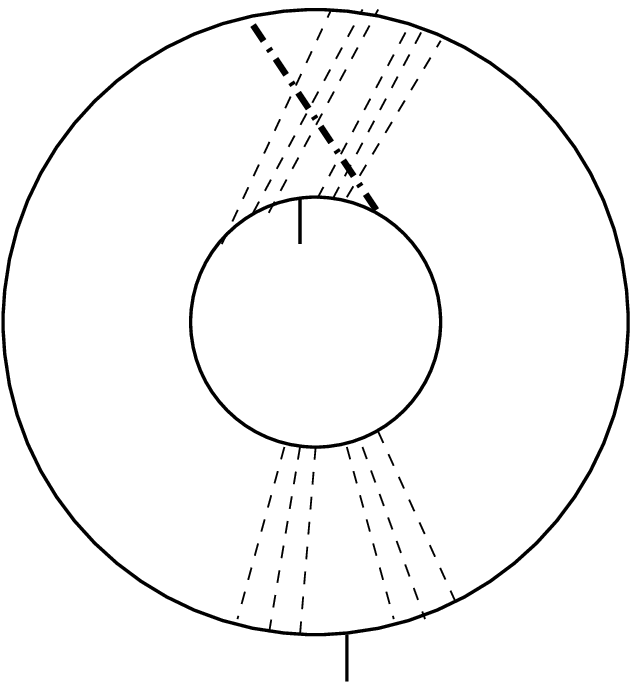}
\hspace*{3mm}
\caption{\label{nonstandard} Examples of diagrams which do not contain the effective interaction $\tilde{U}$. The left diagram has an external current vertex adjacent to the diffuson crossing gauge field line; the right diagram has a velocity vertex splitting the diffuson which crosses the gauge field line.}
\end{figure}

The extra diagrams can be obtained, along with diagrams of type $c$,$d$,$e$, and the part of type $a$ which contains two retarded (two advanced) Green functions at the current vertices (RR part), from the generating functionals shown in Fig.~\ref{functionals}.
The functional without a diffuson crossing the gauge field line gives the exchange diagrams in terms of the bare part of the effective interaction $\tilde{U}$, plus diagrams with the velocity vertices next to the interaction vertices, where the latter are small due to an insufficient number of diffusons. However, the functional with the additional diffuson gives, apart from exchange diagrams in terms of the dressed part of $\tilde{U}$, also relevant diagrams which cannot be classified in terms of $\tilde{U}$ because they do not contain the closed interaction box. In addition, the functional with only the diffuson which crosses the interaction line leads to contributions of the same order too.

We start with the diagrams obtained by inserting two velocity vertices into different bubbles of the second functional of Fig.~\ref{functionals}. There are $25$ possibilities to place the two velocity vertices. In addition, when both vertices are inserted into the same diffuson, the different sub-possibilities of placing them have to be taken into account.
When placing both vertices in the same bubble, there are $15$ possibilities of placing them, which again have to be further distinguished if both vertices are placed in the same diffuson. Of all these possibilities, some contributions are small, some represent the standard diagrams \cite{Altshuler_Aronov_85} in terms of the closed box $\tilde{U}$, and some give a contribution which is of the leading order but cannot be expressed in terms of $\tilde{U}$ (see two examples in Fig.~\ref{nonstandard}).
Also, vertices can be placed splitting the diffuson of the third functional of Fig.~\ref{functionals} (otherwise the contribution is small).

All the remaining contributions which cannot be written in terms of $\tilde{U}$ are only relevant in the case \mbox{$k\ll q$}, since in the opposite case one or more diffusion poles are displaced by $k$. Explicitly evaluating the relevant contributions which do not contain $\tilde{U}$, we find that they cancel and thus do not affect the cancellation of Eq.~(\ref{angular_av}).

From the second functional of Fig.~\ref{functionals}, inserting both velocity vertices into different bubbles so that they split the lower diffuson gives (the part with the dressed gauge field line of) the standard diagrams of type $c$ (if the vertices are not separated by an impurity line) and type $e$ (if the vertices are separated by a part of the diffuson), both in terms of $\tilde{U}$. Similarly, inserting the vertices into the lower diffuson into the same bubble gives part of the standard diagrams of type $d$ and the RR part of type $a$, respectively. Evaluated in the limit \mbox{$k\ll q$}, it can be seen explicitly that these diagrams cancel.

\begin{widetext}
\section{Path integral transformation between weak localization and mesoscopic conductance fluctuations}
\label{path_integral_trafo}
\renewcommand{\theequation}{C.\arabic{equation}}
\setcounter{equation}{0}

In this Appendix we present the details of the path integral transformation which results in Eq.~(\ref{wl_cf_highT}).
For convenience, only the diffuson part of the conductance fluctuations will be considered, assuming that 
the time-reversal symmetry is broken by random or uniform magnetic field, or that harmonics of the Aharonov-Bohm effect are considered, the amplitude of which is determined solely by the diffuson part\cite{TL_diss_06}.

We consider here the weak-coupling situation \mbox{$\lambda^2g \ll 1$}, resulting in \mbox{$l,L_T\ll L_\varphi$}: In this case the regime of interest is \mbox{$T\gg D/L^2$}, and the function (\ref{tilde-delta}) can be approximated by a delta function.
For a static gauge field configuration \mbox{${\bf a}(\br)$}, the amplitude of mesoscopic conductance fluctuations can be written as
\bea
\left\langle\delta g^2\right\rangle
&=&\frac{16\pi D^2}{3TL^4}
   \int {\bf d^dR_1}\int {\bf d^dR_2} \int\limits_0^\infty dt\,
   \int\limits_{\br_1(0)={\bf R_2}}^{\br_1(t)={\bf R_1}}{\cal D}[\br_1(t)]
   \int\limits_{\br_2(0)={\bf R_2}}^{\br_2(t)={\bf R_1}}{\cal D}[\br_2(t)]\phantom{dummy}\nonumber\\
& &\times\:
   \Bigg\langle{\rm exp}\Bigg\{-\int\limits_0^t dt'_1\,
   \bigg[
   \frac{\dot{\br}_1^2(t'_1)}{4D}
    +i\lambda e\,\dot{\br}_1(t'_1)\,\Big[{\bf a}_1[\br_1(t'_1)]-{\bf a}_2[\br_1(t'_1)]
   \Big]
   \bigg]\nonumber\\
& &\qquad\qquad -\int\limits_0^t dt'_2\,
   \bigg[
   \frac{\dot{\br}_2^2(t'_2)}{4D}
    +i\lambda e\,\dot{\br}_2(t'_2)\,\Big[{\bf a}_1[\br_2(t'_2)]-{\bf a}_2[\br_2(t'_2)]
   \Big]
   \bigg]
   \Bigg\}\Bigg\rangle\:\:.
   \label{cf-gf}
\eea
Clearly, a static gauge field, \mbox{${\bf a}_1={\bf a}_2$}, drops out of Eq.~(\ref{cf-gf}). In the opposite limit of the time between measurements being long compared to the gauge field dynamics, correlators between different measurements vanish, \mbox{$\langle{\bf a}_1{\bf a}_2\rangle=0$}. Performing the average over Gaussian variables \mbox{${\bf a}(\br)$} results in $16$ gauge-field induced terms in the exponent. Half of them correlates gauge fields from the same measurement, so that in the static approximation discussed in the main text the result is
\bea
\left\langle\delta g^2\right\rangle
&=&\frac{16\pi D^2}{3TL^4}
   \int {\bf d^dR_1}\int {\bf d^dR_2} \int\limits_0^\infty dt\,
   \int\limits_{\br_1(0)={\bf R_2}}^{\br_1(t)={\bf R_1}}{\cal D}[\br_1(t)]
   \int\limits_{\br_2(0)={\bf R_2}}^{\br_2(t)={\bf R_1}}{\cal D}[\br_2(t)]
   \nonumber\\
& &\times\:{\rm exp}\Bigg\{-\int\limits_0^{t}dt'
   \left[\frac{\dot{\br}_1^2(t')}{4D}+\frac{\dot{\br}_2^2(t')}{4D}\right]\nonumber\\
& &\qquad\qquad{}-\lambda^2e^2\int\limits_0^{t}dt'_1dt'_2\,\bigg[\dot{\br}_1(t'_1)
      \,\Bigl\langle {\bf a}[\br_1(t'_1)]\,{\bf a}[\br_1(t'_2)]\Bigr\rangle\,
      \dot{\br}_1(t'_2)
+\bigl\langle {\bf a}_2[\br_1(t'_1)]\,{\bf a}_2[\br_1(t'_2)]\bigr\rangle
    \Big]\nonumber\\
& & \hspace*{2.5cm}\qquad\qquad{}
+\dot{\br}_2(t'_1)
      \,\Bigl\langle {\bf a}[\br_2(t'_1)]\,{\bf a}[\br_2(t'_2)]\Bigr\rangle\,
      \dot{\br}_2(t'_2)
+\bigl\langle {\bf a}_2[\br_2(t'_1)]\,{\bf a}_2[\br_2(t'_2)]\bigr\rangle
    \Big]\nonumber\\
& & \hspace*{2.5cm}\qquad\qquad{}-\dot{\br}_1(t'_1)
      \,\Bigl\langle {\bf a}[\br_1(t'_1)]\,{\bf a}[\br_2(t'_2)]\Bigr\rangle\,
      \dot{\br}_2(t'_2)
+\bigl\langle {\bf a}_2[\br_1(t'_1)]\,{\bf a}_2[\br_2(t'_2)]\bigr\rangle
    \Big]\nonumber\\
& & \hspace*{2.5cm}\qquad\qquad{}-\dot{\br}_2(t'_1)
      \,\Bigl\langle {\bf a}[\br_2(t'_1)]\,{\bf a}[\br_1(t'_2)]\Bigr\rangle\,
      \dot{\br}_1(t'_2)
+\bigl\langle {\bf a}_2[\br_2(t'_1)]\,{\bf a}_2[\br_1(t'_2)]\bigr\rangle
    \Big]
    \bigg]\Bigg\}\:\:.
\eea

Using the transformation
\be
\br(t')=\left\{\begin{array}{ll}
                    \br_1(t+t')   \:\:,\ \ &-t\le t'\le 0\\
                                              & \\
                    \br_2(t-t')\:\:,\ \ &\:\:\:0\le t'\le t\\
                    \end{array}\right.
\label{join_paths}
\ee
this is equal to the the path-integral representation of the weak-localization correction in a static random gauge field \mbox{${\bf a}(\br)$} after rescaling the gauge field correlator by a factor $2$,
\bea
\left\langle\delta g^2\right\rangle
&=&
\frac{16\pi D^2}{3TL^4}
\int {\bf d^dR}
\int\limits_0^\infty dt
\int\limits_{\br(-t)={\bf R}}^{\br_1(t)={\bf R}}{\cal D}[\br(t)]\:
{\rm exp}\left\{
-\int\limits_{-t}^{t} dt'\:\frac{\dot{\br}^2}{4D}
-\lambda^2e^2\int\limits_{-t}^{t}dt'_1dt'_2\:
\dot{\br}(t'_1)
\,\Bigl\langle
{\bf a}[\br(t'_1)]\,{\bf a}[\br(t'_2)]
\Bigr\rangle\,
\dot{\br}(t'_2)
\right\}\nonumber\\
&=&
\frac{16\pi D^2}{3TL^4}
   \int {\bf d^dR} \int\limits_0^\infty dt
   \int\limits_{\br(-t)={\bf R}}^{\br_1(t)={\bf R}}{\cal D}[\br(t)]\:
{\rm exp}\left\{-\!\int\limits_{-t}^t dt'\,\frac{\dot{\br}^2}{4D}\right\}
   \left\langle
   {\rm exp}
   \left\{
   -2i\lambda e\int\limits_{-t}^t dt'\:
   \dot{\br}(t')\,\frac{1}{\sqrt{2}}{\bf a}[\br(t')]
   \right\}
   \right\rangle\:\:.
\eea
For thermal gauge field fluctuations this is equivalent to rescaling the temperature by a factor of $2$, resulting in Eq.~(\ref{wl_cf_highT}).
The different short-scale cutoffs\cite{Aleiner_Blanter_2002} have been discussed in the main text.

\section{Diffuson self-energy}
\label{app_selfenergies}
\renewcommand{\theequation}{D.\arabic{equation}}
\setcounter{equation}{0}

In this Appendix we calculate the delay-time dependent diffuson self-energy $\Sigma^Z_\eta$ used in Sections~\ref{low_T} and \ref{high_T}
Since in Section~\ref{low_T} the Hartree structure of the contribution to the conductivity correction is ensured by subtracting the disconnected part, here we take into account all possible virtual contributions to the effective interaction box (the diagrams in Fig.~\ref{selfenergy-boxes} plus the vertex part given by the first diagram in Fig.~\ref{boxes}). Similarly to the situation in Appendix~\ref{app-eff-int}, where the diffusons crossing the gauge field line set the low-momentum cutoff, the low-momentum cutoff for the self-energy contributions of Fig.~\ref{selfenergy-boxes} is set by corresponding diagrams with a diffuson covered by the impurity line. In the following, we will therefore only consider the bare boxes with the appropriate cutoff \mbox{$k\gtrsim q$}.

We start with the evaluation of the self-energy of the disconnected diffuson \mbox{$\tilde{{\cal D}}_\infty$} at low frequencies, \mbox{$\omega\ll T_0$}. Because these frequencies are smaller than the characteristic width of the gauge field propagator $T_0$, the interaction vertices may change retarded into advanced Green's functions. As a result, the three contributions shown in Fig.~\ref{selfenergy-boxes} (and the corresponding ones with the gauge field line inserted into the advanced Green's function) are possible. The second and third diagram of Fig.~\ref{selfenergy-boxes} form a Hikami-box 
(see, e.g., Refs. \onlinecite{Adamov} and \onlinecite{Kee_Aleiner_Altshuler_98} for discussion of Hikami-box contributions to the self-energy) 
which combines to one half the second diagram, while the fourth diagram simply acts as a low-$k$ cutoff for the first one, similar to the second and third diagram in Fig.~\ref{boxes} for the first diagram there. We find the following result for the self-energy of the disconnected diffuson,
\bea
\Sigma^Z_\infty
&=&
-2\pi i\nu\,
\frac{1}{\left(2\pi\nu\tau\right)^2}
\int(dp)\int(dk)\:
\lambda^2e^2\:v_\alpha v_\beta
\left[
\int\limits_{-\infty}^{E}\frac{d\epsilon}{2\pi}\:\left(G^R\right)^3G^A
+\frac{1}{2}\int\limits_{E}^{\infty}\frac{d\epsilon}{2\pi}\:\left(G^R\right)^2\left(G^A\right)^2\right.\nonumber\\
& &\left.{}
+\int\limits_{E-\omega}^{\infty}\frac{d\epsilon}{2\pi}\:G^R\left(G^A\right)^3
+\frac{1}{2}\int\limits_{-\infty}^{E-\omega}\frac{d\epsilon}{2\pi}\:\left(G^R\right)^2\left(G^A\right)^2
\right](-i){\rm Re}\,U_{\alpha\beta}(\bk,\epsilon)
\label{Sigma_infty_lowT-raw}
\eea
Similarly to the calculation of the vertex interaction part in Appendix~\ref{app-eff-int}, we get (neglecting $E\ll \omega$)
\bea
\Sigma^Z_\infty
&=&
-2\pi i\nu\,
\frac{i}{\left(2\pi\nu\tau\right)^2}\:
2\pi\nu\tau^3\:
\frac{\lambda^2e^2v_F^2}{2}\:
\int\limits_{-\omega}^{0}\frac{d\epsilon}{2\pi}
\int(dk)\:
{\rm Re}\,\frac{1}{\chi_0k^2-i\sigma_0\epsilon}\:\:
\nonumber\\
&=&
-2\pi i\nu\,
\frac{i}{\left(2\pi\nu\tau\right)^2}\:
2\pi\nu\tau^3\:
\frac{\lambda^2e^2v_F^2}{2}\:
\frac{\omega}{2\pi}\:
\frac{1}{2\pi\chi_0}\:
\frac{1}{2}\left[\ln{\frac{T_0}{\omega}}+1\right]\nonumber \\
&=&
\frac{3}{2\pi}\,\lambda^2g\omega\left[\ln{\frac{T_0}{\omega}}+1\right]\:\:,\quad\omega\ll T_0\:\:.
\label{disconnected-selfenergy-lowT}
\eea
The renormalization self-energy $\Sigma^Z_\infty$ can be cast in the form
\be
\Sigma_\infty^Z=2\pi \nu\int\limits_{0}^{\omega} \frac{d\epsilon}{2\pi}\:
{\rm Re}\,\tilde{U}(\epsilon)\:\:.
\label{Sigma-U-infty}
\ee
with
\be
{\rm Re}{\tilde U}(\epsilon)=\frac{2}{(2\pi \nu)^2 T_0} \int\limits_q^{k_F} k dk \frac{k^2l^2(1+\sqrt{1+k^2l^2})}{[k^2l^2(1+\sqrt{1+k^2l^2})]^2+(2\epsilon/T_0)^2}.
\label{Re-tilde-U-epsilon-k}
\ee
Evaluating the momentum integral in (\ref{Re-tilde-U-epsilon-k}) leads to Eq.~(\ref{Re-tilde-U-epsilon}).

For the fully dressed diffuson $\tilde{{\cal D}_\eta}$, it is straightforward to see that at zero delay time the vertex contribution (first diagram of Fig.~\ref{boxes}, now with the energy transfer through the gauge field line integrated over the interval \mbox{$\left[-\omega,0\right]$} set by the RA channel) exactly cancels with the self-energy (\ref{disconnected-selfenergy-lowT}).

For the fully dressed diffuson at finite delay time, we account for the partial
cancellation with the vertex terms by inserting the factor \mbox{$\left[1-{\rm cos}\,\epsilon\eta\right]$}
[see also Section~\ref{diffuson_dephasing} and Eq.~(\ref{Sigma-U-eta})] into the $\epsilon$-integral in
Eqs.~(\ref{disconnected-selfenergy-lowT}),(\ref{Sigma-U-infty}) and find
\be
\Sigma^Z_\eta
=
\frac{3}{2\pi}\,\lambda^2g
\int\limits_{-\omega}^0 d\epsilon\:
\ln{\frac{T_0}{|\epsilon|}}
\left[1-{\rm cos}\,\epsilon\eta\right]
=
\frac{3}{2\pi}\,\lambda^2g\,\omega
\left[\ln{\frac{T_0}{\omega}}\left(1-\frac{{\rm sin}\,\omega\eta}{\omega\eta}\right)
+1-\frac{{\rm Si}(\omega\eta)}{\omega\eta}
\right]\:\:,\nonumber\\
\label{full-selfenergy-lowT}
\ee
with $\rm Si(x)$ the integral sine function.

We now turn to the situation of high frequencies, \mbox{$\omega\gg T_0$}.
Starting from Eq.~(\ref{Sigma_infty_lowT-raw}), we now allow for the situation of larger energy transfers, thus taking into account contributions from \mbox{$k\gg l^{-1}$}. Neglecting $\omega$ and $\bq$ in the arguments of the Green's functions, we find
\bea
\Sigma^Z_\infty
&=&
-2\pi i\nu\,
\frac{1}{\left(2\pi\nu\tau\right)^2}
\int(dp)\int(dk)\:
\lambda^2e^2\:v_\alpha v_\beta
\left[
\int\limits_{-\infty}^{E}\frac{d\epsilon}{2\pi}\:\left[G^R(E,{\bf p})\right]^2G^R(E-\epsilon,{\bf p-k})\:G^A(E,{\bf p})\right.\nonumber\\
& &{}
+\int\limits_{E}^{\infty}\frac{d\epsilon}{2\pi}\:\left[G^R(E,{\bf p})\right]^2G^A(E-\epsilon,{\bf p-k})\:G^A(E,{\bf p})
-i\tau\int\limits_{E}^{\infty}\frac{d\epsilon}{2\pi}\:\left[G^R(E,{\bf p})\right]^2G^A(E-\epsilon,{\bf p-k})\nonumber\\
& &{}
+\int\limits_{E-\omega}^{\infty}\frac{d\epsilon}{2\pi}\:G^R(E,{\bf p})\left[G^A(E,{\bf p})\right]^2G^A(E-\epsilon,{\bf p-k})
+\int\limits_{-\infty}^{E-\omega}\frac{d\epsilon}{2\pi}\:G^R(E,{\bf p})\:G^R(E-\epsilon,{\bf p-k})\left[G^A(E,{\bf p})\right]^2\nonumber\\
& &{}\left.
+i\tau\int\limits_{-\infty}^{E-\omega}\frac{d\epsilon}{2\pi}\:G^R(E-\epsilon,{\bf p-k})\left[G^A(E,{\bf p})\right]^2
\right](-i){\rm Re}\,U_{\alpha\beta}(\bk,\epsilon)\:\:.
\eea
Using the relation \mbox{$G^RG^A=i\tau\left[G^R-G^A\right]$} and neglecting the energy differences in the arguments of the Green's functions, this can be simplified to the form analogous to Eq.~(\ref{disconnected-selfenergy-lowT}),
\bea
\Sigma^Z_\infty
&=&
-2\pi i\nu\,
\frac{1}{\left(2\pi\nu\tau\right)^2}
\int(dk)\:
\frac{\lambda^2e^2v_F^2}{2}
\left[-2\pi\nu\tau^3\right]
\frac{2}{1+\sqrt{1+k^2l^2}}\nonumber\\
& &
\times
\int\limits_{-\omega}^{0}\frac{d\epsilon}{2\pi}\:
(-i)\,\frac{1}{\sigma(k)}\,\frac{k^2l^2T_0\left(1+\sqrt{1+k^2l^2}\right)/2}{\left[k^2l^2T_0\left(1+\sqrt{1+k^2l^2}\right)/2\right]^2+\epsilon^2}\:\:.
\label{disconnected-selfenergy-highT}
\eea
The factor \mbox{$2/[1+\sqrt{1+k^2l^2}]$} arises from the momentum difference in the combination \mbox{$G^R(p)\,G^A(p-k)$} and is derived in detail in Appendix~\ref{App_interpolation}. It is cancelled by the $k$-dependence of $\sigma(k)$. At \mbox{$\omega\gg T_0$}, Eq.~(\ref{disconnected-selfenergy-highT}) therefore is dominated by the high-momentum part. It can be calculated as follows:
\bea
\Sigma^Z_\infty
&=&
\frac{2\lambda^2}{\nu}
\int\limits_{-\omega}^{0}\frac{d\epsilon}{2\pi}
\int(dk)\:
\frac{\frac{1}{2}k^2l^2T_0\left(1+\sqrt{1+k^2l^2}\right)}{\left[\frac{1}{2}k^2l^2T_0\left(1+\sqrt{1+k^2l^2}\right)\right]^2+\epsilon^2}=
\frac{\lambda^2}{8\pi g\tau}
\int\limits_{-\omega/T_0}^{0}dy
\int\limits_0^\infty dz\:
\frac{\frac{z}{2}\left(1+\sqrt{1+z}\right)}{\left[\frac{z}{2}\left(1+\sqrt{1+z}\right)\right]^2+y^2}\nonumber\\
&=&
\frac{\lambda^2}{8\pi g\tau}\:\frac{\omega}{T_0}
\int\limits_0^\infty d\tilde{z}\:
{\rm arctan}\left\{\frac{2}{\tilde{z}\left(1+\sqrt{1+\tilde{z}\omega/T_0}\right)}\right\}=
\left\{
\begin{array}{ll}
\displaystyle\frac{3}{2\pi}\,\lambda^2g\,\omega\:\pi\left(\frac{4T_0}{\omega}\right)^{1/3}\:\:,&\quad\omega\gg T_0\\[0.3cm]
\displaystyle\frac{3}{2\pi}\,\lambda^2g\,\omega\left[\ln{\frac{T_0}{\omega}}+1\right]\:\:,&\quad\omega\ll T_0\:\:,
\end{array}
\right.
\eea
both confirming Eq.~(\ref{disconnected-selfenergy-lowT}) and providing the behavior for larger $\omega$. Again it is straightforward to see that at zero delay time the vertex contribution exactly cancels Eq.~(\ref{disconnected-selfenergy-highT}).

To calculate the self-energy of the fully dressed diffuson at high frequencies and finite delay time, we again insert the factor \mbox{$\left[1-{\rm cos}\,\epsilon\eta\right]$} into Eq.~(\ref{disconnected-selfenergy-highT}). We first calculate the result for short delay times, \mbox{$\eta\ll 1/\omega$}:
\bea
\Sigma^Z_\eta
&=&
\frac{\lambda^2}{\nu}
\int\limits_{-\omega}^{0}\frac{d\epsilon}{2\pi}\:
\frac{1}{2}\left(\epsilon\eta\right)^2
\int(dk)\:
\frac{\frac{1}{2}k^2l^2T_0\left(1+\sqrt{1+k^2l^2}\right)}{\left[\frac{1}{2}k^2l^2T_0\left(1+\sqrt{1+k^2l^2}\right)\right]^2+\epsilon^2}
\nonumber \\ 
&=&
\frac{\lambda^2}{8\pi g\tau}
\frac{1}{2}\left(T_0\eta\right)^2
\int\limits_{-\omega/T_0}^{0}dy\:y^2
\int\limits_0^\infty dz\:
\frac{\frac{z}{2}\left(1+\sqrt{1+z}\right)}{\left[\frac{z}{2}\left(1+\sqrt{1+z}\right)\right]^2+y^2}\nonumber\\
&\approx&
\frac{\lambda^2}{8\pi g\tau}
\left(T_0\eta\right)^2
\int\limits_{-\omega/T_0}^{0}dy\:y^2
\int\limits_{2^{2/3}\left|y\right|^{2/3}}^\infty dz\:z^{-3/2}\:\:=
\frac{\lambda^2}{8\pi g\tau}
\left(T_0\eta\right)^2
\frac{3}{8}\left(\frac{2 \omega}{T_0}\right)^{8/3}\nonumber\\
&=&
{\cal O}(1)\cdot \lambda^2g\,\omega\left(\omega\eta\right)^2\left(\frac{T_0}{\omega}\right)^{1/3}\:\:,\quad\eta\ll 1/\omega\:\:.
\eea

For long delay times, \mbox{$\eta\gg 1/\omega$}, the long-$\eta$ tail of the self-energy can be obtained by extending the limits of the integration over the cosine contribution from \mbox{$\left[E-\omega,E\right]$} to \mbox{$\left[-\infty,\infty\right]$} and evaluating it using the residue theorem,
\bea
\Sigma^Z_\eta
&=&
\Sigma^Z_\infty
-\frac{\lambda^2}{8\pi g\tau}
\int\limits_0^\infty dz\:
\pi\,{\rm exp}\left\{-\frac{1}{2}\,T_0\eta z\left(1+\sqrt{1+z}\right)\right\}\nonumber\\
&=&
\left\{
\begin{array}{ll}
\Sigma^Z_\infty-3\lambda^2gT_02^{-1/3}\Gamma(5/3)\left(T_0\eta\right)^{-2/3}\:\:,&T_0\eta\ll 1\\[0.2cm]
\Sigma^Z_\infty-3\lambda^2g/\eta\:\:,&T_0\eta\gg 1\:\:.
\end{array}
\right.
\eea
The results of this Appendix are summarized in Eqs.~(\ref{disconnected-selfenergy-main}),(\ref{full-selfenergy-small-omega}), and (\ref{full-selfenergy-main}).
A similar calculation gives also the dephasing part $\Sigma_\eta^\varphi$ of the self-energy, Eqs. (\ref{deph-selfenergy-lowT-low-omega}), (\ref{deph-selfenergy-lowT-high-omega}), and (\ref{deph-selfenergy-highT}). The only difference is in using ${\rm Im}{\tilde U}(\epsilon)$
instead of ${\rm Re}{\tilde U}(\epsilon)$ and in a thermal factor (\ref{coth-tanh}) which is used instead of 
$\tanh([\epsilon+\omega]/2T)-\tanh[\epsilon/2T]$.

\section{Low-temperature Hartree correction}
\label{app-lowT}
\renewcommand{\theequation}{E.\arabic{equation}}
\setcounter{equation}{0}

In this Appendix, we calculate the Hartree conductivity correction at low temperatures, \mbox{$T\ll T_0$}. We start with the contribution of low frequencies, \mbox{$T\lesssim\omega\ll T_0$} (note that the function $f(T_0/\omega)$ drops out exactly):
\bea
\delta\sigma^{\omega<T_0}
&=&
2\sigma_0
\int\limits_{-T_0}^{T_0}\frac{d\omega}{2\pi}\:\frac{\partial}{\partial\omega}\left[\omega\:{\rm coth}\frac{\omega}{2T}\right]
\int\limits_{-\infty}^\infty d\eta
\int(dq)\:
\frac{1}{2\pi\nu}
\left[
\tilde{{\cal D}}_\eta-\tilde{{\cal D}}_\infty
\right]\nonumber\\
&=&
\frac{e^2}{2\pi^3}
\int\limits_0^{T_0}d\omega\:\frac{\partial}{\partial\omega}\left[\omega\:{\rm coth}\frac{\omega}{2T}\right]
\int\limits_0^\infty d\eta
\ln{\frac{\omega+\Sigma^Z_\infty}{\omega+\Sigma^Z_\eta}}\nonumber\\
&\approx&
\frac{e^2}{2\pi^3}
\int\limits_0^{T_0}d\omega\:\frac{\partial}{\partial\omega}\left[\omega\:{\rm coth}\frac{\omega}{2T}\right]
\int\limits_0^{\infty}d\eta
\ln{\frac{\Sigma^Z_\infty}{\Sigma^Z_\infty\left[1-\displaystyle\frac{{\rm sin}\,\omega\eta}{\omega\eta}\right]+\delta\Sigma^Z_\eta}}
\eea
with
\be
\delta\Sigma^Z_\eta
=
\frac{3\lambda^2}{2\pi}g\,\omega\left[\frac{{\rm sin}\,\omega\eta}{\omega\eta}-\frac{{\rm Si}(\omega\eta)}{\omega\eta}\right]\:\:.
\ee
Since $\delta\Sigma^Z_\eta$ is only important for \mbox{$\eta\gtrsim 1/\omega$}, this can be written as
\bea
\delta\sigma^{\omega<T_0}
&\approx&
\frac{e^2}{2\pi^3}
\int\limits_0^{T_0}d\omega\:\frac{\partial}{\partial\omega}\left[\omega\:{\rm coth}\frac{\omega}{2T}\right]
\left[\frac{1}{\omega}
\int\limits_0^\infty dx\:
\ln{\frac{1}{1-\displaystyle\frac{{\rm sin}\,x}{x}}}
+\int\limits_{1/\omega}^{\eta_{\rm max}}d\eta\:\frac{\frac{3}{2\pi}g\omega\:{\rm Si}(\omega\eta)}{\omega\eta\Sigma^Z_\infty}
\right]\nonumber\\
&\approx&
\frac{e^2}{2\pi^3}
\int\limits_0^{T_0}d\omega\:\frac{\partial}{\partial\omega}\left[\omega\:{\rm coth}\frac{\omega}{2T}\right]
\left[\frac{1}{\omega}\:
c_1
+\int\limits_{1/\omega}^{\eta_{\rm max}}\frac{d\eta}{\eta}\:\frac{\pi/2}{\omega\left(1+\ln{\displaystyle\frac{T_0}{\omega}}\right)}
\right]\nonumber\\
&=&
\frac{e^2}{2\pi^3}
\int\limits_0^{T_0}d\omega\:\frac{\partial}{\partial\omega}\left[\omega\:{\rm coth}\frac{\omega}{2T}\right]
\left[\frac{1}{\omega}\:
c_1
+\frac{\pi\ln{g}}{2\,\omega\left(1+\ln{\displaystyle\frac{T_0}{\omega}}\right)}
\right]\nonumber\\
&=&
\frac{e^2}{2\pi^3}
\left[
c_1\cdot\ln{\frac{T_0}{T}}
+\frac{\pi}{2}\ln{g}\ln{\left(\ln{\frac{T_0}{T}}+1\right)}
\right]\:\:,
\eea
resulting in
Eq.~(\ref{delta_sigma_lowfreq}).
Here $\eta_{\rm max}$ is given by the low-$k$ cutoff which is set by the dressed boxes containing extra diffusons, which occurs on the scale \mbox{$Dq^2\sim\Sigma^Z_\infty$}, so that \mbox{$\eta_{\rm max}\sim 1/\left[q^2l^2 T_0\right]\sim 1/\left[\Sigma^Z_\infty\tau T_0\right]\sim 1/\left[g\omega\tau T_0\ln(T_0/\omega)\right]$}.
The numerical constant $c_1$ is given by the integral
\be
c_1=\int\limits_0^\infty dx\:
\ln{\frac{1}{1-\displaystyle\frac{{\rm sin}\,x}{x}}}\:\:.
\label{integral-c1}
\ee

\section{Effective interaction box for unscreened Coulomb interaction}
\label{App_interpolation}
\renewcommand{\theequation}{G.\arabic{equation}}
\setcounter{equation}{0}

In this Appendix we calculate the effective interaction $\tilde{U}$, given by the diagrams in Fig.~\ref{boxes}, for the less singular gauge field propagator (\ref{less_singular_U}) corresponding to unscreened Coulomb interaction between the fermions. As mentioned in the main text, relevant transferred momenta are now \mbox{$k\in\left[l^{-1},k_F\right]$}, so that we only need the bare interaction box. Neglecting the diffusive momentum $q$ and the energy transfer $\epsilon$ through the gauge field line, the effective interaction is
\bea
\tilde{U}
&=&
\frac{1}{\left(2\pi\nu\tau\right)^2}\:
\int(dk)\:
\frac{e^2}{\chi_0\kappa k}
\int(dp)\:v_x^2{\rm sin}^2\phi\:G^R({\bf p})G^R({\bf p-k})G^A({\bf p})G^A({\bf p-k})\nonumber\\
&=&
\frac{1}{\left(2\pi\nu\tau\right)^2}\:
\int(dk)\:
\frac{e^2}{\chi_0\kappa k}
\int(dp)\:v_x^2{\rm sin}^2\phi\:\tau^2\left[G^R({\bf p})G^A({\bf p-k})+G^A({\bf p})G^R({\bf p-k})\right]\nonumber\\
&=&
\frac{1}{\left(2\pi\nu\tau\right)^2}\:
\int(dk)\:
\frac{e^2v_F^2}{\chi_0\kappa k}\:
2\tau^2\:
2\pi i\nu
\int\frac{d\phi}{2\pi}\:\frac{{\rm sin}^2\phi}{(p_Fk/m)\,{\rm cos}\phi+i/\tau}\nonumber \\
&=&
\frac{1}{\left(2\pi\nu\tau\right)^2}\:
\int(dk)\:
\frac{e^2v_F^2}{\chi_0\kappa k}\:
\frac{4\pi\nu\tau^3}{1+\sqrt{1+k^2l^2}}.
\label{interpolation}
\eea
Performing the momentum integration, we get, within the logarithmic accuracy,
\be
\tilde{U}
\approx
\frac{1}{\left(2\pi\nu\tau\right)^2}\:4\pi\nu\tau^3\:\frac{e^2v_F^2}{2\pi\chi_0\kappa l}
\int\limits_{l^{-1}}^{k_F}\frac{dk}{k}
=
\frac{12g}{\pi\nu\kappa l}\ln{g},
\ee
\end{widetext}
which is Eq.~(\ref{tilde-U_for_long_range}) of the main text. The last factor in Eq.~(\ref{interpolation}), interpolating between small and large $k$, is the precise version (for particular disorder scattering model) of the interpolation factor in Eq.~(21) of Ref.~\onlinecite{Woelfle_2000}, where it was derived for the dephasing action.


\begin{thebibliography}{1}

\bibitem{reizer}
T.~Holstein, R.E.~Norton, and P.~Pincus,
Phys.~Rev.~B {\bf 8}, 2649 (1973);
M.~Yu.~Reizer, Phys.~Rev.~B {\bf 40}, 11571 (1989);
G.~Baym, H.~Monien, C.~J.~Pethick, and D.~G.~Ravenhall,
Phys.~Rev.~Lett.\ {\bf 64}, 1867 (1990).

\bibitem{gauge}
D.~V.~Khveshchenko and P.~C.~E.~Stamp, Phys.~Rev.~Lett.\ {\bf 71}, 2118 (1993);
Phys. Rev. B \ {\bf 49}, 5227 (1994);
B.~L.~Altshuler, L.~B.~Ioffe, and A.~J.~Millis, Phys.~Rev.~B {\bf 50}, 14048 (1994);
Y.~B.~Kim, A.~Furusaki, X.-G.~Wen, and P.~A.~Lee,
Phys.~Rev.~B {\bf 50}, 17917 (1994);
A.~Stern and B.~I.~Halperin, Phys.~Rev.~B {\bf 52}, 5890 (1995).

\bibitem{htsc}
G.~Baskaran and P.~W.~Anderson, Phys.~Rev.~B {\bf 37}, 580 (1988);
L.~B.~Ioffe and A.~I.~Larkin, Phys.~Rev.~B {\bf 39}, 8988 (1989);
N.~Nagaosa and P.~A.~Lee, Phys.~Rev.~Lett.\ {\bf 64}, 2450 (1990);
P.~A.~Lee and N.~Nagaosa, Phys.~Rev.~B {\bf 46}, 5621 (1992).

\bibitem{Lopez_Fradkin_91} A.~Lopez and E.~Fradkin,
Phys.~Rev.~B {\bf 44}, 5246 (1991).

\bibitem{Jain_89} J.~K.~Jain,
Phys.~Rev.~Lett.\ {\bf 63}, 199 (1989).

\bibitem{HLR_93} B.~I.~Halperin, P.~A.~Lee, and N.~Read,
Phys.~Rev.~B {\bf 47}, 7312 (1993).

\bibitem{Lopez_Fradkin_review_97} A.~Lopez and E.~Fradkin,
in {\it Composite Fermions}, edited by O.~Heinonen,
World Scientific (1998)
\mbox{[cond-mat/9704055]}.

\bibitem{Simon_review_98} S.~Simon,
in {\it Composite Fermions}, edited by O.~Heinonen,
World Scientific (1998)
\mbox{[cond-mat/9812186]}.

\bibitem{Aronov_Mirlin_Woelfle_93}
A.~G.~Aronov, A.~D.~Mirlin, and P.~W\"olf\/le,
Phys.~Rev.~B {\bf 49}, 16609 (1994).

\bibitem{Aronov_Woelfle_PRL_94}
A.~G.~Aronov and P.~W\"olf\/le,
Phys.~Rev.~Lett.\ {\bf 72}, 2239 (1994).

\bibitem{Aronov_Woelfle_94}
A.~G.~Aronov and P.~W\"olf\/le,
Phys.~Rev.~B {\bf 50}, 16574 (1994).

\bibitem{Aronov_Altshuler_Mirlin_Woelfle_EPL_95}
A.~G.~Aronov, E.~Altshuler, A.~D.~Mirlin, and P.~W\"olf\/le,
Europhys.\ Lett.\ {\bf 29}, 239 (1995).

\bibitem{Aronov_Altshuler_Mirlin_Woelfle_PRB_95}
A.~G.~Aronov, E.~Altshuler, A.~D.~Mirlin, and P.~W\"olf\/le,
Phys.\ Rev.\ B {\bf 52}, 4708 (1995).

\bibitem{Stern_Halperin_95} A.~Stern and B.~I.~Halperin,
Phys.\ Rev.\ B {\bf 52}, 5890 (1995).

\bibitem{Mirlin_Altshuler_Woelfle_95}
A.~D.~Mirlin, E.~Altshuler, and P.~W\"olf\/le,
Annalen der Physik {\bf 5}, 281 (1996).

\bibitem{Lee_Mucciolo_Smith_96}
P.~A.~Lee, E.~R.~Mucciolo, and H.~Smith,
Phys.\ Rev.\ B {\bf 54}, 8782 (1996).

\bibitem{Mirlin_Woelfle_97} A.~D.~Mirlin and P.~W\"olf\/le,
Phys.~Rev.~B {\bf 55}, 5141 (1997).

\bibitem{Woelfle_2000} P.~W\"olf\/le,
Foundations of Physics {\bf 30}(12), 2125 (2000).

\bibitem{TL_diss_06} T.~Ludwig,
Ph.D.~thesis, Universit\"at Karlsruhe (2006).

\bibitem{Khveshchenko_96}
D.~V.~Khveshchenko,
Phys.\ Rev.\ Lett.\ {\bf 77}, 362 (1996).

\bibitem{Rokhinson_Su_Goldman_95}
L.~P.~Rokhinson, B.~Su, and V.~J.~Goldman,
Phys.\ Rev.\ B {\bf 52}, R11588 (1995).

\bibitem{Galitski_05} V.~M.~Galitski,
Phys.~Rev.~B {\bf 72}, 214201 (2005).

\bibitem{Gornyi_Mirlin_Woelfle_01}
I.~V.~Gornyi, A.~D.~Mirlin, and P.~W\"olf\/le,
Phys.~Rev.~B {\bf 64}, 115403 (2001).

\bibitem{Altshuler_Aronov_85}
B.~L.~Altshuler and A.~G.~Aronov, in
{\it Electron-Electron Interaction in Disordered Conductors},
edited by A.~L.~Efros and M.~Pollak, pp.~1-153
(Elsevier, Amsterdam, 1985).

\bibitem{Finkelstein}
A.~M.~Finkel'stein, Zh.~Eksp.~Teor.~Fiz.\ {\bf 84}, 168 (1983) [Sov.~Phys.~JETP {\bf 57}, 97 (1983)];
A.~M.~Finkel'stein, Z.~Phys.~B {\bf 56}, 189 (1984);
A.~M.~Finkel'stein, {\it Electron Liquid in Disordered Conductors}, Vol.~14 of Soviet Scientific Reviews, edited by I.~M.~Khalatnikov (Harwood, London, 1990);
A.~Punnoose and A.~M.~Finkel'stein, Phys.~Rev.~Lett.\ {\bf 88}, 016802 (2001).


\bibitem{Kamenev} A.~Kamenev and A.~Andreev, Phys.\ Rev.\ B {\bf 60}, 2218
(1999).

\bibitem{ZNA} G.~Zala, B.N.~Narozhny, and I.L.~Aleiner,
  Phys. Rev. B {\bf 64}, 214204 (2001).

\bibitem{Gornyi_Mirlin_04}
I.~V.~Gornyi and A.~D.~Mirlin,
Phys.~Rev.~B {\bf 69}, 045313 (2004).

\bibitem{Adamov} Y.~Adamov, I.V.~Gornyi, and A.D.~Mirlin,
Phys. Rev. B \textbf{73}, 045426 (2006)

\bibitem{Rukhadze_Silin_61}
A.~A.~Rukhadze and V.~P.~Silin,
Soviet Physics Uspekhi {\bf 4}, 459 (1961).

\bibitem{AAK_82}
B.~L.~Altshuler, A.~G.~Aronov, and D.~E.~Khmelnitsky,
J.~Phys.~C {\bf 15}, 7367 (1982).
Some errors in numerical coefficients are corrected in Ref.~\onlinecite{AAG_99}.

\bibitem{Stern_Aharonov_Imry_90}
A.~Stern, Y.~Aharonov, and Y.~Imry,
Phys.~Rev.~A {\bf 41}, 3436 (1990).

\bibitem{Gorkov_Larkin_Khmelnitskii_79}
L.~P.~Gor'kov, A.~I.~Larkin, and D.~E.~Khmel'nitskii,
Pis'ma Zh.\ Eksp.\ Teor.\ Fiz.\ {\bf 30}(4), 248 (1979)
[JETP Lett.\ {\bf 30}(4), 228 (1980)].

\bibitem{Altshuler_Khmelnitskii_Larkin_Lee_80}
B.~L.~Altshuler, D.~Khmel'nitzkii and A.~I.~Larkin, and P.~A.~Lee,
Phys.~Rev.~B {\bf 22}, 5142 (1980).

\bibitem{Chakravarty_Schmid_86}
S.~Chakravarty and A.~Schmid,
Phys.~Rep.\ {\bf 140}(4), 193 (1986).

\bibitem{Narozhny_Aleiner_Stern_2001}
B.~N.~Narozhny, I.~L.~Aleiner, and A.~Stern,
Phys.~Rev.~Lett.\ {\bf 86}, 3610 (2001).

\bibitem{AAG_99}
I.~L.~Aleiner, B.~L.~Altshuler, and M.~E.~Gershenson,\\
Waves Random Media {\bf 9}, 201 (1999).

\bibitem{GMP07}
I.V.~Gornyi, A.D.~Mirlin, and D.G.~Polyakov, Phys.\ Rev.\ B
{\bf 75}, 085421 (2007).

\bibitem{vonDelft_etal_05}  F.~Marquardt, J.~von~Delft, R.A.~Smith, and V.~Ambegaokar, Phys. Rev. B {\bf 76}, 195331 (2007);
J.~von~Delft, F.~Marquardt, R.A.~Smith, and V.~Ambegaokar, {\it ibid} 195332 (2007).

\bibitem{Ludwig_Mirlin_2004}
T.~Ludwig and A.~D.~Mirlin,
Phys.~Rev.~B {\bf 69}, 193306 (2004).

\bibitem{Altshuler_85} B.~L.~Altshuler,
Pis'ma Zh.\ Eksp.\ Teor.\ Fiz.\ {\bf 41}, 530 (1985)
[JETP Lett.\ {\bf 41}, 648 (1985)].

\bibitem{Stone_85} A.~D.~Stone,
Phys.~Rev.~Lett.\ {\bf 54}, 2692 (1985).

\bibitem{Lee_Stone_85} P.~A.~Lee and A.~D.~Stone,
Phys.~Rev.~Lett.\ {\bf 55}, 1622 (1985).

\bibitem{Altshuler_Shklovskii_86}
B.~L.~Altshuler, and B.~I.~Shklovskii,
Zh.\ Eksp.\ Teor.\ Fiz.\ {\bf 91}, 220 (1986)
[Sov.\ Phys.\ JETP {\bf 64}, 127 (1986)].

\bibitem{Lee_Stone_Fukuyama_87}
P.~A.~Lee, A.~D.~Stone, and H.~Fukuyama,
Phys.~Rev.~B {\bf 35}, 1039 (1987).

\bibitem{Kane_Serota_Lee_88}
C.~L.~Kane, R.~A.~Serota, and P.~A.~Lee,
Phys.~Rev.~B {\bf 37}, 6701 (1988).

\bibitem{Aleiner_Blanter_2002}
I.~L.~Aleiner and Ya.~M.~Blanter,
Phys.~Rev.~B {\bf 65}, 115317 (2002).

\bibitem{Texier_Montambaux_2005}
C.~Texier and G.~Montambaux,
Phys.~Rev.~ B {\bf 72}, 115327 (2005).

\bibitem{Polyakov_Samokhin_98}
D.~G.~Polyakov and K.~V.~Samokhin,
Phys.~Rev.~Lett.\ {\bf 80}, 1509 (1998)
and private communication.

\bibitem{Kee_Aleiner_Altshuler_98}
H.-Y.~Kee, I.~L.~Aleiner, and B.~L.~Altshuler,
Phys.~Rev.~B {\bf 58}, 5757 (1998).

\bibitem{Narozhny02}
B.~N.~Narozhny, G.~Zala, and I.~L.~Aleiner,
Phys.~Rev.~B \textbf{65}, 180202(R) (2002).

\bibitem{jing91}
T.~W.~Jing, N.~P.~Ong, T.~V.~Ramakrishnan, J.~M.~Tarascon, and K.~Remschnig,
Phys.~Rev.~Lett.\ \textbf{67}, 761 (1991).

\bibitem{CF_data}
R.~R.~Du, H.~L.~Stormer, D.~C.~Tsui, L.~N.~Pfeiffer, and K.~W.~West,
Solid State Commun.\ {\bf 90}, 71 (1994); Phys.~Rev.~Lett.\ {\bf 70}, 2944 (1993);  V.~J.~Goldman, B.~Su, and J.~K.~Jain,
Phys.~Rev.~Lett.~{\bf 72}, 2065 (1994); R.~R.~Du, H.~L.~Stormer, D.~C.~Tsui, A.~S.~Yeh, L.~N.~Pfeiffer, and K.~W.~West, Phys.~Rev.~Lett.\ {\bf 73}, 3274 (1994); D.~R.~Leadley, R.~J.~Nicholas, C.~T.~Foxon, and J.~J.~Harris, Phys.~Rev.~Lett.\ {\bf 72}, 1906 (1994); D.~R.~Leadley, M.~van~der~Burgt, R.~J.~Nicholas, C.~T.~Foxon, and J.~J.~Harris, Phys.~Rev.~B {\bf 53}, 2057 (1996); P.~T.~Coleridge, Z.~W.~Wasilewski, P.~Zawadzki, A.~S.~Sachrajda, and H.~A.~Carmona, Phys.~Rev.~B {\bf 52}, R11603 (1995).

\bibitem{Altshuler_Aronov_Larkin_Khmelnitskii_81}
B.~L.~Al'tshuler, A.~G.~Aronov, A.~I.~Larkin, and D.~E.~Khmel'nitskii,
Zh.~Eksp.~Teor.~Fiz.~{\bf 81}, 768 (1981)
[Sov.~Phys.~JETP {\bf 54}(2), 411 (1981)].

\end{thebibliography}
\end{document}